\ifdefined\ACM

\documentclass[manuscript,screen,review]{acmart}

\else
 \documentclass[12pt]{extarticle}
\usepackage{cuted}%
\ifdefined \POSTER
  \else
\usepackage[utf8]{inputenc}
\ifdefined \bookflag
  \usepackage[a4paper, marginparwidth=100pt, total={5in, 9in}]{geometry}
\else
\usepackage[a4paper, total={6in, 9in}]{geometry}
\fi
\fi

\ifdefined \POSTER \else \usepackage{braket} \fi
\usepackage{xcolor}
\usepackage{amsmath}
\usepackage{amsfonts}
\usepackage{amsthm}
\usepackage{amssymb}
\ifdefined \POSTER \else \usepackage{titling} \fi
\usepackage{hyperref}
\hypersetup{
    colorlinks=true,
    citecolor=blue %
}

\usepackage{authblk}
\usepackage{framed}
\usepackage{aliascnt}
\usepackage{cleveref}
\usepackage{graphicx}
\usepackage{svg}
\usepackage{float}
\usepackage{wrapfig}
\usepackage{tikz}
\usetikzlibrary{quantikz2}
\usetikzlibrary{patterns, shapes.arrows}
\usepackage{adjustbox}
\usepackage{tkz-graph}
\usetikzlibrary{decorations.pathmorphing}
\usepackage{tkz-berge}
\usepackage[linesnumbered]{algorithm2e}
\usepackage{multicol}
\usepackage[backend=biber,style=alphabetic,sorting=ynt]{biblatex}
\usepackage{pgfplots}
\usepackage{setspace}
\usepackage{etoc}
\usepackage{pgfgantt}
\usepackage{pdfpages}
\usepackage{subcaption}
\ifdefined \POSTER \else \usepackage[activate={true,nocompatibility}, final, tracking=true, kerning=true, spacing=true, factor=1100, stretch=10, shrink=10]{microtype} \fi

\usepackage{colonequals}
\usepackage{pifont}
\usepackage{booktabs}

\ifdefined \POSTER \else \DeclareUnicodeCharacter{2212}{−} \fi
\usepgfplotslibrary{groupplots,dateplot}
\pgfplotsset{compat=newest}

\newcommand{\mynewtheorem}[2]{
    \newaliascnt{#1}{theorem}
    \newtheorem{#1}[#1]{#2}
    \aliascntresetthe{#1}
}
\newcommand{\mynewtheoremstar}[2]{  
  \newtheorem*{#1}{#2}

}
\ifdefined \POSTER \else \newtheorem{theorem}{Theorem}[section] \fi
\ifdefined \POSTER \else \mynewtheorem{corollary}{Corollary} \fi
\ifdefined \POSTER \else \mynewtheorem{definition}{Definition} \fi
\ifdefined \POSTER \else \mynewtheorem{example}{Example} \fi
\ifdefined \POSTER \else \mynewtheorem{claim}{Claim} \fi
\ifdefined \POSTER \else \mynewtheorem{proposition}{Proposition} \fi
\ifdefined \POSTER \else  \fi
\ifdefined \POSTER \else \mynewtheoremstar{claim*}{Claim} \fi
\ifdefined \POSTER \else \mynewtheorem{fact}{Fact} \fi
\ifdefined \POSTER \else \mynewtheorem{remark}{Remark} \fi
\ifdefined \POSTER \else \mynewtheoremstar{theorem*}{Theorem} \fi
\ifdefined \POSTER \else \mynewtheorem{lemma}{Lemma} \fi
\ifdefined \POSTER \else \mynewtheorem{conjecture}{Conjecture} \fi
\ifdefined \POSTER \else \mynewtheorem{exercise}{Exercise} \fi
\ifdefined \POSTER \else \mynewtheorem{solution}{Solution} \fi
\ifdefined \POSTER \else \mynewtheorem{openproblem}{Open problem} \fi
\ifdefined \POSTER \else \crefname{lemma}{Lemma}{Lemmas} \fi
\ifdefined \POSTER \else \Crefname{algocf}{Algorithm}{Algorithms} \fi
\ifdefined \POSTER \else \Crefname{sub@algcof}{Algorithm}{Algorithms} \fi
\ifdefined \POSTER \else \hypersetup{colorlinks=true} \fi

\ifdefined \POSTER \else \mynewtheorem{problem}{Problem} \fi
\ifdefined \POSTER \else \crefname{problem}{Problem}{Problems} \fi

\ifdefined \POSTER \else  \fi
\ifdefined \POSTER \else \crefname{oproblem}{Open-Problem}{Open Problems} \fi

\usepackage{cancel}
\usepackage{tikz-3dplot}
\usepackage{subcaption}
\fi
\usepackage{physics} 

\begin{document}
\newcommand{\commentt}[1]{\textcolor{blue}{ \textbf{[COMMENT]} #1}}
\newcommand{\inr}[1]{\textcolor{red}{ #1}}
\newcommand{\inb}[1]{{\leavevmode\color{blue}#1}}
\newcommand{\ing}[1]{{\leavevmode\color{green}#1}}
\newcommand{\iny}[1]{{\leavevmode\color{yellow}#1}}
\newcommand{\inbr}[1]{{\leavevmode\color{brown}#1}}
\newcommand{\ctt}[1]{\commentt{#1}}
\newcommand{\prb}[1]{ \mathbf{Pr} \big[ #1 \big]}
\newcommand{\prbm}[2]{ \mathbf{Pr}_{ #2 }\big[ #1 \big]}
\newcommand{\prbc}[3]{ \mathbf{Pr}_{ #2 }\big[ #1 | #3 \big]}
\newcommand{\prbcprb}[3]{ \prbc{#2}{#1}{#3} \cdot \prb{#3} } 
\newcommand{\expp}[1]{ 
 \mathbf{E}\ifx\@currenvir\enumerate[{#1}]\else \left[ {#1} \right]\fi}

\newcommand{\exppm}[2]{ \mathbf{E}_{#2}\left[ {#1} \right]}
\newcommand{\onotation}[1]{\(\mathcal{O} \left( {#1}  \right) \)}
\newcommand{\ona}[1]{\onotation{#1}}
\newcommand{\PSI}{{\ket{\psi}}}
\newcommand{\xij} { X_{ij} } 
\newcommand{\Cpa}{[n, \rho n, \delta n]}
\newcommand{\Jvv}{ \bar{J_{v}} } 
\newcommand{\Cvv}{ \tilde{C_{v}} } 

\newcommand{\Gz}{ G_{z}^{\delta} } 
\newcommand{ \Tann } {  \mathcal{T}\left( G, C_0 \right) }
\newcommand{\ireducable}{ireducable \hyperref[ire]{[\ref{ire}]} }
\newcommand{\cutUU}{E(U_{-1} \bigcup U_{+1} ,U)} 
\newcommand{\wcutUU}{w\left( E(U_{-1} \bigcup U_{+1} ,U)  \right)}
\newcommand{\testgo}{  \mathcal{T}\left(J, q , C_{0}\right) } 

\newcommand{\duC}{\left( C_{A}^{\perp}\otimes C_{B}^{\perp} \right)^{\perp}}
\newcommand{\duduC}{\left( C_{A}\otimes C_{B}\right)^{\perp}}
  
\newcommand*{\Czdu}{C_{Z}^\perp }

\newcommand*{\EPR}{\ket{\mathbf{EPR}}}
\newcommand*{\EPRD}{\bra{\mathbf{EPR}}}

\newcommand*{\image}{ \text{Im }  }

\newcommand*{\Trr}{\mathbf{Tr} }
\newcommand*{\Su}{\text{Supp } }

\title{ Local Quantum Fault Tolerance Without Expansion  }
\author{David Ponarovsky\footnote{David.Ponarovsky@mail.huji.ac.il}}
\affil{School of Computer Science, Hebrew University}
\maketitle

\newcommand*{\xb}{\mathbf{x}}

\ifdefined\SWEEPHEA
\else 
  \newcommand*{\SWEEPHEA}{}
\newcommand{\History}[1]{H_{#1}}
  \newcommand{\Time}[1]{T(#1)}
  \newcommand{\Excuse}[1]{\mathbf{Excuse}(#1)}
  \newcommand{\Excusei}[1]{\mathbf{Excuse}_{i}(#1)}
  \newcommand{\Size}[1]{\mathbf{Size}(#1)}
  \newcommand{\Span}[1]{\mathbf{Span}(#1)}
  \newcommand{\pre}[1]{\text{pre}(#1)}
  \newcommand{\poli}[1]{\text{pole}_{i}(#1)}
  \newcommand{\prei}[1]{\text{pre}_{i}(#1)}
  \newcommand{\predd}[1]{\text{pre}_{d+1}(#1)}
  \newcommand{\rotf}[2]{\theta_{#1 ,#2}}
  \newcommand*{\rotfs}{\rotf{v}{S^{\prime}}}
  \newcommand{\rotfv}[2]{\mathbf{\rotf{#1}{#2} }}
  \newcommand*{\rotfvs}{\mathbf{\rotfs{} }}
\fi

\begin{abstract}
We study fault-tolerant quantum computation with a \emph{fixed, local error-correction rule} and a geometrically simple code. Our construction uses the cubical $(d,d')$-Toric code, with $d',d-d'\geq 2$ (in particular, the 4D Toric code), and tolerates sufficiently weak local-stochastic  noise in the circuit model. Classical processing is assumed noiseless and has bounded depth per physical time step. Unlike constructions whose local decoders require a hierarchy of auxiliary classical states and scale-dependent communication delays, our error-correction cycle acts directly on the current syndrome, has constant depth, and is repeated unchanged during idle logical computation.

Our proof works with errors and syndromes as chains in a cubical chain complex. It combines geometric shrinking under the sweep rule with a space--time explanation-tree analysis inspired by Toom's theorem, without relying on an expansion property of the code complex. We establish three results:
\begin{enumerate}
    \item \textbf{An explicit higher-dimensional sweep rule.} We give a local procedure for choosing the sweep correction on cubical $(d,d')$-Toric complexes, resolving the higher-dimensional local-existence question of Kubica and Preskill \cite{Kubica_2019}.
    \item \textbf{Fault-tolerant computation with a simple code and local correction.} We prove a circuit-level threshold theorem for repeated noisy sweep cycles interleaved with logical operations. In particular, the 4D Toric code supports a fixed, constant-depth syndrome-based correction cycle without a multiscale classical decoder or expansion-based encoding. The construction is not claimed to have constant total-depth overhead.
    \item \textbf{Ideal decoding under local-stochastic noise.} Starting from a local-stochastic Pauli error of rate $p$ below a threshold, $n^{d+1}$ \emph{noiseless} sweep cycles on a side-length-$n$ $(d,d')$-Toric code recover the encoded state with probability at least $1-O(\operatorname{poly}(n))p^{\Omega(n)}$. This extends earlier sweep-decoding guarantees for independent errors \cite{Kubica_2019}.
\end{enumerate}
\end{abstract}

\tableofcontents
\newpage
\section{Introduction}

\subsection{Background and motivation}

A fault-tolerant quantum computer must protect its logical state while gates and error correction are themselves subject to faults. An important design choice is not just the quantum code but the \emph{classical and quantum mechanism used to keep errors under control}. A conceptually attractive target is a regular code for which each correction round consists of the same bounded-depth local procedure, determined only by the syndrome currently observed. Such a procedure neither stores a growing decoding history nor organizes the correction into successively larger scales.

The quantum threshold theorem established the feasibility of fault-tolerant computation below a constant noise threshold \cite{aharonov1999faulttolerant,Ninequ}. A major subsequent direction reduced the qubit overhead using high-rate quantum LDPC codes and, in several constructions, expander-based complexes \cite{gottesman2014faulttolerant,grospellier:tel-03364419,Fawzi_2018,nguyen2025quantumfaulttoleranceconstantspace}. Expansion supplies powerful control of errors and their syndromes, but it is a substantive structural assumption. We pursue a complementary question: \emph{Can a simple, nonexpanding topological code support a local correction cycle within a rigorous fault-tolerant computation?} We answer this question affirmatively for the cubical $(d,d')$-Toric code when $d',d-d'\geq 2$. We do not aim to optimize the total depth or the space overhead of the simulation.

The Toric code is particularly suited to this question. Its qubits, stabilizer checks, and logical operators are described by the cells of a regular cubical complex. Unlike a code whose analysis requires expansion, it admits a direct topological description: errors are chains, their syndromes are boundaries, and logical faults correspond to nontrivial homology classes. Our proof exploits this description rather than trying to maintain a strong probabilistic invariant on the entire physical error after every round. The chain-complex viewpoint is especially useful in proving the correctness\footnote{Namely, that the `bottom-left` corner of the error can suggest a flip that satisfies the sweep-rule constraints.} of the sweep rule: treating the local syndrome as a boundary chain, we construct a correction chain whose boundary matches the prescribed syndrome and whose support satisfies the local sweep constraints. An induction on face dimension yields an explicit algorithm, establishing the higher-dimensional existence claim of Kubica and Preskill.

\paragraph{Toom-type recovery for the 4D Toric code.}
Local recovery for the 4D Toric code was already proposed in the literature. Dennis, Kitaev, Landahl, and Preskill \cite{Dennis_2002} discussed anisotropic local dynamics inspired by Toom's rule. Ahn studied 4D Toom correction \cite{Ahn2004}; we discuss gaps in her proof in a companion paper. More recently, Drechsler, Oreg, and Aharonov \cite{DrechslerOregAharonov2026} studied quantum memories in an asynchronous framework; their 4D Toom-memory application relies on Ahn's threshold analysis. 

Kubica and Preskill \cite{Kubica_2019} introduced the \emph{sweep rule}, extending Toom's oriented local dynamics to topological codes, and proved ideal-decoding results for independent noise. Their formulation also identifies an existence question for certain higher-dimensional local corrections. We work with the ordinary cubical Toric complex, give an explicit local implementation of the sweep rule, and analyze it both under continuously occurring faults and as a noiseless decoder for initially correlated local-stochastic errors.

\paragraph{Related local fault-tolerant architectures and their classical hardware.}
Other cellular-automaton approaches include the works of Balasubramanian, Davydova, and Lake and of D\"unnweber, Styliaris, and Trivedi \cite{Balasubramanian_2026,dünnweber2026quantummemoryautonomouscomputation}. In particular, \emph{Selub, Bhardwaj, and Lake} \cite{SelubBhardwajLake2026} have already achieved fault-tolerant quantum computation based on a \emph{local decoder for a topological code}: their two-dimensional surface-code architecture has time-translation-invariant local decoding dynamics and bounded-speed classical communication. Its constant-resource-density realization, however, implements a multiscale hierarchy with auxiliary classical state and classical wires whose lengths grow with the decoding scale. The propagation delays along those wires \emph{function as distributed timers}: rather than storing a large timer in each processor, the hardware uses physical signal travel to implement the longer periods of the higher-level updates. For this scheduling task, a long delay line is the hardware analogue of a stored timer: the requisite waiting time is encoded in wire length rather than counter bits. Thus the timing resource is relocated into spatial communication hardware, not eliminated. This is an architectural tradeoff, not a claim that their decoder has growing local Boolean-circuit depth.

\emph{Our correction cycle requires neither such distributed timers nor a multiscale communication network.} Each local correction is a fixed function of the presently measured syndrome on a constant-size neighborhood, and all locations execute the same finite schedule in each sweep cycle. In particular, an idle logical identity is implemented by repeated copies of this cycle, with no persistent decoder messages or hierarchy-dependent clocks. We emphasize the precise scope of the comparison: these statements concern the \emph{repeated error-correction procedure on the native 4D Toric lattice}. They do not assert a fully autonomous or clock-free device, remove classical processing for logical gates, or eliminate the potential physical cost of embedding four-dimensional connectivity in lower-dimensional hardware. In this respect, the 2D physical locality achieved by \cite{SelubBhardwajLake2026} is a genuine advantage of their construction.

These results also sharpen our mathematical contribution. One can obtain a local decoder for a nonexpanding topological code by supplying additional hierarchical decoding hardware. Our result shows that \emph{expansion is unnecessary even when the correction rule is a single fixed, local, syndrome-based map without such a hierarchy}, provided the code geometry is allowed to be four-dimensional. The proof mechanism is different from expansion-based estimates: it tracks the combinatorial histories of syndromes viewed as boundaries in a chain complex.

\subsection{Our results}

\paragraph{Fault-tolerant computation with repeated local correction.}
Fix integers $d,d'$ satisfying $d',d-d'\geq 2$. The $(d,d')$-Toric code is the CSS code associated with the cubical chain groups of a side-length-$n$ $d$-dimensional torus. Each encoded block undergoes a constant-depth \emph{sweep cycle}, consisting of parallel local updates in the two CSS sectors. The update rule and the finite sequence making up a cycle are independent of the elapsed time, the block size, and the number of preceding correction rounds.

We use the circuit model with local-stochastic Pauli noise on quantum locations and reliable classical bits and logic. More precisely, if $\mathcal N_{\mathcal E}$ is a distribution over fault sets with rate $p$, then for every predetermined finite set $S$ of space--time locations,
\begin{equation*}
   \prbm{S\subseteq\operatorname{Loc}(\mathcal E)}{\mathcal E\sim\mathcal N_{\mathcal E}}\leq p^{|S|}.
\end{equation*}
This condition allows correlations between faults; independence is not assumed. At any physical time step, only constant-depth noiseless classical computation is allowed.

\begin{theorem*}[Informal version of \Cref{theorem:main}]
There is a constant $p_0>0$ such that for local-stochastic Pauli noise of rate $p<p_0$, every ideal circuit of width $w$ and depth $\mu$ has a 4D-Toric-code-encoded simulation using side-length-$n$ blocks whose logical failure probability is at most
\begin{equation*}
  O\!\left(\operatorname{poly}(w,\mu,n)\right)p^{\Omega(n)}.
\end{equation*}
For sufficiently large $n$, the encoded circuit has depth $\Theta\!\bigl(\mu n+\operatorname{polylog}n\bigr)$ and width $\Theta\!\bigl((w+w\mu)n^8\bigr)$. After fault-tolerant initialization, the encoded blocks use the 4D Toric code, and each intervening correction cycle has constant depth and applies the same local sweep rule.
\end{theorem*}

The stated costs include the implementation of logical operations; in particular, gate teleportation may need a classical decoding computation of nonconstant total duration. Our claim of constant depth is therefore \emph{about the recurring correction cycle}, not the entire simulation of each logical gate. Unlike the local-decoding constructions in which code expansion is the essential geometric input, our theorem uses a nonexpanding cubical chain complex. The correction dynamics require no scale-dependent auxiliary control network.

\paragraph{An explicit higher-dimensional sweep rule.}
The sweep rule, introduced by Kubica and Preskill \cite{Kubica_2019}, is an oriented local correction procedure for topological codes. At each vertex $v$, it reads the syndrome $\sigma$ on the positive $(d'-1)$-faces rooted at $v$ and searches for a correction chain $\tilde z$ supported on the positive $d'$-faces rooted there, satisfying
\begin{equation*}
  (\partial^-\tilde z)|_{v+}=\sigma|_{v+},
  \qquad
  \mathbf{Str}_{0+}(v,\tilde z)\subseteq\mathbf{Str}_{0+}(v,\sigma|_{v}).
\end{equation*}
The first condition ensures that the correction cancels the syndrome seen at $v$, while the second confines its positive vertex support to that of the observed syndrome. If an admissible chain is found, the rule flips its qubits; otherwise, no correction is applied at $v$. All vertices execute this fixed local rule in parallel.
The nontrivial question is whether the boundary and support conditions can be satisfied simultaneously in higher dimensions. We prove the following local-existence claim:
\begin{claim*}[Informal version of \Cref{claim:findcor}]
  For $d^{\prime}\ge 2$, whenever a $d'$-chain $z$ rooted at $v$ satisfies $(\partial^- z)|_{v+}=\sigma|_{v+}$, there is also a rooted $d'$-chain $\tilde z$ satisfying \emph{both} sweep-rule conditions above. 
\end{claim*}

Our inductive proof gives an explicit algorithm for constructing $\tilde z$, rather than searching over all local corrections. In fixed dimension, its computation takes constant time and can be implemented by bounded-depth classical processing.

In particular, at a corner of an error component---a \emph{trailing vertex} in the terminology of Kubica and Preskill---the error faces meeting the corner lie in its positive neighborhood and provide the rooted chain required by the claim. The sweep rule can therefore flip qubits in that neighborhood so as to erase the syndrome at the corner, moving the error boundary away from the trailing vertex.
\paragraph{Ideal decoding of local-stochastic errors.}
Our third result concerns the same correction rule used \emph{without faults during decoding}. It is separate from the above threshold theorem, where noise may occur between and during repeated correction cycles.

\begin{theorem*}[Informal version of \Cref{theorem:idealdec}]
Fix $d,d'$ with $d',d-d'\geq2$. There is a constant $p_0>0$ such that a codeword of the side-length-$n$ $(d,d')$-Toric code, corrupted by a local-stochastic Pauli error of rate $p<p_0$, is recovered after $n^{d+1}$ noiseless sweep cycles with probability at least
\begin{equation*}
  1-O\!\left(\operatorname{poly}(n)\right)p^{\Omega(n)}.
\end{equation*}
\end{theorem*}

The earlier analysis in \cite{Kubica_2019} treats independent initial errors. Our extension allows correlated local-stochastic errors; the crucial probabilistic estimates follow from counting fault explanations rather than assuming error independence.

\subsection{Discussion and scope}

The main message is a separation of two often-confounded ingredients of local quantum decoding: \emph{geometric expansion of the code} and \emph{the ability to control errors by local updates}. For the 4D Toric code, the latter is provable without the former. Moreover, the decoder's rule has no persistent auxiliary state, so there is no need to store large local timers or implement them by scale-dependent wire delays. The construction still relies on a synchronous constant-depth correction cycle, ideal classical computation, and a four-dimensional lattice geometry; none of these hardware assumptions is eliminated by the theorem.

The proof may also be of independent interest. It translates an idea from reliable classical cellular automata---explaining a large error by a space--time dependency tree---into homological language. In a CSS chain complex, errors of large Hamming weight may be stabilizers, while a syndrome is a boundary; thus the relevant objects to track are not the erroneous qubits themselves but the evolving supports and geometric spans of syndrome chains. We use this distinction to make a first-passage argument for logical errors and to count the fault histories that could produce them.

\section{Outline of the Proof}

Our threshold proof combines two ingredients: the deterministic shrinking of syndrome components under the sweep rule and a probabilistic bound on the fault histories that could produce a large component. We do not attempt to correct every physical error before the next logical operation, or to maintain a prescribed distribution of the residual error after every round. Instead, we show that a logical failure would require a large intermediate syndrome, and that such a syndrome has an unlikely space--time explanation. We outline these steps before giving their formal versions in the following sections.

\paragraph{Proof strategy and Toom's argument.}
A standard fault-tolerance strategy is to identify tolerated error configurations and prove that correction preserves this class. Concatenated-code constructions, such as \cite{aharonov1999faulttolerant}, do so recursively across encoding levels, with level-dependent correction schedules. In quantum LDPC constructions such as \cite{grospellier:tel-03364419}, one can instead control a local-stochastic distribution of effective Pauli errors. For the Toric code, however, local corrections depend on neighboring syndrome checks and introduce spatial dependencies that make such a per-round invariant difficult to establish without expansion. We therefore trace a potentially dangerous syndrome \emph{backward} through the noisy computation and ask how many faults are required to explain it.

The model for this approach is Toom's north-east-center majority rule. Under noiseless updates, finite droplets of incorrect classical bits erode in a preferred direction. Berman and Simon \cite{Berman} exploit this orientation by tracing a wrong bit at time $t+1$ to wrong predecessors at time $t$, unless a fault created it. Backward \emph{arrows} record these dependencies, \emph{forks} connect interacting branches, and terminal fault witnesses become \emph{leaves}. The resulting space--time graph, or \emph{Toom contour}, explains a late error by earlier faults; directional shrinking forces a large error to have many leaves while keeping the explanation's size under control. We adapt this principle to \emph{syndrome boundaries}, rather than to erroneous qubits. For related developments of Toom contours, see \cite{swart2026peierlsboundstoomcontours}.

\paragraph{The sweep rule and geometric shrinking.}
Let $\mathcal T_n^d$ be the side-length-$n$ cubical Toric complex, with fixed $d,d'$ satisfying $d',d-d'\geq 2$. Qubits occupy $d'$-faces. In one CSS sector, a Pauli error is a binary $d'$-chain $e$ with syndrome $\sigma=\partial^-e$, where $\partial^-$ denotes the cubical boundary operator. A correction chain $c$ changes the error to $e+c$ and the syndrome by $\partial^-c$; the other sector is treated on the dual complex. This representation lets us formulate correction in terms of local boundaries, while recognizing that errors differing by stabilizers need not be distinguished.

At each vertex $v$, the sweep rule of Kubica and Preskill \cite{Kubica_2019} seeks a chain $c_v$ on the positive $d'$-faces rooted at $v$ such that
\begin{equation*}
  (\partial^-c_v)|_{v+}=\sigma|_{v+},
  \qquad
  \textbf{Str}_{0+}(v,c_v)\subseteq\textbf{Str}_{0+}(v,\sigma|_{v}).
\end{equation*}
The first condition cancels the locally observed positive syndrome; the second prevents the correction from extending beyond its prescribed positive support (\Cref{def:positiveNei}). If no admissible chain exists, the vertex makes no correction. All vertices update in parallel, and a complete sweep cycle applies the rule in both CSS sectors. A conceptual rule of this kind is not yet an algorithm: the required local correction must exist wherever shrinking needs it. We resolve the higher-dimensional existence question raised in \cite{Kubica_2019} by constructing $c_v$ inductively over face dimension. Since $d$ is fixed, the procedure examines a constant-size neighborhood and requires bounded-depth classical processing (\Cref{def:sweeprule} and \Cref{subsection:shrink}).

To quantify shrinking, we work in the infinite cubical cover $\mathcal T_\infty^d$, with vertex set $\mathbb Z^d$, where finite syndrome components have unambiguous coordinate extrema. Define
\begin{equation*}
  L_i(v)=v_i\quad (1\leq i\leq d),
  \qquad L_{d+1}(v)=-\sum_{i=1}^{d}v_i,
  \qquad
  \operatorname{Size}(A)=\sum_{i=1}^{d+1}\max_{v\in A}L_i(v)
\end{equation*}
for a finite nonempty vertex set $A$. These $d+1$ \emph{potential walls} enclose a simplex-like region, whose extremal witnesses are called \emph{poles}. In two dimensions the region is a triangle, as illustrated below. The size is translation invariant, and it measures geometric span rather than Hamming weight.

\begin{figure}[H]

\begin{center}
  \ifdefined\POSTER  
    \begin{tikzpicture}[scale=3.2]
  \else
  \begin{tikzpicture}[scale=1.5]
  \fi

  \draw[blue!20!black,fill=blue!70!black!70,rounded corners=1, opacity=0.5]%
  (1.9,0.9) 
  rectangle (2.1,2.1);

  \draw[blue!20!black,fill=blue!70!black!70,rounded corners=1, opacity=0.5]%
  (2.9,0.9) 
  rectangle (3.1,2.1);
  \draw[blue!20!black,fill=blue!70!black!70,rounded corners=1, opacity=0.5]%
  (1.9,1.9) 
  rectangle (2.1,3.1);

  \draw[blue!20!black,fill=blue!70!black!70,rounded corners=1, opacity=0.5]%
  (1.9,0.9) 
  rectangle (3.1,1.1);

  \draw[blue!20!black,fill=blue!70!black!70,rounded corners=1, opacity=0.5]%
  (1.9,1.9) 
  rectangle (3.1,2.1);
  \draw[blue!20!black,fill=blue!70!black!70,rounded corners=1, opacity=0.5]%
  (2.9,0.9) 
  rectangle (4.1,1.1);
  \draw[dashed] (-1,3) -- (5.1,3) node[right] { $\max_{v \in z }L_{1}(v)$ };
  \draw[dashed] (4,6) -- (4,-2) node[below] { $\max_{v \in z} L_{2}(v)$ };
  \draw[dashed] (-1,4) -- (5,-2) node[right] { $\max_{v \in z} L_{3}(v)$ };

  \draw[step=1cm,gray,very thin] (0,0) grid (5,5);
  \foreach \x in {0,1,2,3,4,5}
  \foreach \y in {0,1,2,3,4,5}
  {
  \node[draw,circle,inner sep=1pt,fill] at (\x,\y) {};
}

\end{tikzpicture}
\end{center}
\caption{
A graphical example of the maximal values of the potential walls. Here we consider the $(2,1)$-Toric code, namely the bits are set on the edges where the $Z$-stabilizers (parity checks) are set on the vertices. The grid is a finite $6\times 6$ patch of the infinite 2D plane. The edges marked in blue correspond to a bits assignment $z$. Each of the dashed lines marks the line of vertices whose $L_{i}$ value converges with the maximal value reached by a vertex in $z$. Moving into the infinite plane enables us to determine those lines and draw the triangle that bounds $z$.
}
\label{fig:potwall}
\end{figure}

The oriented sweep advances one potential wall without moving the others outward: in a complete noiseless sweep cycle, the appropriate extremal potential improves by at least one unit, while the remaining maxima do not increase. Consequently, a finite syndrome component disappears after a number of ideal cycles controlled by its geometric size. For our noisy analysis, the important implication is the reverse one: a syndrome that survives a sweep must have predecessors with sufficiently large combined geometric extent, including a strict one-unit gain in the shrinking direction. These statements concern \emph{finite components in the infinite cover}; periodic lifts of nontrivial toric cycles cannot be treated as finite components (\Cref{subsection:infinitetoric,subsection:shrink}).

\paragraph{The barrier argument.}
Geometric shrinking alone does not imply fault tolerance, because a nontrivial logical Pauli can have \emph{zero syndrome}. We therefore use a geometric \emph{barrier argument}, related to earlier work on topological memories \cite{Dennis_2002,BravyiTerhal2009,BombinEtAl2013}. A transition from a trivial error to a nontrivial logical class, effected through bounded-support elementary operations, must pass through an error whose syndrome has a connected component of geometric span $\Omega(n)$. Otherwise the small syndrome components can be filled locally without changing the logical class. The formal argument uses reduced errors and bounds how many components one local operation can merge (\Cref{claim:barrier}  and \Cref{coro:barrier}).

To detect this intermediate event, we refine each circuit or fault layer into elementary steps in an order compatible with the circuit's causal structure. This ordering is only a proof device; it does not serialize the physical computation. We consider the \emph{first barrier crossing}, when a connected syndrome component reaches the forbidden scale. It may occur partway through a layer, even if the syndrome at the end of that layer is small or vanishes. It therefore suffices to bound the probability of such a crossing, rather than the probability of a final nontrivial cycle directly.

\paragraph{Toom contours and explanation trees.}
The histories to be counted include more than sweep cycles and fault layers. After fault-tolerant initialization using standard methods \cite{aharonov1999faulttolerant,gottesman2014faulttolerant,grospellier:tel-03364419}, the encoded circuit interleaves sweeps with logical operations from the universal set $\mathbf X,\mathbf Z,\mathbf{CX},\mathbf H,\mathbf S,\mathbf T$. Logical Paulis and controlled-NOT are transversal; the other gates use encoded-state gate teleportation \cite{Zhou_2000}. During any classical feed-forward delay, the output register continues to undergo ordinary sweep cycles. The relevant $X$-basis measurement uses a fold-transversal Hadamard \cite{Breuckmann_2024} on a register that is then discarded. These gadgets permit controlled propagation of Pauli errors and syndromes without requiring the same geometric estimate at every type of layer: sweeps give strict shrinking, logical operations control propagation, and faults introduce new witnesses.

Fix a realization of the circuit faults and measurement outcomes. At each time boundary, represent the $X$- and $Z$-components of the propagated Pauli error by chains in the primal and dual infinite covers. A \emph{syndrome witness} $(v,t)$ records a lattice vertex belonging to the support of a syndrome boundary at time $t$, with its register and CSS sector also recorded. We connect witnesses by time-respecting \emph{arrows} whenever an update can propagate syndrome influence between them; bounded-length \emph{forks} join interacting predecessor histories, and witnesses first produced by faults or noisy initialization serve as \emph{leaves}. The resulting space--time graph $\mathcal H$ tracks syndrome histories, not physical error supports: a large stabilizer error, for example, need not produce a witness. The figure illustrates one sweep and its arrows and forks.

\begin{figure}[h]
  \centering
  \ifdefined \POSTER \begin{tikzpicture}[scale=4] \else \begin{tikzpicture}[scale=1.5] \fi

\tikzset{xyp/.style={canvas is xy plane at z=#1}}
\tikzset{xzp/.style={canvas is xz plane at y=#1}}
\tikzset{yzp/.style={canvas is yz plane at x=#1}}

\draw[green!20!black,fill=green!70!black!70,rounded corners=1,opacity=0.5, xzp=0]%
(-0.1,2.1) rectangle (1.1,0.9);

\draw[green!20!black,fill=green!70!black!70,rounded corners=1,opacity=0.5, xzp=1]%
(0.9,2.1) rectangle (2.1,0.9);

      \draw[green!20!black,fill=green!70!black!70,rounded corners=1,opacity=0.5, yzp = 0]%
      (-0.1,2.1) rectangle (1.1, 0.9);

      \draw[green!20!black,fill=green!70!black!70,rounded corners=1,opacity=0.5, xyp = 2]%
      (-0.1,-0.1) rectangle (1.1, 1.1);

      \draw[green!20!black,fill=green!70!black!70,rounded corners=1,opacity=0.5, yzp = 1]%
      (-0.1,2.1) rectangle (1.1, 0.9);

      \draw[blue, thick]  (0,1,2) -- (1,1,2); %
      \draw[blue, thick] (0,1,1) -- (0,1,2) -- (0,1,1) -- (0,0,1) -- (1,0,1) -- (1,1,1) -- (2,1,1) -- (2,1,2) -- (1,1,2);

      \node[below , scale=0.7] at (0,1,2) { $(v_{1}, t)$ } ;
      \node[below , scale=0.7] at (0,1,1) {$(v_{2}, t)$} ;
      \node[below , scale=0.7] at (0,0,1) {$(v_{3}, t)$} ;
      \node[below , scale=0.7] at (1,0,1) {$(v_{4}, t)$} ;
      \node[below , scale=0.7] at (1,1,1) {$(v_{5}, t)$} ;
      \node[below , scale=0.7] at (2,1,1) {$(v_{6}, t)$} ;
      \node[below , scale=0.7] at (2,1,2) {$(v_{7}, t)$} ;
      \node[below , scale=0.7] at (1,1,2) {$(v_{8}, t)$} ;
\foreach \x in {0,1,2} {
  \foreach \y in {0,1,2} {
    \draw[gray] (\x,\y,0) -- (\x,\y,2);
  }
}

\foreach \x in {0,1,2} {
  \foreach \z in {0,1,2} {
    \draw[gray] (\x,0,\z) -- (\x,2,\z);
  }
}

\foreach \y in {0,1,2} {
  \foreach \z in {0,1,2} {
    \draw[gray] (0,\y,\z) -- (2,\y,\z);
  }
}

\foreach \x in {0,1,2} {
  \foreach \y in {0,1,2} {
    \foreach \z in {0,1,2} {
      \fill (\x,\y,\z) circle (1pt);
    }
  }
}

\draw (2.5,1) node { $\rightarrow$ };
\end{tikzpicture} \ifdefined \POSTER \begin{tikzpicture}[scale=4] \else \begin{tikzpicture}[scale=1.5] \fi
\tikzset{xyp/.style={canvas is xy plane at z=#1}}
\tikzset{xzp/.style={canvas is xz plane at y=#1}}
\tikzset{yzp/.style={canvas is yz plane at x=#1}}

\draw[green!20!black,fill=green!70!black!70,rounded corners=1,opacity=0.5, xzp=0]%
(-0.1,2.1) rectangle (1.1,0.9);

\draw[green!20!black,fill=green!70!black!70,rounded corners=1,opacity=0.5, xzp=1]%
(0.9,2.1) rectangle (2.1,0.9);

      \draw[green!20!black,fill=green!70!black!70,rounded corners=1,opacity=0.5, yzp = 0]%
      (-0.1,2.1) rectangle (1.1, 0.9);

      \draw[green!20!black,fill=green!70!black!70,rounded corners=1,opacity=0.5, xyp = 2]%
      (-0.1,-0.1) rectangle (1.1, 1.1);

      \draw[green!20!black,fill=green!70!black!70,rounded corners=1,opacity=0.5, yzp = 1]%
      (-0.1,2.1) rectangle (1.1, 0.9);

\draw[green!20!black,fill=green!70!black!70,rounded corners=1,opacity=0.5, xzp=1]%
(-0.1,2.1) rectangle (1.1,0.9);

      \draw[green!20!black,fill=green!70!black!70,rounded corners=1,opacity=0.5, xyp = 1]%
      (-0.1,-0.1) rectangle (1.1, 1.1);

      \draw[blue, thick]  (1,1,1) -- (2,1,1) -- (2,1,2) -- (1,1,2) -- (1,1,1);

      \node[below , scale=0.7] at (0,1,2) { $(v_{1}, t+1)$ } ;
      \node[below , scale=0.7] at (0,1,1) {$(v_{2}, t+1)$} ;
      \node[below , scale=0.7] at (0,0,1) {$(v_{3}, t+1)$} ;
      \node[below , scale=0.7] at (1,0,1) {$(v_{4}, t+1)$} ;
      \node[below , scale=0.7] at (1,1,1) {$(v_{5}, t+1)$} ;
      \node[below , scale=0.7] at (2,1,1) {$(v_{6}, t+1)$} ;
      \node[below , scale=0.7] at (2,1,2) {$(v_{7}, t+1)$} ;
      \node[below , scale=0.7] at (1,1,2) {$(v_{8}, t+1)$} ;

\foreach \x in {0,1,2} {
  \foreach \y in {0,1,2} {
    \draw[gray] (\x,\y,0) -- (\x,\y,2);
  }
}

\foreach \x in {0,1,2} {
  \foreach \z in {0,1,2} {
    \draw[gray] (\x,0,\z) -- (\x,2,\z);
  }
}

\foreach \y in {0,1,2} {
  \foreach \z in {0,1,2} {
    \draw[gray] (0,\y,\z) -- (2,\y,\z);
  }
}

\foreach \x in {0,1,2} {
  \foreach \y in {0,1,2} {
    \foreach \z in {0,1,2} {
      \fill (\x,\y,\z) circle (1pt);
    }
  }
}
\end{tikzpicture}

\ifdefined \POSTER \begin{tikzpicture}[scale=4] \else \begin{tikzpicture}[scale=1.5] \fi

\tikzset{xyp/.style={canvas is xy plane at z=#1}}
\tikzset{xzp/.style={canvas is xz plane at y=#1}}
\tikzset{yzp/.style={canvas is yz plane at x=#1}}

\node(A) at (0,1,2) {} ;
\node(B) at (0,1,1) {} ;
\node(C) at (0,0,1) {} ;
\node(D) at (1,0,1) {} ;
\node(E) at (1,1,1) {} ;
\node(F) at (2,1,1) {} ;
\node(G) at (2,1,2) {} ;
\node(H) at (1,1,2) {} ;

      \fill (0,1,2) circle(1pt)  ;
      \fill (0,1,1) circle(1pt) ;
      \fill (0,0,1) circle(1pt) ;
      \fill (1,0,1) circle(1pt) ;
      \fill (1,1,1) circle(1pt) ;
      \fill (2,1,1) circle(1pt) ;
      \fill (2,1,2) circle(1pt) ;
      \fill (1,1,2) circle(1pt) ;

\foreach \x in {0,1,2} {
  \foreach \y in {0,1,2} {
    \foreach \z in {0,1,2} {
      \node at (\x,\y,\z) { } ;
    }
  }
}

\node[below , scale=0.7](I) at (5,1,1) {$(v_{5}, t+1)$} ;
\node[below , scale=0.7](J) at (6,1,1) {$(v_{6}, t+1)$} ;
\node[below , scale=0.7](K) at (6,1,2) {$(v_{7}, t+1)$} ;
\node[below , scale=0.7](L) at (5,1,2) {$(v_{8}, t+1)$} ;
 
\draw[blue,very thin,-> ]  (A) to [out=50, in=130] (I); 
\draw[blue,very thin,-> ]  (A) to [out=50, in=130] (L);
\draw[blue,very thin,-> ]  (B) to [out=50, in=130] (I); 
\draw[blue,very thin,-> ]  (B) to [out=50, in=130] (L);
\draw[blue,very thin,-> ]  (C) to [out=50, in=130] (I); 
\draw[blue,very thin,-> ]  (D) to [out=50, in=130] (I); 
\draw[blue,very thin,-> ]  (D) to [out=50, in=130] (L);
\draw[blue,very thin,-> ]  (E) to [out=50, in=130] (J); 
\draw[blue,very thin,-> ]  (E) to [out=50, in=130] (L);
\draw[blue,very thin,-> ]  (E) to [out=50, in=130] (K);

\draw[blue,very thin,-> ]  (H) to [out=50, in=130] (I); 
\draw[blue,very thin,-> ]  (H) to [out=50, in=130] (K);
\draw[blue,very thin,-> ]  (H) to [out=50, in=130] (J); 

\draw[blue,very thin,-> ]  (F) to [out=50, in=130] (I); 
\draw[blue,very thin,-> ]  (F) to [out=50, in=130] (L);
\draw[blue,very thin,-> ]  (F) to [out=50, in=130] (K); 

\draw[blue,very thin,-> ]  (G) to [out=50, in=130] (I); 
\draw[blue,very thin,-> ]  (G) to [out=50, in=130] (L);
\draw[blue,very thin,-> ]  (G) to [out=50, in=130] (J); 

\draw[green,very thin,<-> ]  (A) to (B); 
\draw[green,very thin,<-> ]  (A) to (C); 
\draw[green,very thin,<-> ]  (A) to (D); 
\draw[green,very thin,<-> ]  (A) to (E); 
\draw[green,very thin,<-> ]  (A) to  (H);
\draw[green,very thin,<-> ]  (A) to  (G);
\draw[green,very thin,<-> ]  (A) to  (F);

\draw[green,very thin,<-> ]  (B) to (C); 
\draw[green,very thin,<-> ]  (B) to (D); 
\draw[green,very thin,<-> ]  (B) to (E); 
\draw[green,very thin,<-> ]  (B) to  (H);
\draw[green,very thin,<-> ]  (B) to  (G);
\draw[green,very thin,<-> ]  (B) to  (F);

\draw[green,very thin,<-> ]  (C) to (D); 
\draw[green,very thin,<-> ]  (C) to (E); 
\draw[green,very thin,<-> ]  (C) to  (H);
\draw[green,very thin,<-> ]  (C) to  (G);
\draw[green,very thin,<-> ]  (C) to  (F);

\draw[green,very thin,<-> ]  (D) to (E); 
\draw[green,very thin,<-> ]  (D) to  (H);
\draw[green,very thin,<-> ]  (D) to  (G);
\draw[green,very thin,<-> ]  (D) to  (F);

\draw[green,very thin,<-> ]  (E) to  (H);
\draw[green,very thin,<-> ]  (E) to  (G);
\draw[green,very thin,<-> ]  (E) to  (F);

\draw[green,very thin,<-> ]  (F) to  (H);
\draw[green,very thin,<-> ]  (F) to  (G);

\draw[green,very thin,<-> ]  (H) to  (G);

      \fill (5,1,1) circle(1pt) ;
      \fill (6,1,1) circle(1pt) ;
      \fill (6,1,2) circle(1pt) ;
      \fill (5,1,2) circle(1pt) ;

\node[below , scale=0.7] at (0,1,2) { $(v_{1}, t)$ } ;
\node[below , scale=0.7] at (0,1,1) {$(v_{2}, t)$} ;
\node[below , scale=0.7] at (0,0,1) {$(v_{3}, t)$} ;
\node[below , scale=0.7] at (1,0,1) {$(v_{4}, t)$} ;
\node[below , scale=0.7] at (1,1,1) {$(v_{5}, t)$} ;
\node[below , scale=0.7] at (2,1,1) {$(v_{6}, t)$} ;
\node[below , scale=0.7] at (2,1,2) {$(v_{7}, t)$} ;
\node[below , scale=0.7] at (1,1,2) {$(v_{8}, t)$} ;

      \foreach \x in {4,5,6} {
  \foreach \y in {0,1,2} {
    \foreach \z in {0,1,2} {
      \node at  (\x,\y,\z) { };
    }
  }
}
\end{tikzpicture}

\caption{
Above: The application of a sweep-cycle, where the green faces represent the bits assignment $z$, and the blue edges correspond to unsatisfied checks $\partial^{-} z$. The vertices witnessing a syndrome, namely belonging to $\partial^{-}z$, are labeled as $v_{1}, v_{2}, \ldots, v_{8}$. Notice that the leftmost bottom-inner vertex is not witnessing a syndrome even though it belongs to $z$. Below: The Toom's contour derived from the single computation layer that matches the cycle. Arrows are colored in blue, while forks are colored in green. Note that $(v_{1}, t+1)$ doesn't belong to the Toom's contour (since $v_{1}$ doesn't see an unsatisfied check at time $t+1$).
}
  \label{fig:toomscontours}
\end{figure}

We organize the witnesses of $\mathcal H$ into time \emph{slices}. The size of a slice is determined by its selected poles for the $d+1$ potentials. Starting from the poles of a late slice $K$, we follow backward arrows to predecessor slices $K_0,\ldots,K_k$ and add only the forks needed to connect them. Choosing poles rather than every syndrome witness keeps this \emph{predecessor decomposition} small. Distinct predecessor components have disjoint backward histories, and their geometric sizes satisfy, schematically,
\begin{equation*}
  \sum_{j=0}^{k}s_j+\sum_{j=1}^{k}\operatorname{Size}(F_j)
  \ \geq\
  \begin{cases}
    s+1, & \text{if $K$ follows a sweep cycle},\\
    s,   & \text{for other ordinary layer types},
  \end{cases}
\end{equation*}
where $s$ is the selected-pole size of $K$, $s_j$ are predecessor sizes, and $F_j$ are connecting forks. The strict improvement after a sweep is the bridge between deterministic shrinking and probabilistic fault counting (\Cref{subsection:spacetimegraph}).

Iterating these decompositions backward yields a connected \emph{explanation tree} rooted at the relevant poles and terminating at fault leaves. The induction simultaneously controls the number $m$ of marked leaves and the numbers of arrows $A$ and forks $F$. Its layer-dependent invariant has the schematic form
\begin{equation*}
  |A|\leq\alpha(m-c_r)-\beta s,
  \qquad |F|\leq m-1,
\end{equation*}
for constants $\alpha,\beta>0$ and a last-layer constant $c_r$. Thus a slice of size $s$ requires $\Omega(s)$ leaves, and the explanation cannot have arbitrarily many edges relative to its leaves (\Cref{subsection:explantion}).

There is one essential modification at the first barrier crossing. Because the crossing may occur within a partially executed layer, the \emph{analysis contour} uses the actual intermediate error at its final time and adds short connections between witnesses meeting a common syndrome check. These links ensure that the large connected component appears as a connected contour slice. They are added \emph{only in the last layer}, so that the predecessor structure elsewhere remains unchanged; they weaken the final estimate by only a dimension-dependent constant (\Cref{subsection:toricencoded,subsection:spacetimegraph,subsection:explantion}).

\paragraph{Counting explanations and obtaining the threshold.}
The preceding explanations live in the infinite cover, whereas faults occur at physical locations on the finite torus. Different leaves may therefore project to the \emph{same} location modulo $n$, and local-stochastic noise charges that location only once. As in the combinatorial selection argument of \cite{Berman}, we use a bounded-degree tree-separator lemma to extract a subtree of sufficiently small spatial spread, retaining a controlled ratio of leaves to edges. Its marked leaves yield distinct physical fault locations up to a fixed multiplicity. In this way, a large slice certifies $\Omega(s)$ distinct faults, with the stronger edge-sensitive bound needed to sum over explanations. The separator argument appears in the appendix.

For a fixed root, the number of connected explanations with $q$ edges in the bounded-degree space--time graph is at most exponential in $q$. On the other hand, local-stochastic noise bounds the probability of faults at any prescribed set of $r$ distinct locations by $p^r$. Since the selected explanations require $r=\Omega(s+q)$ such locations, summing over their possibilities gives an exponentially small probability for a slice of geometric size $s$ when $p$ is sufficiently small. A logical failure requires a first barrier crossing with $s=\Omega(n)$; taking a union bound over registers and intermediate steps gives
\begin{equation*}
  \Pr(\text{logical failure})
   \leq O\!\left(\operatorname{poly}(w,\mu,n)\right)p^{\Omega(n)},
\end{equation*}
as claimed in \Cref{theorem:main}.

\paragraph{Ideal decoding.}
The same mechanism applies when one local-stochastic Pauli error is sampled initially and all subsequent sweep cycles are noiseless. The contour leaves then come from the initial error alone. With high probability no syndrome component reaches the geometric barrier, so shrinking removes every relevant component while the barrier argument prevents a change of logical class. After $n^{d+1}$ ideal sweep cycles the encoded state is recovered with probability at least
\begin{equation*}
  1-O\!\left(\operatorname{poly}(n)\right)p^{\Omega(n)},
\end{equation*}
as stated in \Cref{theorem:idealdec}. In both results, the cubical geometry supplies local shrinking and the barrier, while Toom-type histories make the large-syndrome event unlikely without an expansion assumption or a per-round local-stochastic invariant.

\section{Notations and Preliminaries}\label{sec:nota}

\paragraph{Sets and indices.}
For a positive integer $n$, let $[n]=\{1,2,\ldots,n\}$. We write $|A|$ for the cardinality of a set $A$, and $\mathbf{1}_i$ for the vector with a one in coordinate $i$ and zeros elsewhere. An indexed family is written $\{u_i\}_{i=1}^k$ (or $\{u_i\}_{i=1}$ when the upper limit is not relevant); we omit an index range when it is understood. We use $\{a:\,a\in A,\,a\in B\}$ for the intersection $A\cap B$.

When a set of directions $S$ is fixed, we use the following shorthand for its subsets:
\begin{equation*}
  S'/g := S'\setminus\{g\},\qquad
  S/S' := S\setminus S',\qquad
  S'\cup g := S'\cup\{g\}.
\end{equation*}
In particular, $S/g=S\setminus\{g\}$. Products of elements of an abelian group are written multiplicatively; for a subset $J$ of commuting generators, $Jv$ means $(\prod_{g\in J}g)v$. These conventions are used in the definition of the cubical Toric complex and in the subsequent local arguments.

\paragraph{Chains and restrictions.}
For a set $\mathcal F$ of cells, $(\mathcal F;\mathbb F_2)$ denotes the $\mathbb F_2$-vector space of \emph{finite formal sums} of cells in $\mathcal F$; if $\mathcal F$ is finite, this is naturally $\mathbb F_2^{\mathcal F}\cong\mathbb F_2^{|\mathcal F|}$. We identify a cell with its corresponding basis chain when no confusion arises. The support of a chain is the collection of cells with nonzero coefficients. We use $f\in z$ for a cell in the support of a chain $z$ when $f$ has the same dimension as $z$, and also use the customary incidence notation for lower-dimensional faces of supported cells; in particular, $v\in z$ means that a vertex $v$ belongs to some cell supporting $z$. The cubical boundary and coboundary operators are written $\partial^{-}$ and $\partial^{+}$, respectively, and are defined in \Cref{def:asontoric}.

The notation $z|_D$ denotes restriction of a chain or assignment $z$ to the specified set of coordinates $D$. When $D$ denotes a vertex, a local face family, or the positive faces of a vertex, the precise restriction is specified by its type. In particular, the vertex restrictions $|_v$ and $|_{v+}$ are defined in \Cref{def:asontoric}; $z|_{/v}$ denotes the complementary restriction $z+z|_v$ (addition is over $\mathbb F_2$). These conventions also apply to finite-support chains on the infinite covering lattice.

\paragraph{Quantum notation.}
Unless stated otherwise, all quantum systems consist of qubits. We write $\mathcal H_2$ for the one-qubit Hilbert space, $\mathcal H_{2^n}=\mathcal H_2^{\otimes n}$ for $n$ qubits, and $L(\mathcal H)$ for the linear operators on $\mathcal H$. Pure states are denoted by $\ket\psi$ and $\bra\psi$; the corresponding density operator is $\ket\psi\!\bra\psi$, and $\rho$ usually denotes a general density operator. If $A$ is an operation, $A\circ\rho$ denotes its action on $\rho$; for a unitary $A$, this is $A\rho A^\dagger$.

The letters $X$ and $Z$ usually denote the single-qubit Pauli matrices. We write $\mathcal P^{\otimes n}$ for the $n$-qubit Pauli group and omit the superscript when its size is clear. For $P\in\mathcal P^{\otimes n}$, the $X$- and $Z$-components are denoted $X(P)$ and $Z(P)$, respectively, so that $P=X(P)Z(P)$ up to an overall phase. We denote controlled-NOT and controlled-$Z$ by $\mathbf{CX}$ and $\mathbf{CZ}$, and the single-qubit gates by
\begin{equation*}
 \mathbf H=\frac{1}{\sqrt 2}\begin{bmatrix}1&1\\1&-1\end{bmatrix},\qquad
 \mathbf S=\begin{bmatrix}1&0\\0&i\end{bmatrix},\qquad
 \mathbf T=\begin{bmatrix}1&0\\0&e^{i\pi/4}\end{bmatrix}.
\end{equation*}
For a gate $U$, $\ket U$ denotes a resource state used in its gate-teleportation implementation; such a state need not be unique. We write $\prb{A}$ for the probability of an event $A$. For group elements $g_1,\ldots,g_\ell$, $\langle g_1,\ldots,g_\ell\rangle$ is the subgroup they generate.

\paragraph{CSS codes.}
We use the standard conventions for CSS codes \cite{Calderbank_1996,Steane_1996}. For a linear subspace $A\subseteq\mathbb F_2^n$, $\ket A:=|A|^{-1/2}\sum_{a\in A}\ket a$. A CSS code is specified by linear spaces $\mathcal C_X,\mathcal C_Z\subseteq\mathbb F_2^n$ satisfying $\mathcal C_Z^\perp\subseteq\mathcal C_X$; its $X$-stabilizer vectors constitute $\mathcal C_Z^\perp$, and its computational-basis logical labels lie in $\mathcal C_X/\mathcal C_Z^\perp$. In particular, the logical zero is $\ket{\mathcal C_Z^\perp}$. We write $x_1\sim_{\mathcal C_X}x_2$ when $x_1+x_2\in\mathcal C_Z^\perp$, and analogously define $\sim_{\mathcal C_Z}$ with the roles of $X$ and $Z$ interchanged.

\ifdefined\FULL

\input{shrink_v1.tex}
\fi

\label{subsection:model}
\newcommand{\comDAG}{\pi_{C[\mathcal{E}]}} 
\paragraph{The Computation Model.} Informally, a quantum circuit is a placement of a collection of quantum gates, taken from a finite set, at different space-time locations. Each such location records when the gate is applied and on which qubits it acts. Formally:

\begin{definition}[Quantum Circuit]
  \label{def:quantcirc}
  A \emph{space-time location} is a pair $l=(t,S)$, where $t$ is a time coordinate and $S=(s_{1},\ldots,s_{\Delta})$ is an ordered tuple of qubit indices. For a finite collection of quantum gates 
  \begin{equation*}
    \begin{split}
      \mathcal{G} \subset \Big\{ g \colon L(\mathcal{H}_{2}^{\otimes \Delta}) \to L(\mathcal{H}_{2}^{\otimes \Delta}) \Big\},
    \end{split}
  \end{equation*}
  and integers $w,\mu\in\mathbb{N}$, a quantum circuit $C$ of depth $\mu$ and width $w$ is a collection of tuples
  \begin{equation*}
    \begin{split}
      C=\{u_i=(g_i,l_i)\}_i,
    \end{split}
  \end{equation*}
  where each $g_i\in\mathcal{G}$ and each location $l_i=(l_{i,1},l_{i,2})$ belongs to $[\mu]\times [w]^{\Delta}$. We call the tuples $u_i$ \textit{the gates of $C$}. They must satisfy the following compatibility condition: if two gates have the same time coordinate, then their supports are disjoint. Namely, for $u_i=(g_i,l_i),u_j=(g_j,l_j)\in C$, the equality $l_{i,1}=l_{j,1}$ implies that the sets of qubit indices appearing in $l_{i,2}$ and $l_{j,2}$ are disjoint. We denote by
  \begin{equation*}
    \begin{split}
      \operatorname{Loc}(C) \colonequals \{l_i : u_i=(g_i,l_i)\in C\}
    \end{split}
  \end{equation*}
  the set of locations occupied by the gates of $C$. We define the \textit{$t$-layer} of $C$ to be its restriction to gates at time $t$:
  \begin{equation*}
    \begin{split}
      C|_{t} \colonequals \{u_i=(g_i,l_i)\colon u_i\in C \text{ and } l_{i,1}=t\}.
    \end{split}
  \end{equation*}
\end{definition}
Later, we will restrict $\mathcal{G}$ to contain $\mathbf{X}, \mathbf{Z}, \mathbf{CX}, \mathbf{H}, \mathbf{S}, \mathbf{T}$, and measurements. To define the action of a quantum circuit $C$ on a given mixed state, we first associate $C$ with a directed acyclic graph.
\begin{definition}[Associated Computation DAG]
  \label{def:steps}
  Let $C=\{u_i\}_i$ be a quantum circuit with depth $\mu$ and width $w$. We define the associated computation DAG of $C$, denoted by $\pi_C$, to be the directed acyclic graph $\pi_C=(\mathcal{U},E)$ as follows:
  \begin{enumerate}
    \item The vertex set is the gate set of $C$, namely $\mathcal{U}=\{u_i\}_i$.
    \item For $u_i=(g_i,l_i)$ and $u_j=(g_j,l_j)$, there is a directed edge $(u_i,u_j)\in E$ if and only if $j$ is one of the minimizers of
      \begin{equation*}
        \begin{split}
          \big\{ l_{k,1} : (g_k,l_k)\in \mathcal{U},\ l_{k,1}>l_{i,1},\ l_{k,2}\cap l_{i,2}\neq\emptyset \big\}.
        \end{split}
      \end{equation*}
      Equivalently, $u_j$ is one of the earliest later gates whose support intersects the support of $u_i$. There may be more than one such minimizer, so the out-degree may exceed one.
  \end{enumerate}

  Suppose that $J=(j_1,j_2,\ldots,j_k)$ is an ordering of the gate indices that agrees with a topological order induced by $\pi_C$. Then we define the associated partial circuits $C|_J=\{C|_{J,\rightarrow i}\}_{i=1}^{k}$ by
  \begin{enumerate}
    \item $C|_{J,\rightarrow1}=\{u_{j_1}\}$.
    \item For $1\le i<k$, set $ C|_{J,\rightarrow i+1}=C|_{J,\rightarrow i}\cup\{u_{j_{i+1}}\}$.
  \end{enumerate}
We call $C|_{J,\rightarrow i}$ the partial computation circuit up to step $i$. When the order itself does not play a role, except for being time-respecting (namely, a gate $u$ that is applied before gate $u^{\prime}$ has a lower step than $u^{\prime}$'s step in $J$), we will omit $J$ from the subscript and briefly write $C|_{\rightarrow i}$ to describe a partial circuit up to that step.
\end{definition}

\begin{definition}
  Let $\rho$ be a density matrix over $w$ qubits, and let $C = \{u_{i}\}_{i=1}$ be a quantum circuit with depth $\mu$ and width $w$. For any density matrix $\sigma$, quantum gate $g \in \mathcal{G}$, and location $l$, the application of the gate-location tuple $(g,l)$ on $\sigma$, denoted by $(g,l) \circ \sigma$, is the application of $g$ on $\sigma$ where wires are ordered such that the first qubit in the input to $g$ is $l_{2,1}$, the second is $l_{2,2}$, and so on. The result of the application of $C$ on $\rho$, denoted by $C \circ \rho$, is given by iterative applications of the gates $\{u_{i}\}_{i=1}$ on $\rho$, according to a topological order induced by $\pi_{C}$.

We say that $\rho_{1}, \rho_{2}, \ldots, \rho_{k}$ is a computation prefix if $\rho_{1} = \rho$ and $\rho_{i+1} = u^\prime_{i} \circ \rho_{i}$, where $\{u^{\prime}_{i}\}_{i=1}$ is a prefix of a topological ordering of $\{u_{i}\}_{i=1}$. If $\{u^{\prime}_{i}\}_{i=1}$ contains all the gates in $\{u_{i}\}_{i=1}$, then we call $\rho_{1}, \rho_{2}, \ldots, \rho_{k}$ a running.
\end{definition}
For a quantum circuit $C$, we would like to discuss its action in the presence of different outcomes of faults. The following definition encodes the running in the presence of a specific set of faults $\mathcal{E}$.
\begin{definition}[$C$ subjected to $\mathcal{E}$ every $m$ computational layers]
  Let $C=\{u_i\}_i$ be a quantum circuit with depth $\mu$ and width $w$, let $m\in\mathbb{N}$, and set
  \begin{equation*}
    \begin{split}
      \mu_m \colonequals \left\lceil \frac{\mu}{m} \right\rceil.
    \end{split}
  \end{equation*}
  Let $\mathcal{L}\subset [\mu_m]\times [w]^{\Delta}$ be a set of locations, and let $\mathcal{E}$ be a quantum circuit, such that $\text{Loc}(\mathcal{E}) = \mathcal{L}$. We name $\mathcal{E}$ \textit{subsets of faults}, and we name the layers of $\mathcal{E}$ \textit{fault layers}.
  The circuit \textit{$C$ subjected to $\mathcal{E}$ every $m$ computational layers}, denoted by $C[\mathcal{E}]_{m}$, is obtained by setting the first layer to be the first layer of $\mathcal{E}$, then placing in order the layers of $C$ and inserting a fault layer before each successive block of at most $m$ layers of $C$. Thus, the first fault layer precedes the first block of $C$, the second follows its first $m$ layers, and the third follows its first $2m$ layers, provided a further block remains. Concretely, each gate $u_i=(g_i,(t_i,S_i))$ of $C$ is moved to the time coordinate
  \begin{equation*}
    \begin{split}
      t_i + \left\lceil \frac{t_i}{m} \right\rceil,
    \end{split}
  \end{equation*}
  while each fault gate $(g,(t,S))\in\mathcal{E}$ is placed at time coordinate
  \begin{equation*}
    \begin{split}
      (t-1)(m+1)+1.
    \end{split}
  \end{equation*}
When $m=1$, the gates occupy the even time coordinates, and the fault layers occupy the odd time coordinates. When the subscript parameter is omitted, we mean that the noise layers are injected between any two logical layers, and the logical layers are clear from the context. We denote the associated computation DAG of $C[\mathcal{E}]_{m}$ by $\comDAG$.
\end{definition}

\begin{remark}
  Formally, a fault circuit is just a particular quantum circuit whose gates are interpreted as faults rather than as computational operations. Thus the distinction between a quantum circuit and a fault circuit is semantic rather than structural: both are circuits in the sense of \Cref{def:quantcirc}, but they play different roles in the analysis.
\end{remark}

\begin{example}
  When $m=3$, the `computational` layers are sent to the times
  \begin{equation*}
    \begin{split}
      2,3,4,6,7,8,10,11,12,\ldots,
    \end{split}
  \end{equation*}
  while the fault layers are sent to the times
  \begin{equation*}
    \begin{split}
      1,5,9,13,\ldots.
    \end{split}
  \end{equation*}
\end{example}
Later, all the quantum registers will be encoded via stabilizer codes. Then, following the common approach in the literature, we define the fault layers to end with stabilizer measurements, thereby forcing the faults to be Paulis (see \cite{gottesman1997stabilizercodesquantumerror}). Hence, we confine ourselves to noise channels of only this type.
\begin{definition}[Noise channel and local stochastic noise channel]
  \label{def:localstoc}
  A \textit{noise channel} $\mathcal{N}_{\mathcal{E}}$ is a distribution over sets of faults and their locations. Let $p\in(0,1)$, we say that the noise channel $\mathcal{N}_{\mathcal{E}}$ is \textit{local stochastic with rate $p$} if for every finite set of locations $S\subseteq \mathbb{N}\times \mathbb{N}$,
  \begin{equation*}
    \begin{split}
      \prbm{S\subseteq \operatorname{Loc}(\mathcal{E})}{\mathcal{E}\sim \mathcal{N}_{\mathcal{E}}}
      \le p^{|S|}.
    \end{split}
  \end{equation*}
Through this work, quantum registers will be encoded in CSS codes, fault layers are followed by stabilizer measurements, so the sampled faults will be assumed to be Pauli.
\end{definition}
So, $\mathcal{N}_{\mathcal{E}}$ defines a probability space. The fault set $\mathcal{E}$ is an atomic outcome, and for a given circuit $C$, the subjected circuit $C[\mathcal{E}]$ is a random variable. Equipped with this notation, we are allowed to assume having constant depth cycles that don't observe faults. The following definition considers the effective circuit when the running is restricted such that the outcomes of the measurement are fixed.

\begin{definition}
  Let $C$ be a quantum circuit that contains $l$ measurements. For possible outcomes $\alpha \in \mathbb{F}^{l}_{2}$, let $C[,\alpha]$ be the circuit obtained by conditioning that $\alpha_{1},\ldots,\alpha_{l}$ were measured. We extend the notation for the subjected circuit by denoting $C[\mathcal{E},\alpha]$ as the subjected circuit where the measurements are assumed to have outcome $\alpha_{1},\ldots,\alpha_{l}$.
\end{definition}

There might be multiple different ways to present a conditioned circuit with a collection of gates and their locations. Since we would like to keep the step-order invariant to a conditioning of the measurement outcome, we choose a representation that substitutes the original gates with gates that act on the same location. The following remark explains how to do that explicitly.
\begin{remark}
  \label{remark:condmeas}
  Consider the presentation of a conditioned circuit as a subset of gates acting on locations. For an outcome $\alpha \in \mathbb{F}_{2}^{(\cdot)}$ for which a conditional gate (for example, a $\mathbf{CCZ}$, where the control bits are conditioned on measurement outcomes) in the original circuit $C$ is mapped to a Pauli in $C[,\alpha]$, we substitute this gate with the Pauli on the target tensored with identity acting on the control locations. Thus, the order of the steps, the locations of each gate, and the computation graph of $C[,\alpha]$ are the same as for $C$ (and independent of the outcomes). We summarize this by briefly saying that the locations are invariant to the conditioning of measurements. 
\end{remark}

In our case, the conditioned circuit $C[\mathcal{E}, \alpha]$ is a Clifford + projections circuit; in particular, it induces a `structured trajectory` tracking on the accumulated error. This brings us to define, for a given set of faults $\mathcal{E}$ and outcomes $\alpha$, a `propagation` that describes the evolution of the faults through the circuit.
\begin{definition}[Propagation]
  \label{def:propagation}
Let $C[\mathcal{E}]$ be a subjected quantum circuit at depth $\mu$, and let $\alpha \in \mathbb{F}_{2}^{(\cdot)}$ be a possible outcome vector for $C[\mathcal{E}]$. Suppose that $C$ is a Clifford+measurements circuit, so $C[\mathcal{E},\alpha]$ is a Clifford + projections circuit. We define the propagation through the conditioned circuit $C[\mathcal{E},\alpha]$ to be the tuple $(\mathbf{P},\mathbf{a})$ such that $\mathbf{P}$ consists of $\mu+1$ Paulis $\mathbf{P} \colonequals P_{1}, \ldots, P_{\mu+1}$ and $\mathbf{a}$ consists of $\mu+1$ measurements outcome vectors $\mathbf{a} \colonequals \alpha^{1}, \alpha^{2}, \ldots, \alpha^{\mu+1}$ satisfying that $P_{1} = \mathcal{E}|_{1}, \alpha^{1}=\alpha$ and for any $i \in [2, \mu+1]$:
  \begin{equation*}
    \begin{split}
      P_{i}C[,\alpha^{i}]|_{\rightarrow i} =  C[\mathcal{E},\alpha]|_{ i} P_{i-1} C[,\alpha^{i-1}]|_{\rightarrow i-1} = C[\mathcal{E},\alpha]|_{\rightarrow i}
    \end{split}
  \end{equation*}
We extend the notation to step-propagation by considering the propagation of the circuit obtained from $C[\mathcal{E},\alpha]$ by flattening the circuit such that each gate is applied at the time that equals its step. 
\end{definition}

\begin{remark}
Notice that, for two different measurement outcomes $\alpha,\alpha^{\prime}$, the conditioned $C[,\alpha]$ and $C[,\alpha^{\prime}]$ might differ also by their Clifford gates (and not only by the projections).
\end{remark}

\section{The $(d,d^{\prime})$-Dimensional Toric Code}

We use the cubical Toric code of Kitaev \cite{Kitaev1997QuantumCA,Dennis_2002}. In two dimensions, qubits are associated with edges of a square lattice on the torus, and stabilizer checks with its vertices and plaquettes. More generally, a $(d,d')$-Toric code places qubits on $d'$-dimensional cells of a $d$-dimensional cubical Toric complex, and checks on cells of dimensions $d'-1$ and $d'+1$. We first describe the cells and their chain spaces using the rooted-face notation needed for our local decoder. We retain the same conventions, in particular $\rotfv{v}{S'}$ for basis chains and $(\mathcal F_i;\mathbb F_2)$ for their span, in the sections that follow.

\begin{definition}[$d$-Dimensional Toric Complex]
\label{def:toric_complex}
Let $d,n$ be positive integers and set
\begin{equation*}
  \mathcal G_n^d:=\mathbb Z_n^d,
  \qquad S:=\{\mathbf 1_i:i\in[d]\},
\end{equation*}
where $\mathbf 1_i$ is the $i$th standard generator. We write the group law multiplicatively when acting on vertices: $gv$ is the translate of $v$ by $g$, and $S'v=(\prod_{g\in S'}g)v$ for $S'\subseteq S$.

The \emph{$d$-dimensional cubical Toric complex} $\mathcal T_n^d$ has vertex set $\mathcal F_0=\mathcal G_n^d$. For each $i\in\{0,\ldots,d\}$, an $i$-cell is specified by a root $v\in\mathcal G_n^d$ and a direction set $S'\subseteq S$ with $|S'|=i$. We denote this cell by
\begin{equation*}
  \rotf{v}{S'}=\theta_{v,S'},
  \qquad V(\theta_{v,S'})=
  \left\{\left(\prod_{g\in I}g\right)v:I\subseteq S'\right\},
\end{equation*}
where $V(\theta_{v,S'})$ is its set of vertices. The root and direction set specify the cell (including its incidences), not merely the displayed set of vertices. Let $\mathcal F_i$ be the set of $i$-cells; thus $\mathcal T_n^d=(\mathcal F_0,\ldots,\mathcal F_d)$ with the standard cubical incidence structure. We call $n$ its \emph{side length} and $d$ its \emph{dimension}.
\end{definition}

    \begin{figure}[h]

\begin{center}
  \ifdefined\POSTER
    \begin{tikzpicture}[scale=3.2]
    \else
    \begin{tikzpicture}[scale=1.5]
    \fi

  \draw[blue!20!black,fill=blue!70!black!70,rounded corners=1, opacity=0.5]%
  (0.9,-0.1) node[below, opacity=1] {$\rotfv{v}{\{ g_{1}\} } $}
  rectangle (1.1,1.1);

  \draw[blue!20!black,fill=blue!70!black!70,rounded corners=1, opacity=0.5]%
  (1.9,2.9) node[below, opacity=1] {$\rotfv{u}{\{ g_{2} \} }$} 
  rectangle (3.1,3.1);

  \draw[blue!20!black,fill=blue!70!black!70,rounded corners=1, opacity=0.5]%
  (3.9,1.9) node[below, opacity=1] {$\rotfv{w}{\{g_{1}, g_{2} \} }$} 
  rectangle (5.1,3.1);

  \draw[step=1cm,gray,very thin] (0,0) grid (5,5);
  \foreach \x in {0,1,2,3,4,5}
  \foreach \y in {0,1,2,3,4,5}
  {
  \node[draw,circle,inner sep=1pt,fill] at (\x,\y) {};
}
\draw[gray,very thin,-> ]  (0,0) to [out=50, in=130] (0,5);
\draw[gray,very thin,-> ]  (1,0) to [out=50, in=130] (1,5);
\draw[gray,very thin,-> ]  (2,0) to [out=50, in=130] (2,5);
\draw[gray,very thin,-> ]  (3,0) to [out=50, in=130] (3,5);
\draw[gray,very thin,-> ]  (4,0) to [out=50, in=130] (4,5);
\draw[gray,very thin,-> ]  (5,0) to [out=50, in=130] (5,5);
\draw[gray,very thin,-> ]  (0,5) to [out=130, in=220] (5,5);
\draw[gray,very thin,-> ]  (0,4) to [out=130, in=220] (5,4);
\draw[gray,very thin,-> ]  (0,3) to [out=130, in=220] (5,3);
\draw[gray,very thin,-> ]  (0,2) to [out=130, in=220] (5,2);
\draw[gray,very thin,-> ]  (0,1) to [out=130, in=220] (5,1);
\draw[gray,very thin,-> ]  (0,0) to [out=130, in=220] (5,0);

\end{tikzpicture}
\end{center}
\caption{ Here we consider a 2D Toric complex. We name the set of generators $S= \{ g_{1},g_{2} \}$, where $g_{1}$ is the generator that sends us one step in the north direction, and $g_{2}$ sends us in the east direction. For vertices $v, u, w$, we mark in blue three faces that are rooted at them. }
\label{fig:Toric}
\end{figure}

\begin{definition}[Chains and Assignments on a Toric Complex]
\label{def:asontoric}
Let $\mathcal T=\mathcal T_n^d$. For each $i$, we denote the space of $i$-dimensional chains by
\begin{equation*}
  C_i(\mathcal T):=(\mathcal F_i;\mathbb F_2)
  =\operatorname{Span}_{\mathbb F_2}\{\rotfv{v}{S'}:
  v\in\mathcal G_n^d,\ S'\subseteq S,\ |S'|=i\}.
\end{equation*}
Here $\rotfv{v}{S'}$ is the standard basis chain corresponding to the geometric face $\rotf{v}{S'}$. Since $\mathcal F_i$ is finite, $C_i(\mathcal T)$ is canonically identified with both $\mathbb F_2^{\mathcal F_i}$ and $\mathbb F_2^{|\mathcal F_i|}$. An \emph{assignment} on the $i$-faces is simply a chain $z\in C_i(\mathcal T)$; we use these terms interchangeably. If $z$ has support on the distinct cells $\theta_{v_j,S'_j}$, then
\begin{equation*}
  z=\sum_{j=1}^{m}\rotfv{v_j}{S'_j}.
\end{equation*}
A basis face is regarded as a chain whenever it occurs in a sum or under a linear operator. We write $f\in z$ when $f$ is a cell supporting $z$, and $v\in z$ when $v$ is a vertex of one of its supporting cells. More generally, the incidence notation $f\in z$ for a lower-dimensional face means that $f$ lies in a cell supporting $z$; this is different from saying that its coefficient in the boundary is nonzero. We write $V(z)$ for the set of vertices of the supporting cells.

For $i>0$, define the \emph{boundary operator} $\partial^-:C_i(\mathcal T)\to C_{i-1}(\mathcal T)$ on basis chains by
\begin{equation*}
  \partial^-\rotfv{v}{S'}
  :=\sum_{g\in S'}\left(
       \rotfv{v}{S'/g}+\rotfv{gv}{S'/g}\right).
\end{equation*}
For $i<d$, the \emph{coboundary operator} $\partial^+:C_i(\mathcal T)\to C_{i+1}(\mathcal T)$ is
\begin{equation*}
  \partial^+\rotfv{v}{S'}
  :=\sum_{g\in S/S'}\left(
       \rotfv{v}{S'\cup g}+\rotfv{g^{-1}v}{S'\cup g}\right).
\end{equation*}
Both operators are extended linearly to chains. We use $S'/g=S'\setminus\{g\}$, $S/S'=S\setminus S'$, and $S'\cup g=S'\cup\{g\}$ throughout, as in \Cref{sec:nota}.

For a vertex $v$, define the \emph{restriction at $v$}, $z|_v$, to retain the cells of $z$ having $v$ among their vertices:
\begin{equation*}
  \left.\rotfv{u}{S'}\right|_v
  :=\begin{cases}
      \rotfv{u}{S'},&v\in V(\theta_{u,S'}),\\
      0,&\text{otherwise}.
    \end{cases}
\end{equation*}
The \emph{positive restriction} $z|_{v+}$ retains only cells \emph{rooted} at $v$:
\begin{equation*}
  \left.\rotfv{u}{S'}\right|_{v+}
  :=\begin{cases}
      \rotfv{u}{S'},&u=v,\\
      0,&\text{otherwise}.
    \end{cases}
\end{equation*}
The maps $|_v$ and $|_{v+}$ extend linearly; the complementary restriction is $z|_{/v}:=z+z|_v$. If $D$ is a specified set of face coordinates, $z|_D$ likewise retains the terms of $z$ supported in $D$. When a local code is written in place of $D$, this notation refers to its underlying face coordinates.

Two distinct $i$-faces are \emph{adjacent} if they share an $(i-1)$-face. A nonzero $i$-chain is \emph{connected} if the adjacency graph on its supporting $i$-faces is connected; its connected components are the chains corresponding to the components of this graph. An $i$-chain $z$ is \emph{rooted at $v$} if all of its supporting $i$-faces have root $v$, that is,
\begin{equation*}
  z\in\operatorname{Span}_{\mathbb F_2}
       \{\rotfv{v}{S'}:S'\subseteq S,\ |S'|=i\}.
\end{equation*}
\end{definition}

We are ready to define the code in this notation.

\begin{definition}[$(d,d')$-Toric Code]
  Let $1\le d' < d$. The side-length-$n$ $(d,d')$-Toric code is the CSS code associated with $\mathcal T_n^d$ whose physical qubits correspond to $\mathcal F_{d'}$. As in the rest of the paper, the check families are indexed by $\mathcal F_{d'-1}$ and $\mathcal F_{d'+1}$ (called the $X$-checks ($Z$-stabilizers) and $Z$-checks ($X$-stabilizers), respectively, in our convention). A lower-dimensional check is supported on the qubits incident to its $(d'-1)$-cell; an upper-dimensional check is supported on the qubits in the boundary of its $(d'+1)$-cell. The corresponding check matrices are the incidence matrices of $\partial^-$ and $\partial^+$.
\end{definition}

\begin{claim}[CSS structure]
\label{claim:csscode}
The $(d,d')$-Toric code is a CSS code.
\end{claim}
\begin{proof}
Give each chain space its standard inner product, in which the basis faces are orthonormal. Directly from the definitions, $\partial^+$ is the transpose of $\partial^-$ in consecutive dimensions:
\begin{equation*}
  \bigl\langle\partial^+\rotfv{v}{A},\rotfv{u}{B}\bigr\rangle
  =\bigl\langle\rotfv{v}{A},\partial^-\rotfv{u}{B}\bigr\rangle.
\end{equation*}
Indeed, either coefficient is one precisely when there is $g\in B$ such that $A=B/g$ and either $u=v$ or $u=g^{-1}v$ (equivalently, $v=gu$).

Moreover, the boundary of a boundary vanishes. On expanding $\partial^-\partial^-\rotfv{v}{S'}$, every term indexed by two distinct deleted directions $g,g'\in S'$ occurs twice, once in each order, so over $\mathbb F_2$ it cancels. Hence $(\partial^-)^2=0$; by transposition, $(\partial^+)^2=0$ as well. In particular, the check matrices associated with the two adjacent cell dimensions are orthogonal, and the corresponding Pauli stabilizers commute, as required for a CSS code.
\end{proof}

The chain formulation will be used for both errors and their syndromes: in the relevant sector, a qubit error is a $d'$-chain and its syndrome is the corresponding boundary chain. We next describe the local vertices visible from a given root.

\begin{definition}[Positive Neighborhood]
\label{def:positiveNei}
For a vertex $v$ and a chain $z$, the \emph{positive one-step neighborhood of $v$ in $z$} is
\begin{equation*}
  \textbf{Str}_{0+}(v,z)
  :=V(z)\cap\{gv:g\in S\}.
\end{equation*}
Thus $\textbf{Str}_{0+}(v,z)$ consists of those vertices supporting $z$ that can be reached from $v$ by one step in a positive generator direction. This is the local neighborhood used in the subsequent sweep-rule and reduced-error arguments.
\end{definition}

\subsection{Sweep Rule}

The sweep-rule is the core of our construction. A decoding cycle begins with each vertex $v$ reading the syndrome on the $X$-checks rooted at $v$ (the positive $X$-checks of $v$) and, based on what it sees, applying a Pauli correction supported on the qubits rooted at $v$. The sweep-rule defines the correction to be over the positive qubits such that it will turn off the positive syndrome. It's not hard to see that for the $(2,2)$-Toric complex, the Toom's and the sweep rules are the same. Compared to other correction cycles on topological codes, the main advantage of the sweep is that it is independent of time.

The invention of the sweep-rule is attributed to Aleksander Kubica and John Preskill \cite{Kubica_2019}. For further development of the sweep-rule, we refer to the work of Michael Vasmer, Dan Browne, and Aleksander Kubica \cite{Vasmer_2021}, which generalizes the rule beyond standard lattices (specifically the Rhombic Dodecahedral lattice).

\begin{definition}[Sweep Rule]
  \label{def:sweeprule}
  Let $\sigma$ be a $d^{\prime}-1$-dimensional chain. The \textit{sweep rule decoding procedure} at a vertex $v \in \mathcal{G}_{n}^{d}$ is defined as follows: Measure the checks rooted at $v$ and look for a rooted chain $\tilde{z}$ at $v$ such that: 
  \begin{enumerate}
    \item The vertices in $\tilde{z}$ are contained in the vertices of $\sigma|_{v+}$, namely: $ \mathbf{Str}_{0+}(v, \tilde{z}) \subseteq \mathbf{Str}_{0+}(v, \sigma|_{v})$ %
    \item The boundary of $\tilde{z}$ agrees with $\sigma$ on the checks rooted at $v$, namely: $(\partial^{-}\tilde{z})|_{v+} = \sigma|_{v+}$.
  \end{enumerate}
  If such an assignment exists, flip $\tilde{z}$.
\end{definition}

The first property is analogous to the second property of 'causal structure' defined in \cite{Kubica_2019}\footnote{In their language: $\diamondsuit(\tilde{z}) = \diamondsuit(\sigma|_{v})$.}. There, they conjectured that vertices that see a syndrome only on their positive checks can suggest a correction that is 'supported' on the vertices of their unsatisfied checks. For lower dimensions, they mentioned that the property can be easily checked. Here, we provide a proof for any dimensions where the face dimension is greater than one. We use the language of cubical complexes; however, since the 'positive' local view at a vertex is identical for both cubical and simplicial complexes, generalizing the proof for simplicial complexes is a matter of adjusting notations. By that, we affirmatively answer their conjecture. Moreover, the proof will provide us with an algorithm to find $\tilde{z}$, which is significantly better than running all options.

\begin{claim}
  \label{claim:findcor}
  For $d^{\prime}\ge 2$, consider a nontrivial $d^{\prime}-1$-dimensional chain $\sigma$. If there is a $d^{\prime}$-dimensional chain $z$ that is rooted at $v$ and whose boundary agrees with $\sigma|_{v+}$, then there is a $d^{\prime}$-dimensional chain $\tilde{z}$ rooted at $v$ that satisfies the sweep-rule conditions.
\end{claim}
\begin{proof} 
We prove it by induction on $d^{\prime}$. Without loss of generality, we can assume that $\sigma$ has support only over the positive faces of $v$, namely $\sigma|_{v+} = \sigma$.
  \begin{enumerate}
    \item Base. Consider $d^{\prime} = 2$ and let $d$ be any integer equal to or greater than $3$. We argue that the support size of $\sigma$ has to be even. To see that, denote the faces in $z$ by $\rotfv{v}{S^{\prime}_{i}}$, where $S^{\prime}_{i} = \{g_{i,1},g_{i,2}\}$. Using the fact that $\sigma$ is the boundary of $z$ at $v$, we have:
      \begin{equation*}
        \begin{split}
          \sigma = (\partial^{-}(z))|_{v} = \sum_{ \rotfvs{} \in z}\sum_{ g \in S^{\prime} } \rotfv{v}{S^{\prime } / g } = \sum_{ i } \rotfv{v}{ \{g_{i,1}\} }+ \rotfv{v}{ \{g_{i,2}\} } 
        \end{split}
      \end{equation*}
So the number of terms in the sum is even, and since a term is zeroed out only if it appears an even number of times, we have that after excluding the vanishing terms, $\sigma$ has an even number of faces. Since the number of faces is even, there is a perfect matching between them. It's easy to see that the $2$-chain $\tilde{z}$, each of whose faces corresponds to the union of the two edges in a match, satisfies the sweep-rule conditions.
    \item Assume the correctness of the claim for every face dimension lower than $d^{\prime}$.   Now for a face dimension $d^{\prime}$ and an arbitrary generator $g \in S$ such $gv \in \sigma$, divide $\sigma$ into a sum of two chains, the first contains the faces of $\sigma$ that are supported on the vertex $gv$ while the second contains the complement. Denote them by $\sigma|_{gv}$ and $\sigma|_{/ gv}$, so $\sigma = \sigma|_{gv} + \sigma|_{/ gv}$. 

      Now let $ \sigma^{*}$ be the chain obtained from $\sigma|_{gv}$ by removing from any face of it the vertex $gv$, thus $\sigma^{*}$ is a $d^{\prime}-2$-dimensional chain. Moreover, let $z^{*}$ be the $d^{\prime}-1$-dimensional chain obtained from $z|_{gv}$ by excluding $gv$ from its faces. Observe that 
      \begin{equation*}
        \begin{split}
          (\partial^{-} z^{*})|_{v} =  \sum_{\rotfvs \in z|_{gv}}\sum_{g^{\prime} \in S^{\prime}/g}\rotfv{v}{S^{\prime}/\{g,g^{\prime}\}} = \sum_{\rotfvs \in (\partial^{-} z)|_{gv}}\rotfv{v}{S^{\prime}/\{g\}} = \sigma^{*}
        \end{split}
      \end{equation*}
      Thus, by the induction assumption, there is a $d^{\prime}-1$-dimensional chain $\tilde{z}^{*}$ such that: 
  \begin{enumerate}
    \item $ \mathbf{Str}_{0+}(v, \tilde{z}^{*}) \subseteq \mathbf{Str}_{0+}(v, \sigma^{*}|_{v})$  
    \item $(\partial^{-}\tilde{z}^{*})|_{v+} = \sigma^{*}|_{v+}$
  \end{enumerate}
The first condition implies that $gv \notin \tilde{z}^{*}$. Hence, we can define the $d^{\prime}$-dimensional chain $\tilde{z}$ by adding the vertex $gv$ to each face of $\tilde{z}^{*}$. From the second it follows that: 
\begin{equation*}
  \begin{split}
    (\partial^{-} \tilde{z})|_{v} &= \bigl(\partial^{-} \bigl(  \sum_{\rotfvs \in \tilde{z}^{*} } \rotfv{v}{S^{\prime}\cup \{g\}}  \bigr)\bigr)|_{v} =  \sum_{\rotfvs \in \tilde{z}^{*} } \bigl( \sum_{g^{\prime} \in S^{\prime}/g } \rotfv{v}{S^{\prime} / \{g^{\prime}\} \cup \{g\}}  +   \rotfv{v}{S^{\prime}} \bigr)\\
    &=  \sum_{\rotfvs \in (\partial^{-}\tilde{z}^{*})|_{v} } \rotfv{v}{S^{\prime}\cup \{g\}} + \tilde{z}^{*} = \sigma|_{gv} + \tilde{z}^{*} 
  \end{split}
\end{equation*}
Therefore, 
\begin{equation*}
  \begin{split}
    \sigma + (\partial^{-} \tilde{z})|_{v} &= \sigma|_{gv} + \sigma|_{/ gv} + \sigma|_{gv} + \tilde{z}^{*} \\ 
    &= \sigma|_{/ gv} + \tilde{z}^{*}
  \end{split}
\end{equation*}
Thus we reduce the question to handle the $d^{\prime}-1$-dimensional chain $\sigma + (\partial^{-} \tilde{z})|_{v}$. However, since $gv \notin  \sigma + (\partial^{-} \tilde{z})|_{v}$ it follows that: 
\begin{equation*}
  \begin{split}
    |\mathbf{Str}_{0+}(v, \sigma + (\partial^{-} \tilde{z})|_{v} \, )| \le  |\mathbf{Str}_{0+}(v, \sigma )| - 1
  \end{split}
\end{equation*}
Thus, we can repeat recursively until the resulting $d^{\prime}-1$ chain is 'supported' on fewer than $d^{\prime}$ vertices. In that case, the chain must be trivial. That is because the resulting chain is the $\partial^{-}$-boundary of the rooted chain $z + \tilde{z}$ at vertex $v$ restricted to the positive faces of $v$. So, any nontrivial $d^{\prime}-1$ dimensional face in the resulting $d^{\prime}-1$ chain introduces $d^{\prime}$ vertices, one of them is $v$ itself, and the other $d^{\prime}-1$ are vertices in $\mathbf{Str}_{0+}(v, \sigma + (\partial^{-} \tilde{z})|_{v} \, )$.
  \end{enumerate}
\end{proof}
The algorithm given in the proof of \Cref{claim:findcor} for finding $\tilde{z}$ is as follows: given $\sigma$, pick a $g \in S$ such that $gv \in \sigma$. Then, define $\sigma^{*}$ by removing $gv$ from each face of $\sigma|_{gv}$. Let $\tilde{z}^{*}$ be the result of the recursive call on $\sigma^{*}$, and return '$\tilde{z}^{*} \cup \{gv\}$' plus the recursive call on $\sigma|_{/gv} + \tilde{z}^{*}$.

We stop the recursion whenever either $d^{\prime}=2$ (returning the perfect-matching 2-chain from the base case of \Cref{claim:findcor} ) or $|\mathbf{Str}_{0+}(v, \sigma|_{v})| < d^{\prime}-1$ (returning $0$). If the $\partial^{-}$-boundary of the returned chain doesn't match $\sigma|_{v+}$, then there is no chain that satisfies the sweep rule constraints. Otherwise, the returned chain is a valid sweep-rule suggestion.

The work in each recursion step consists of decomposing $\sigma$ into $\sigma|_{gv}$ and $\sigma|_{/gv}$ and summing ${ d \choose d^{\prime} }$-length vectors. We summarize by introducing the following claim which gives the running time of the algorithm in the RAM computation model.
\begin{claim}
  Let $ T(d, d', k) : \mathbb{N}^{\times 3} \rightarrow \mathbb{R} $ be the upper bound function for the running time of the algorithm given in \Cref{claim:findcor}, where $d$ is the dimension of the skeleton, $d^{\prime}$ is the face dimension, and $k$ is the number of vertices supporting the local view, namely $k = |\mathbf{Str}_{0+}(v, \sigma|_{v} )|$. Then:
  \begin{equation*}
    \begin{split}
      T(d, d', k) = O\bigl( 2^{k} \max_{d^{\star}\le d^{\prime}}{ d \choose d^{\star} } \bigr)
    \end{split}
  \end{equation*}
\end{claim}
\begin{proof}
  In each recursive call, the vertex support of $\sigma$ matches a different subset of the vertices in $\mathbf{Str}_{0+}(v, \sigma|_{v})$; thus, the number of recursive calls is at most the number of subsets, namely $2^{k}$. The decomposition of $\sigma$ into $\sigma|_{gv}$ and $\sigma|_{/gv}$, and the summation $\sigma|_{/gv} + \tilde{z}^{*}$ are linear in the vector's length, namely linear in $\max_{d^{\star} \le d^{\prime} } {d \choose d^{\star}}$. So we bound the total work by the number of recursive steps multiplied by the maximal work at each recursive step. That becomes at most:
\begin{equation*}
  \begin{split}
    T(d,d^{\prime},k) \le O\bigl( 2^{k} \max_{d^{\star} \le d^{\prime}} { d \choose d^{\star} } \bigr) 
  \end{split}
\end{equation*}
\end{proof}

\begin{figure}[h]
  \centering
  \ifdefined\POSTER
    \begin{tikzpicture}[scale=4]
    \else
    \begin{tikzpicture}[scale=1.5]
    \fi

\tikzset{xyp/.style={canvas is xy plane at z=#1}}
\tikzset{xzp/.style={canvas is xz plane at y=#1}}
\tikzset{yzp/.style={canvas is yz plane at x=#1}}

\draw[blue!20!black,fill=blue!70!black!70,rounded corners=1,opacity=0.5, xzp=1]%
(0.9,1.1) rectangle (2.1,0.9);

      \draw[blue!20!black,fill=blue!70!black!70,rounded corners=1,opacity=0.5, xzp = 1]%
      (0.9,1.1) rectangle (1.1,-0.1);

\draw[blue!20!black,fill=blue!70!black!70,rounded corners=1,opacity=0.5, xzp=0]%
(-0.1,2.1) rectangle (1.1,1.9);

      \draw[blue!20!black,fill=blue!70!black!70,rounded corners=1,opacity=0.5, xzp = 0]%
      (-0.1,2.1) rectangle (0.1, 0.9);

      \draw[blue!20!black,fill=blue!70!black!70,rounded corners=1,opacity=0.5, xyp = 2]%
      (-0.1,-0.1) rectangle (0.1, 1.1);

      \node[below] at (1,1,1) { $v$ } ;
      \node[below = 2pt] at (0,0,2) { $u$ } ;

\foreach \x in {0,1,2} {
  \foreach \y in {0,1,2} {
    \draw[gray] (\x,\y,0) -- (\x,\y,2);
  }
}

\foreach \x in {0,1,2} {
  \foreach \z in {0,1,2} {
    \draw[gray] (\x,0,\z) -- (\x,2,\z);
  }
}

\foreach \y in {0,1,2} {
  \foreach \z in {0,1,2} {
    \draw[gray] (0,\y,\z) -- (2,\y,\z);
  }
}

\foreach \x in {0,1,2} {
  \foreach \y in {0,1,2} {
    \foreach \z in {0,1,2} {
      \fill (\x,\y,\z) circle (1pt);
    }
  }
}

\draw (2.5,1) node { $\rightarrow$ };
\end{tikzpicture} \ifdefined\POSTER \begin{tikzpicture}[scale=4] \else \begin{tikzpicture}[scale=1.5] \fi

\tikzset{xyp/.style={canvas is xy plane at z=#1}}
\tikzset{xzp/.style={canvas is xz plane at y=#1}}
\tikzset{yzp/.style={canvas is yz plane at x=#1}}

      \draw[green!20!black,fill=green!70!black!70,rounded corners=1,opacity=0.5, xzp = 1]%
      (0.9,1.1) rectangle (2.1,-0.1);

      \node[below] at (1,1,1) { $v$ } ;
      \node[below = 2pt] at (0,0,2) { $u$ } ;

\foreach \x in {0,1,2} {
  \foreach \y in {0,1,2} {
    \draw[gray] (\x,\y,0) -- (\x,\y,2);
  }
}

\foreach \x in {0,1,2} {
  \foreach \z in {0,1,2} {
    \draw[gray] (\x,0,\z) -- (\x,2,\z);
  }
}

\foreach \y in {0,1,2} {
  \foreach \z in {0,1,2} {
    \draw[gray] (0,\y,\z) -- (2,\y,\z);
  }
}

\foreach \x in {0,1,2} {
  \foreach \y in {0,1,2} {
    \foreach \z in {0,1,2} {
      \fill (\x,\y,\z) circle (1pt);
    }
  }
}
\end{tikzpicture}
\caption{
An example of applying the sweep rule within the $(3,2)$-Toric code. In the $(3,2)$-Toric code, bits are placed on the squares and $Z$-stabilizers (parity checks) on the edges. The view is restricted to two vertices, $u$ and $v$. On the left, the unsatisfied positive checks of $u$ and $v$ are highlighted in blue. On the right, the bits to be flipped are marked in green. Given the syndrome, vertex $v$ flips one of its positive faces, whereas for $u$, there is no subset of positive faces that, when flipped, satisfies local checks. Therefore, $u$ flips no bit.
}
  \label{fig:sweep}
\end{figure}

Later (see \Cref{def:trans} and \Cref{def:htrans}), we will see that an isomorphism between the $(d,d^{\prime})$-Toric code and its dual, which is the $(d,d-d^{\prime})$-Toric code, can be realized by transversal Hadamards followed by the following permutation over the qubits.
  \begin{equation*}
    \begin{split}
      \phi(\rotfv{v}{S^{\prime}}) \rightarrow \rotfv{ S^{\prime}v }{S/S^{\prime}} 
    \end{split}
  \end{equation*}
A full decoding cycle consists of four steps: first, applying the sweep rule regarding the $X$-syndrome; next, moving to the dual code; then applying the sweep rule again; and finally getting back to the code.
\begin{definition}[Sweep cycle]
We define the \textit{sweep cycle} to be the following procedure:
\begin{enumerate}
    \item For each vertex, perform the sweep rule.
    \item Apply the Hadamard transform $\mathbf{H}^{\otimes\mathcal{F}_{d^\prime}^{n}}$ followed by the $\phi$-transpose.
    \item Again, apply the sweep rule at each vertex.
    \item Reverse stage 2 by reordering the qubits according to the $\phi^{-1}$-transpose, and then apply the Hadamard transform $\mathbf{H}^{\otimes\mathcal{F}_{d^\prime}^{n}}$.
\end{enumerate}
\end{definition}

\subsection{Infinite Sides Torics}
\label{subsection:infinitetoric}
The Toric skeleton provides a notation for positive directions. However, to demonstrate that an `error cluster` shrinks as a result of the decoding cycle, or at least does not expand too much during the operation of the logical operator, we will need a way to mark extremities (for example, the northernmost point). To achieve this, we will analyze the view over the finite Toric as a projection derived from a view over the infinite plane. For this purpose, given our initial (subjected) quantum circuit, we construct a representation in which any gate that originally acted on a quantum register, namely on the finite Toric, is mapped to infinite translations of it, each acting on qubits on the infinite plane. Using the Wrapping Lemma, proved in \cite{schumacher2004reversiblequantumcellularautomata}, and combining that the mapping respects a periodic condition, the finite and the infinite presentations are isomorphic.

The following definition extends the notion of location to differentiate between a coordinate that encodes a register and a coordinate of the location inside the register. This will be useful, as in the infinite plane representation, each register will be mapped to a unique infinite register. %
\begin{definition}[Toric register coordinate system]
  Let $C=\{u_i\}_i$ be a quantum circuit of width $w$ with registers $\mathcal{R}=R_{1},\ldots,R_{k}$, each encoded in the $(d,d^{\prime})$-Toric code, induced by a $d$-dimensional Toric with a side length $n$. We define the \textit{Toric register coordinate system} to be the coordinate system that encodes spatial location with two coordinates as follows: In the first coordinate, the register number is recorded, and in the second, the qubit is encoded by the face associated with it. We call the first the \textit{register coordinate} and the second the \textit{inner coordinate}.
\end{definition}

We will need two types of gate translation. The first is the standard spatial translation, which means that any of the inner coordinates of the gate are advanced by the same vector. The second is \textit{rooted translation}. Here, we assume that the locations of the gate are given in a relative-to-root fashion, meaning that there is a root vertex, and all the locations are given by positive paths from the root. Then, we first translate the root and then advance the inner coordinate of any of the other locations to be on the path from the translated root. When considering a product gate, the translations act the same.   
\begin{definition}[Translation by vertex $v$]
  Let $u$ be a gate, and denote the inner coordinates of $u$'s locations by $\rotf{v_{1}}{J_{1}}, \ldots, \rotf{v_{\Delta}}{J_{\Delta}}$. The translation of $u$ by vertex $v$ is defined to be the gate obtained from $u$ by replacing the inner coordinates of its locations by: $\rotf{v v_{1}}{J_{1}}, \ldots, \rotf{v v_{\Delta}}{J_{\Delta}}$, where the multiplication $v \cdot v_{i}$ is with respect to the infinite cover $\mathcal{G}^{d}$.
  Furthermore, for a $d$-dimensional Toric complex with side length $n$, we define a translation by an integer vector $a \in \mathbb{Z}^{d}$ as a translation by the vertex:
   \begin{equation*}
     \begin{split}
       (g_{1}^{n})^{a_{1}} (g_{2}^{n})^{a_{2}} \cdots (g_{d}^{n})^{a_{d}}
     \end{split}
   \end{equation*}
    We denote the translation by an integer vector $a$ as $a * u$.
\end{definition}

\begin{definition}[Rooted gate at vertex $v$]
  Let $u$ be a quantum gate that acts on a register whose qubits are associated with a Toric complex at dimension $d$. We say that a gate $u$ is rooted at vertex $v$ if the inner coordinate of its location can be encoded by lists of pairs $(I_{1}, J_{1}), (I_{2}, J_{2}), \ldots, (I_{l}, J_{l})$ such that $I_{i} \in S^{d}, J_{i} \subset S$, and the $i$th qubit on which it acts is in the face $\rotf{ I_{i} v }{  J_{i}  }$.
\end{definition}

\begin{definition}[Root translation in Register System] 
  For a gate $u$ that is a rooted gate at vertex $v$, we define its translation by a vertex $v^{\prime}$ to be the gate obtained by copying $u$ and replacing its inner location encoding the root by $v^{\prime}v$. 

Similarly to the standard translation by an integer vector, we extend the action to a root translation by an integer vector $a \in \mathbb{Z}^{d}$, as the root translation by the vertex:
   \begin{equation*}
     \begin{split}
       (g_{1}^{n})^{a_{1}} (g_{2}^{n})^{a_{2}} \cdots (g_{d}^{n})^{a_{d}}
     \end{split}
   \end{equation*}
   We denote the root translation by an integer vector $a$ as $a *_{r} u$.
\end{definition}

Now we are ready to give an explicit description of the mapping. Given a quantum circuit $C$, we name its presentation over the infinite plane the \textit{$\infty$-lift of $C$}.
\begin{definition}
  \label{def:extension}
  Let $n$ be an integer, and let $C = \{u_{i}\}_{i=1}$ be a quantum circuit with width $w$ and quantum registers $\mathcal{R} = R_{1}, \ldots, R_{k}$ such that each register $R_{i}$ encodes a $(d,d^{\prime})- $Toric code with side length $n$. We define the \textit{$\infty$-lift of $C$}, denoted by $C^{\infty}$, as follows:
  \begin{enumerate}
    \item Each register $R_{i}$ in $\mathcal{R}$ is mapped to a register $R^{\prime}_{i}$, whose qubits are associated with the $d^{\prime}$-dimensional faces of the infinite $d$-dimensional Toric. 
    \item The gate $u_{i} \in C$ is mapped to an infinite collection of gates, each being a rooted translation by an integer vector, as follows:
      \begin{equation*}
        \begin{split}
          u_{i}  \mapsto  \bigcup_{ a \in \mathbb{Z}^{d} } \left\{ a *_{r} u_{i} \right\} 
        \end{split}
      \end{equation*}
  \end{enumerate}
  The restriction of the qubits in $C^{\infty}$'s registers to finite Toric complexes with side length $n$ will be denoted by $\cdot |_{[w]}$, and it includes all the faces whose root vertex has powers (the powers of $g_{1}, \ldots, g_{d}$) in the range $[0, n-1]$. For a set of faults $\mathcal{E} = \{ f_{i} \}_{i=1}$, and measurements outcomes $\alpha \in \mathbb{F}_2^{(\cdot)}$, we denote by $C^{\infty}[\mathcal{E},\alpha]$ the \textit{$\infty$-lift of $C[\mathcal{E},\alpha]$}. 
  \paragraph{ }
We define the $\infty$-lift of a Pauli operator $P \in \mathcal{P}^{(\cdot)}$ to be the result of lifting the circuit composed only of $P$. Denote it by $P^{\infty}$. Finally, the $\infty$-lifting of a chain $x \in (\mathcal{F}^{n}_{d^{\prime}}; \mathbb{F}_{2})$ is the chain in the infinite complex $(\mathcal{F}_{d^{\prime}}; \mathbb{F}_{2})$ obtained by taking all the translations of $x$ by integer vectors.
\end{definition}

The infinite presentation of a circuit agrees with it in terms of action. We give, without proof, an adaptation of the Wrapping Lemma, proved in \cite{schumacher2004reversiblequantumcellularautomata}.
\begin{claim}[ Adaptation of \cite{schumacher2004reversiblequantumcellularautomata} ]
The $\infty$-lifting is an isomorphism. In particular, consider a conditioned subjected quantum circuit $C[\mathcal{E},\alpha]$, such that it has a propagation $P_{1},P_{2},\ldots$, it holds that:
  \begin{equation*}
    \begin{split}
    P_{i}^{\infty} =  (C[\mathcal{E},\alpha]|_{ i} P_{i-1})^{\infty} =   C[\mathcal{E},\alpha]|_{ i}^{\infty} P_{i-1}^{\infty} 
    \end{split}
  \end{equation*}
\end{claim}

The computation DAG lifts naturally to the $\infty$-lift gates that arise as translates of the same local gate act on disjoint translated supports inside different periods, so they may be executed simultaneously. The same periodic lifting applies to the underlying geometric objects, passing from the finite Toric complex to its infinite cover.

\paragraph{}
We close that subsection by introducing the \textit{potential walls}. Analyzing the infinite presentation enables us to bound a subset of errors by 'triangles'. A potential wall is a linear function $\mathbb{R}^{d} \rightarrow \mathbb{R}$. For the subset of vertices (that might be the subset that supports some error), the maximal value of the potential wall on it is achieved at the point of tangency between the subsets and its bounding triangle. With the assistance of potential walls, we can discuss the endpoint of a bunch of qubits and the 'size' of the (hyper)triangle that bounds them. Soon, the qubits will be associated with the support of an error, and showing that an action decreases the maximums of the potential walls implies that the error shrinks. 
The next definition explains how the embedding of vertices of the infinite grid in $\mathbb{R}^{d}$ is handled.
\begin{definition}[The standard embedding]
  Consider the $d$-dimensional infinite grid $\mathcal{G}^{d}$. The \textit{standard embedding} $\mathcal{G}^{d} \rightarrow \mathbb{R}^{d}$ is given by fixing a vertex in $\mathcal{G}^{d}$ to be sent to the origin and then mapping the generators of the grid to the standard basis. We extend notation as follows: for a vertex $v \in \mathcal{G}^{d}$, a generator $g \in S$, and a coordinate $i \in [d]$, such that under the embedding $g \rightarrow i$, we denote by $v_{i} = v_{g}$ the $i$'th coordinate of the image of $v$.
\end{definition}

Now we are ready to define the \textit{potential walls}:
\begin{definition}[Potential Walls]
  We define the \textit{potential walls} to be the $d+1$ functions $L_{1},L_{2},..,L_{d+1}$, from the vertices of $\mathcal{G}^{d}$ to the reals as follows:
  \begin{enumerate}
    \item For any coordinate $i \in [d]$, we define $L_{i} \colon \mathcal{G}^{d} \rightarrow \mathbb{R}$ as $L_{i}(v) \colonequals v_{i}$. In addition, for the generator $g \in S$ which is mapped to $i$, i.e $g \rightarrow i $, we identify $L_{g} = L_{i}$. 
    \item We define $L_{d+1} \colon \mathcal{G}^{d} \rightarrow \mathbb{R}$ by $L_{d+1}(v) \colonequals -\sum_{i = 1}^{d}v_{i}$.
  \end{enumerate}
  Consider an assignment $z$ over the faces. For a vertex $v \in z$, and a potential wall $l \in \{ L_{1},..,L_{d+1}\} $, we say that $v$ is a \textit{local maximum} of $l$ in $z$ if for any neighbor $u$ of $v$, such that $u \in z$, we have $l(v) \ge l(u)$. When the assignment $z$ is clear from the context, we might omit it and briefly say that $v$ is a local maximum of $l$. In addition we define the function $\Size{ \cdot } : 2^{\mathcal{G}^{d}} \rightarrow \mathbb{R} $, to act on a finite subset of points $A \subset \mathcal{G}^{d}$, as follows:   
  \begin{equation*}
    \begin{split}
      \Size{A} \colonequals \sum_{g \in S}\max_{v \in A}L_{g}(v) + \max_{v \in A}L_{d+1}(v) 
    \end{split}
  \end{equation*}
and extend the function to act over assignment on faces $z$, by taking its action on the set of vertices that support $z$.
\end{definition}
The following claim shows that size has a similar behavior to a metric:
\begin{claim}
  \label{claim:tringinq}
  Let $A,B \subset \mathbb{R}^{d}$ be finite subsets such that $A \cap B \neq \emptyset$ then: 
  \begin{equation*}
    \begin{split}
      \Size{A\cup B} \le \Size{A} + \Size{B}
    \end{split}
  \end{equation*}
\end{claim}
\begin{proof}
  Let $p$ be a point in $A\cap B$. Then: 
  \begin{equation*}
    \begin{split}
      \Size{A\cup B} &= \sum_{i \in [d+1]}\max_{v \in A\cup B} L_{i}(v) \\
      & =  \sum_{ i \in [d+1]} \max \bigl\{ \max_{v \in A } L_{i}(v) , \max_{v \in B } L_{i}(v) \bigr\}  \\
      & =  \sum_{ i \in [d+1]}  \max_{v \in A } L_{i}(v) + \max_{v \in B } L_{i}(v) -  \min \{   \max_{v \in A } L_{i}(v), \max_{v \in B } L_{i}(v) \} \\
      & \le  \sum_{ i \in [d+1]}  \max_{v \in A } L_{i}(v) + \max_{v \in B } L_{i}(v) -  L_{i}(p)  \\
      & \le \Size{A} + \Size{B}   - \sum_{ i \in [d+1]}  L_{i}(p)  =  \Size{A} + \Size{B} 
    \end{split}
  \end{equation*}
Where in the last pass, we used that the size of a single point is zero.
\end{proof}

\subsection{Reduced Error}
We represent Pauli errors as chains on the $d^{\prime}$-dimensional faces. Their syndromes are the corresponding boundary chains. In particular, $X$-errors can be characterized as $d^{\prime}$-dimensional chains $z \in C_{d^{\prime}}(\mathcal T_\infty^d)$ such that $\partial^{-} z \neq 0$. In this section, we assume dealing with the $\infty$-lift of a circuit (see \Cref{def:extension}), so the cubical complex is the $d$-dimensional grid with infinite side length. The \textit{reduced} form of an error is its representative, a translation by a stabilizer, which minimizes its size. Here we prove that the sizes of a reduced error and its boundary (derived from $\partial^{-}$) are the same\footnote{In fact, the hypertriangles bounding them are the same.}. In particular, they share poles of the potential walls. That will come into use in analyzing the behavior of the sweep rule, as it is defined to decide on correction upon the syndrome it sees.

For this purpose we introduce two local codes that describe local views of vertices. They are the \textit{stabilizers code at vertex} $v$ and the \textit{small code at vertex} $v$. The first corresponds to stabilizers rooted at the vertex, while the second is the collection of local views on faces rooted at $v$ that satisfy the checks rooted at $v$. Then, in \Cref{claim:agreement}, we consider a vertex whose $X$-checks don't admit a syndrome, meaning that its local view is in its \textit{small code}, and show that there is a codeword of its \textit{stabilizers code} that agrees with its positive faces. Addition by that stabilizer yields a new chain excluding $v$. If $v$ were a unique maximum of $L_{d+1}$, then the vertices on the yielded chain would have a lower value of $L_{d+1}$. Subsequently, since the exclusion decreases the size, it follows that the `corner`—the vertex achieving the maximal $L_{d+1}$ value—of a reduced error sees a syndrome.

\begin{definition}[Stabilizers Code at Vertex $v$]
  We define \textit{stabilizer code at vertex} $v$, denoted by $\text{Stab}_{v}$, as the restriction of the $(d,d^{\prime})$-Toric code to the $d^{\prime}$-dimensional faces such that their vertices are at a distance of at most one positive step in each direction from $v$. That is equivalent to the union of all the $d^{\prime}$-dimensional faces rooted at $v$ with the $d^{\prime}$-dimensional faces rooted at each of $v$'s siblings $gv$, excluding faces containing $g$ as their generator. Namely:
  \begin{align*}
    \text{Stab}_{v} \colonequals  \bigl\{ z : \partial^{-} z  &= 0 \text{ and } \\
    & z \in \operatorname{Span}_{\mathbb F_2} \big(  \big\{  \rotfvs{} : S^{\prime} \subseteq S, |S^{\prime}| = d^{\prime}  \big\} \\
   & \ \ \ \ \ \ \ \ \ \ \ \ \  \bigcup_{g \in S} \big\{ \rotfv{gv}{S^{\prime}} : S^{\prime} \subseteq S/g , |S^{\prime}|=d^{\prime} \big\} \big) \bigr\}
  \end{align*}
\end{definition}
The checks of the code, are of the following three types: 
\begin{enumerate}
  \item Checks that are rooted at $v$, each identified by choosing $d^{\prime}-1$ generators from $S$. Namely: 
    \begin{equation*}
      \begin{split}
        \big\{ \rotfvs{}: S^{\prime} \subseteq S , |S^{\prime}|=d^{\prime}-1  \big\} 
      \end{split}
    \end{equation*}
    For each check, the bits that connect to it, are given by restricting the coboundary of the check to bits domain of $\text{Stab}_{v}$:  
    \begin{equation*}
      \begin{split}
        \big( \partial^{+} \rotfvs{} \big)|_{\text{Stab}_{v}} & =  \big(   \sum_{g \in S/S^{\prime}} \rotfv{v}{S^{\prime} \cup g} + \sum_{g \in S/S^{\prime}} \rotfv{g^{-1}v}{S^{\prime}\cup g}  \big)|_{\text{Stab}_{v}} =     \sum_{g \in S/S^{\prime}} \rotfv{v}{S^{\prime} \cup g} 
      \end{split}
    \end{equation*}
  \item Checks are rooted at the positive sibling of $v$. Consider such a sibling, let's say $gv$ (namely the vertex that is given by stepping in the direction $g \in S$ from $v$). Notice that any face rooted at $gv$ that contains $ggv$ cannot be contained in faces rooted at $v$. Thus, the checks of $gv$ correspond to choosing subsets from $S/g$. Namely:
    \begin{equation*}
      \begin{split}
      \bigcup_{g \in S}\big\{ \rotfv{gv}{S^{\prime}}: S^{\prime} \subseteq S/g , |S^{\prime}|=d^{\prime}-1  \big\}
      \end{split}
    \end{equation*}
    And each check is supported by:
    \begin{equation*}
      \begin{split}
        \big( \partial^{+} \rotfv{gv}{S^{\prime}} \big)|_{\text{Stab}_{v}} & = 
        \big(   \sum_{g^{\prime} \in S/S^{\prime}} \rotfv{gv}{S^{\prime} \cup g^{\prime}} + \sum_{g^{\prime} \in S/S^{\prime}} \rotfv{g^{\prime -1}gv}{S^{\prime}\cup g^{\prime}} \big)|_{\text{Stab}_{v}} \\
        &  =    \bigl(\sum_{ \substack{ g^{\prime} \in S/S^{\prime} \\ g^{\prime } \neq g }} \rotfv{g v}{S^{\prime} \cup g^{\prime}} \bigr) + \rotfv{v}{S^{\prime}\cup g}
      \end{split}
    \end{equation*}
  \item Checks whose roots are at a two-step distance from $v$, namely vertices of the form $g^{\prime}gv$ where $g, g^{\prime} \in S$, and $g \neq g^{\prime}$. By the arguments presented in the second type, the checks are the $d^{\prime}-1$ faces which do not have the generators on the path leading from $v$ to them, i.e., $g, g^{\prime}$, in their generator set.
    \begin{equation*}
      \begin{split}
        \bigcup_{\substack{g,g^{\prime}\in S\\g\ne g^{\prime}}}\big\{ \rotfv{g^{\prime}gv}{J}: J \subseteq S/\{g,g^{\prime}\}, |J|=d^{\prime}-1 \big\}
      \end{split}
    \end{equation*}
  And for each check, its support is:
    \begin{equation*}
      \begin{split}
        \bigl(\partial^+\rotfv{g'g v}{J}\bigr)|_{\text{Stab}_v}
        &=\left(\sum_{h\in S/J}\bigl(\rotfv{g'gv}{J\cup h}
                +\rotfv{h^{-1}g'gv}{J\cup h}\bigr)\right)|_{\text{Stab}_v}\\
        &=\rotfv{gv}{J\cup g'}+\rotfv{g'v}{J\cup g}
      \end{split}
    \end{equation*}
\end{enumerate}

\begin{definition}[Small Code at Vertex $v$]
For a vertex $v$, we denote by $\mathcal{C}_{v}$ the \textit{small code at vertex $v$}, which consists of all the $d^{\prime}$-dimensional chains rooted at $v$ that satisfy the checks rooted at $v$, namely:
  \begin{align*}
    \mathcal{C}_{v} \colonequals  \bigl\{ z : (\partial^{-} z)|_{v+}  &= 0 \text{ and } \\
    & z \in \operatorname{Span}_{\mathbb F_2} \big(  \big\{  \rotfvs{} : S^{\prime} \subseteq S, |S^{\prime}| = d^{\prime}  \big\} \big) \bigr\} 
  \end{align*}
The checks of the code are exactly the first type checks of $\text{Stab}_{v}$.
\end{definition}
\begin{claim}
  \label{claim:agreement}
  Let $v$ be a vertex. For any $d^{\prime}\ge 2$, $d\ge d^{\prime}+1$, and any $z \in \mathcal{C}_{v}$, there is a stabilizer $w \in \text{Stab}_{v}$ such that $z = w|_{v}$; namely, $z$ and $w$ agree on the faces rooted at $v$.
\end{claim}

\begin{proof}
  By induction on the dimensions $d, d'$. We extend the notation of the stabilizer and the small codes at vertex $v$ to $\mathcal{C}_{v}^{J}, \text{Stab}_{v}^{J}$, where $J$ is a subset of $S$, to refer to the codes derived from the skeleton generated by $J$. Showing that the belonging $z\in \mathcal{C}_{v}^{J}$ implies the existence of $w \in \text{Stab}_{v}^{J}$ such that $w|_{v} =z|_{v}$ proves the claim.
  \begin{enumerate}
    \item Base. For the base cases, let $d$ be any number greater than or equal to three, and let $d^{\prime}$ be two. First, observe that, in the case $d^{\prime}=2$, the code $\mathcal{C}_{v}^{J}$ is a Tanner code, whose variables correspond to pairs of the form $\{vg_{i}, vg_{j} \}$ and checks are associated with vertices of the form $\{vg_{i} \}$. Therefore, any codeword is a cycle. Let:
      \begin{equation*}
        \begin{split}
          \{vg_{1},vg_{2}\},\{vg_{2},vg_{3}\},..,\{vg_{k},vg_{1}\}
        \end{split}
      \end{equation*}
    Be the cycle corresponding to $z$. Namely:
      \begin{equation*}
        \begin{split}
          z = \rotfv{v}{\{g_{1},g_{2}\}} + \rotfv{v}{\{g_{2},g_{3}\}} + ..+ \rotfv{v}{\{g_{k},g_{1}\}}
        \end{split}
      \end{equation*}
    And consider the stabilizer $w$ defined as:
      \begin{equation*}
        \begin{split}
          w =  \partial^{-}\left(  \rotfv{v}{\{g_{1},g_{2},g_{3}\}} + \rotfv{v}{\{g_{1},g_{3},g_{4}\}} + .. +  \rotfv{v}{\{g_{1},g_{i},g_{i+1}\}} + .. + \rotfv{v}{\{g_{1},g_{k-1},g_{k}\}} \right)
        \end{split}
      \end{equation*}
    Clearly, $w \in \text{Stab}_{v}^{J}$, and in addition its restriction to $v$ equals:
      \begin{equation*}
        \begin{split}
          w|_{v} &=  \left( \sum_{i=2}^{k-1}\partial^{-}\rotfv{v}{\{g_{1},g_{i},g_{i+1}\}}  \right)|_{v} = \sum_{i=2}^{k-1} \rotfv{v}{\{g_{i},g_{i+1}\}}  + \rotfv{v}{\{g_{1},g_{i}\}}  + \rotfv{v}{\{g_{1},g_{i+1}\}} \\
          & = \left(\sum_{i=2}^{k-1} \rotfv{v}{\{g_{i},g_{i+1}\}} \right) + \rotfv{v}{\{g_{1},g_{2}\}}  + \rotfv{v}{\{g_{k},g_{1}\}} = z
        \end{split}
      \end{equation*}
      
    \item Assume the correctness of the claim for dimensions lower than $d, d^{\prime}$. Consider a generators subset $J \subseteq S$ and let $z \in \mathcal{C}_{v}^{J}$. In addition, consider a generator $g \in J$.
      
      Decompose $z = z|_{gv} + z|_{/gv}$ where $z|_{gv}$ and $z|_{/gv}$ are the chains consisting of all the faces in $z$ that have support on $gv$ and its complement. Define $z^{*}$ to be the $d^{\prime}-1$-dimensional chain obtained by excluding $gv$ from any face of $z|_{gv}$. Notice that $(\partial^{-} z^{*})|_{v}$ has to be zero since it contains all the faces of $\partial^{-}z$ that contain $gv$ (up to the removal). In other words, $z^{*} \in \mathcal{C}_{v}^{J/g}$. Thus, we can use the induction assumption and infer the existence of $w^{*} \in \text{Stab}_{v}^{J/g}$ such that $z^{*} = w^{*}|_{v}$.

      Now, since $w^{*} \in \text{Stab}_{v}^{J/g}$, then $\partial^{-}w^{*} = 0$. And since the local complex is contractible, $w^{*}$ can be decomposed into a sum of stabilizers rooted in $v$. Let $\rotfv{v}{S^{\prime}_{1}}, \rotfv{v}{S^{\prime}_{2}}, \rotfv{v}{S^{\prime}_{3}}, \ldots, \rotfv{v}{S^{\prime}_{l}}$ be the $d^{\prime}$-dimensional cells whose boundaries are the $d^{\prime}-1$-dimensional stabilizer chains composing $w^{*}$. Namely:
      \begin{equation*}
        \begin{split}
          w^{*} = \sum_{i} \partial^{-} \rotfv{v}{S^{\prime}_{i}} 
        \end{split}
      \end{equation*}
Since those stabilizers are basis elements of $\text{Stab}_{v}^{J/g}$, we have that $g \notin S^{\prime}_{i}$ for any $i$. Hence, we can define the extended $d^{\prime}$-dimensional chain $w$ to be the chain obtained by adding $g$ to each of the stabilizers deriving $w^{*}$. Namely:
      \begin{equation*}
        \begin{split}
          w =  \sum_{i} \partial^{-} \rotfv{v}{S^{\prime}_{i} \cup g} 
        \end{split}
      \end{equation*}
      However, $\rotfv{v}{S^{\prime}_{1} \cup g}, \ldots, \rotfv{v}{S^{\prime}_{l} \cup g}$ are $d^{\prime}+1$-dimensional chains corresponding to the stabilizers, which are basis elements of $\text{Stab}_{v}^{J}$. Therefore, $w \in \text{Stab}^{J}_{v}$. In addition, opening up the boundary operator yields:
      \begin{equation*}
        \begin{split}
          w|_{v} &=  \sum_{i} \bigl(\partial^{-} \rotfv{v}{S^{\prime}_{i} \cup g} \bigr)|_{v} = \sum_{i} \bigl(\sum_{g^{\prime} \in S^{\prime}_{i}/g}\rotfv{v}{S^{\prime}_{i}/g^{\prime} \cup g} + \rotfv{v}{S^{\prime}_{i}}   \bigr)\\
          &= \sum_{\rotfvs{}\in w^{*}|_{v} }\rotfv{v}{S^{\prime} \cup g} + \sum_{i} \rotfv{v}{S^{\prime}_{i}} = z|_{gv} + \sum_{i} \rotfv{v}{S^{\prime}_{i}} 
        \end{split}
      \end{equation*}
Thus, adding $w|_{v}$ to $z$ results in:
      \begin{equation*}
        \begin{split}
          z + w|_{v} = z|_{/gv} + \sum_{i} \rotfv{v}{S^{\prime}_{i}} 
        \end{split}
      \end{equation*}
      Which is a $d^{\prime}$-dimensional chain, supported on vertices of the form $g^{\prime}v$ for $g^{\prime}\in J/g$, namely a $d^\prime$-dimensional chain on the complex induced by the $(d-1)$-dimensional subcomplex. In addition, since the boundary operator can only advance the root of a face forward, we have that $\partial^{-} w|_{v} =  (\partial^{-} w)|_{v}$ (as $w|_{v}=w|_{v+}$) and therefore $\partial^{-}( z + w|_{v}) = \partial^{-}z + (\partial^{-}w)|_{v} = 0$.
      
To summarize, we have both $z + w|_{v} \in \mathcal{C}_{v}^{J/g}$ and $z+w|_{v}$ sits on a $d-1$-dimensional skeleton. Therefore, we can repeat in a recursive manner, with each step resulting in a $d^{\prime}$-dimensional chain in a skeleton with dimension decreased by one. We stop when $d = d^{\prime}$; in that case, the resulting chain must be zero.
  \end{enumerate}
\end{proof}

We would like to differentiate chains whose projection onto the finite Toric with side length $n$ closes a non-trivial cycle from those which don't. For this, we fix a rough threshold via the size function. Chains which have size less than $n$ will be called finite. Observe that a chain might have a projection that doesn't close a non-trivial loop on the Toric and still be marked, according to our definition, as infinite. So, the classification is rough, yet it will turn out to be sufficient.
\begin{definition}
  \label{def:finiteerr}
  Let $z$ be a $d^{\prime}$-dimensional chain. We will say that a chain $z$ is finite if:
  \begin{equation*}
    \begin{split}
        \Size{z} \le n-1     %
    \end{split}
  \end{equation*}
Similarly, $z$ is infinite if its size is greater than or equal to $n$.
\end{definition}

\begin{remark}
Observe that if $\Size{z} < n$, then the projection of $z$ on the $n$-side length Toric does not close a non-trivial loop.
\end{remark}

The next claim states that the translation by a stabilizer that minimizes its size of a finite chain admits a boundary with size equal to itself.
\begin{claim}
  \label{claim:reducederr}
  Let $x \in C_{d^{\prime}}(\mathcal T_\infty^d)$ be a $d^{\prime}$-dimensional finite chain, consider the stabilizer  $z \in \mathcal{C}_{Z}^{\perp}$ which when added to $x$ minimizes the size , Namely: 
  \begin{equation*}
    \begin{split}
      z =  \arg \min_{z \in \mathcal{C}_{Z}^{\perp}} \Size{ x + z }  
    \end{split}
  \end{equation*}
  Then: 
  \begin{equation*}
    \begin{split}
      \Size{ x + z } = \Size{ \partial^{-}( x + z) } 
    \end{split}
  \end{equation*}
\end{claim}
\begin{proof}
  First, observe that since the vertex domain of $\partial^{-} x$ is contained in the vertex domain of $x$, it follows that for any $x$ the inequality $\Size{x} \ge \Size{\partial^{-} x}$ holds. Combining this with the fact that $\partial^{-} z = 0$ for any $z \in \mathcal{C}_{Z}^{\perp}$, we have that $\Size{x + z} \ge \Size{\partial^{-}(x + z)}$. Now suppose that $\Size{x} > \Size{\partial^{-} x}$; that implies the existence of $v \in x/\partial^{-} x$ such that $v$ is a maximum of one of the potential walls in $x$ and $(\partial^{-} x)|_{v} = 0$. Let's split into cases, depending on the potential wall type.
 
  \begin{enumerate}

    \item If $v$ is a maximum of $L_{d+1}$, then by \Cref{claim:agreement}, there exists a stabilizer $z$ rooted at $v$, such that $z|_{v}=z|_{v+}=x|_{v+}=x|_{v}$ (using that $v$ is a local maximum of $L_{d+1}$ in $x$ to infer $x|_{v}=x|_{v+}$). Therefore, $(x+z)|_{v} = 0$, meaning adding $z$ to $x$ excludes $v$. In addition, since the rooted stabilizer $z$ is supported only on the vertices in $\textbf{Str}_{0+}(v,x|_{v})$, the addition cannot introduce a vertex with a higher $L_{g}$ value than the maximal $L_{g}$ value in $\textbf{Str}_{0+}(v,x|_{v})$ into $x$. So, repeating the process until excluding all the vertices that get the maximal $L_{d+1}$-value in $x$, and summing the stabilizers, results in a stabilizer $z$ for which $\max_{v \in x + z}L_{d+1}(v) < \max_{v \in x}L_{d+1}(v)$.

      \item If, for some generator $g \in S$, $v$ is a maximum of $L_{g}$, then without loss of generality, pick $v$ among the maximum points of $L_{g}$ with the highest value of $-L_{d+1}$ (maximize $\sum_{g \in S} L_{g}(v)$). Since stepping through the positive direction can only increase the value of $-L_{d+1}$, it follows that $v$ is also a local max of $-L_{d+1}$.
        Now consider the skeleton system where the directions are flipped, and $\rotfs{}$ is set to be the face containing the vertices that are reached by the selection of reverse steps from $S^{\prime}$ (i.e., $g^{-1}v \in \rotfs{}$). Similarly, for the sake of the proof, let's present the following notation for the positive neighborhood (\Cref{def:positiveNei}) in the reversed system:
        \begin{equation*}
          \begin{split}
            \textbf{Str}_{0-}(v,x) \colonequals \big\{ u \in x \colon \exists g \in S \text{ such }  u = g^{-1}v   \big\}
          \end{split}
        \end{equation*}
So, $\textbf{Str}_{0-}$ plays the same role as $\textbf{Str}_{0+}$ does in the original skeleton.
In that system, we can apply \Cref{claim:agreement} (in the reversed direction) and get that there is a stabilizer $z$ such that adding it to $x$ excludes $v$ from $x$. By the same argument as above, no vertex with a higher $L_{g^{\prime}}$-value (although $L_{d+1}$ can increase, that's fine\footnote{ If it happens, we will apply the first case in a future exclusion, after we run the second case to completion. Since the first case can't increase the maximal $L_{g}$-value for any $g\in S$, it can't undo previous applications of the second case. }) than what it already has in $\textbf{Str}_{0-}(v,x)$ is introduced into $x$. Repeating the process excludes all the vertices with the maximal value of $L_{g}$ in $x$, so eventually, we end up with $x+z$, such that $\max_{v \in x +z }L_{g}(v) < \max_{v \in x}L_{g}(v)$.
  \end{enumerate}
We keep excluding vertices until there is no potential wall $L_{g}$ and a global maximizer vertex of $L_{g}$ such that $v$ doesn't belong to the $\partial^{-}$-boundary. So, in the end, we get the desired equality.

\end{proof}

\begin{figure}[h]
  \centering
  \ifdefined\POSTER
  \begin{tikzpicture}[scale=5]
  \else
  \begin{tikzpicture}[scale=1.5]
  \fi

\tikzset{xyp/.style={canvas is xy plane at z=#1}}
\tikzset{xzp/.style={canvas is xz plane at y=#1}}
\tikzset{yzp/.style={canvas is yz plane at x=#1}}

\draw[blue!20!black,fill=blue!70!black!70,rounded corners=1,opacity=0.5, xzp=0]%
(-0.1,2.1) rectangle (1.1,0.9);

      \draw[blue!20!black,fill=blue!70!black!70,rounded corners=1,opacity=0.5, yzp = 0]%
      (-0.1,2.1) rectangle (1.1, 0.9);

      \draw[blue!20!black,fill=blue!70!black!70,rounded corners=1,opacity=0.5, xyp = 2]%
      (-0.1,-0.1) rectangle (1.1, 1.1);
      
      \draw[blue!20!black,fill=blue!70!black!70,rounded corners=1,opacity=0.5, yzp = 1]%
      (-0.1,2.1) rectangle (1.1, 0.9);

      \node[below = 3pt] at (0,0,2) { $v$ } ;

\foreach \x in {0,1,2} {
  \foreach \y in {0,1,2} {
    \draw[gray] (\x,\y,0) -- (\x,\y,2);
  }
}

\foreach \x in {0,1,2} {
  \foreach \z in {0,1,2} {
    \draw[gray] (\x,0,\z) -- (\x,2,\z);
  }
}

\foreach \y in {0,1,2} {
  \foreach \z in {0,1,2} {
    \draw[gray] (0,\y,\z) -- (2,\y,\z);
  }
}

\foreach \x in {0,1,2} {
  \foreach \y in {0,1,2} {
    \foreach \z in {0,1,2} {
      \fill (\x,\y,\z) circle (1pt);
    }
  }
}

\draw (2.5,1) node { $\rightarrow$ };
\end{tikzpicture}  \ifdefined\POSTER \begin{tikzpicture}[scale=5] \else \begin{tikzpicture}[scale=1.5] \fi

\tikzset{xyp/.style={canvas is xy plane at z=#1}}
\tikzset{xzp/.style={canvas is xz plane at y=#1}}
\tikzset{yzp/.style={canvas is yz plane at x=#1}}

\draw[green!20!black,fill=green!70!black!70,rounded corners=1,opacity=0.5, xzp=1]%
(-0.1,2.1) rectangle (1.1,0.9);

      \draw[green!20!black,fill=green!70!black!70,rounded corners=1,opacity=0.5, xyp = 1]%
      (-0.1,-0.1) rectangle (1.1, 1.1);

      \node[below = 3pt] at (0,0,2) { $v$  } ;
      \node[below] at (0,1,2) { $u$ } ;
      \node[below] at (1,0,1) { $w$ } ;

\foreach \x in {0,1,2} {
  \foreach \y in {0,1,2} {
    \draw[gray] (\x,\y,0) -- (\x,\y,2);
  }
}

\foreach \x in {0,1,2} {
  \foreach \z in {0,1,2} {
    \draw[gray] (\x,0,\z) -- (\x,2,\z);
  }
}

\foreach \y in {0,1,2} {
  \foreach \z in {0,1,2} {
    \draw[gray] (0,\y,\z) -- (2,\y,\z);
  }
}

\foreach \x in {0,1,2} {
  \foreach \y in {0,1,2} {
    \foreach \z in {0,1,2} {
      \fill (\x,\y,\z) circle (1pt);
    }
  }
}
\end{tikzpicture}
\caption{
Consider the $(3,2)$-Toric code, with qubits and $X$-checks set on the 2- and 1-dimensional faces. In blue, we have a 2-dimensional chain where the vertex achieving the maximum of $L_{d+1}$, denoted by $v$, does not see any unsatisfied checks in its positive local view. The green faces represent a 2-dimensional chain, which is the \textbf{reduced} form of the blue. In particular, any local maximum of $L_{d+1}$ (vertices $u$ and $w$) sees a non-trivial syndrome in its positive local view. Namely, the maximal points of $L_{d+1}$ in the chain are also the maximal points in the boundary $\partial^{-}$.
}
  \label{fig:edge}
\end{figure}

We close the section by defining the \textit{reduced} form of a finite $d^{\prime}$-dimensional chain. 
\begin{definition}
  \label{def:reduced}
Let $x \in C_{d^{\prime}}(\mathcal T_\infty^d)$ be a $d^{\prime}$-dimensional finite chain. We will say that $x$ is reduced if, for any potential wall $L_{i}$, the maximizing vertices of $L_{i}$ in $x$ belong to $\partial^{-} x$.
  In particular, we have that $\Size{x} = \Size{\partial^{-} x}$ and that the reduced form of a given chain is obtained by the construction presented in the proof of \Cref{claim:reducederr}. 
\end{definition}
We emphasize that the reduced form is not unique. Throughout the paper, every time we choose a reduced form of a chain, we mean obtaining it from the given algorithm presented in \Cref{claim:reducederr}, so the reduced form is effectively unique.

\subsection{Shrinking}
\label{subsection:shrink}
We are about to prove that the sweep rule shrinks finite errors.  
The following claim relates the local maximum points of the potential walls $L_{1}, \ldots, L_{d}$ in an assignment $z$ to the structure of the syndrome they observe. Later, we will use that relation to infer that the decoding step of the sweep rule can't spread the error to a vertex\footnote{To faces whose vertices have higher values of $L_{1}, \ldots, L_{d}$ than the current global maximum values.} with a higher value than the current global maximum of $L_{1}, \ldots, L_{d}$. In other words, the error does not spread in the positive directions.

\begin{claim}
  \label{claim:edge_pole}
Let $z$ and $\xi$ be $d^{\prime}$-dimensional finite chains such that $\xi$ results from the application of the sweep rule on $z$.
For any generator $g \in S$, the maximal values of $L_{g}$ in $\xi$ and $\partial^{-} \xi$ are lower than the maximal values in $z$ and $\partial^{-} z$. Namely:
\begin{equation*}
  \begin{split}
    & \max \left\{ L_{g}(u) : u \in \xi \right\}  \le  \max \big\{ L_{g}(u) : u \in z \big\} \\
    & \max \left\{ L_{g}(u) : u \in \partial^{-} \xi \right\}  \le  \max \big\{ L_{g}(u) : u \in \partial^{-} z \big\} 
  \end{split}
\end{equation*}

\end{claim}
\begin{proof}
  First, we argue that for a vertex $v$ in $\partial^{-} z$, and a generator $g \in S$, such that $v$ is a local maximum of $L_{g}$ in either $z$ or $\partial^{-} z$, it implies that $gv \notin \partial^{-} z$. The contradiction where $v$ is a maximum of $L_{g}$ in $z$ follows immediately, since containment in $\partial^{-} z$ implies containment in $z$. Where $v$ is a maximum in $\partial^{-} z$, suppose through contradiction that $gv \in \partial^{-} z$, then
  \begin{equation*}
    \begin{split}
      L_{g}(gv) = (gv)_{g} = 1 + v_{g} = 1 + L_{g}(v)
    \end{split}
  \end{equation*}
In contradiction to $v$ being a local maximum of $L_{g}$ in $\partial^{-} z$.

For a vertex $v \in \partial^{-}z$ denote the correction it applies by $\varphi_{v} \in C_{d^{\prime}}(\mathcal T_\infty^d) $, so 
\begin{equation*}
  \begin{split}
    \xi = \sum_{v \in \partial^{-}z}\varphi_{v} + z \ \ \text{ and } \ \ \partial^{-}\xi = \sum_{v \in \partial^{-}z}\partial^{-}\varphi_{v} + \partial^{-}z
  \end{split}
\end{equation*}
So for any vertex $u \in \xi/z$, there is a vertex $v \in \partial^{-}z$ such that $u \in \varphi_{v}$. Moreover, since $\varphi_{v}$ is a rooted chain, it follows that $L_{g}(u) \le L_{g}(v) + 1$, where the equality holds when $g$ belongs to the generator's subset of the face $\varphi_{v}$. Similarly for $u \in \partial^{-}\xi/\partial^{-}z$. 

Now, combining with the sweep rule's first condition, which restricts the correction applied by a vertex to have a vertex support that is contained in the positive syndrome it sees, we have that for any vertex $v$ which is a local maximum of $L_{g}$ (in either $z$ or $\partial^{-}z$), any $u \in \varphi_{v}$ satisfies $L_{g}(u) = L_{g}(v)$. In other words, $\xi$ and $\partial^{-}\xi$ don't contain vertices of the form $gv$ where $v$ is a local maximum of $L_{g}$ in $z$ and $\partial^{-}z$, and the desired inequalities follow.
\end{proof}

Next, we prove that for a finite $X$-error, the sweep-cycle advances the corner (the maximum of $L_{d+1}$) of its $\partial^{-}$-boundary. We assume we are given a finite error, so its $\partial^{-}$-boundary corner sees a syndrome, and show that the sweep rule application excludes that corner. In other words, the sweep rule shrinks the size of the $\partial^{-}$-boundaries of $X$-errors.

\begin{claim}
  \label{claim:sweepshrink}
Let $z$ be a finite $d^{\prime}$-dimensional chain, and let $\xi$ be the result of applying the sweep rule to it. Then:
  \begin{enumerate}
    \item For any generator $g \in S$: $\max \big\{ L_{g}(v) : v \in \partial^{-} z \big\} \ge \max \big\{ L_{g}(v) : v \in \partial^{-} \xi \big\}  $
    \item And for $L_{d+1}$ : $ \max \big\{ L_{d+1}(v)  : v \in \partial^{-} z \big\} \ge \max \big\{  L_{d+1}(v) : v \in \partial^{-}\xi \big\} + 1$
  \end{enumerate}
  In particular, $\Size{ \partial^{-}\xi} \le \Size{\partial^{-}z} - 1$.
\end{claim}
\begin{proof}
  The first inequality was proved in \Cref{claim:edge_pole}, so it's left to prove the second. Denote by $\varphi$ the applied correction and by $v$ the vertex on which $L_{d+1}$ gets the maximal value in $\partial^{-} z$, if there is more than one vertex then choose $v$ arbitrarily. Since $v$ is a maximizer vertex of $L_{d+1}$ in $\partial^{-} z$, it's also a maximizer vertex of $L_{d+1}$ in the reduced form of $z$ ( \Cref{def:reduced} ). Thus, the two follows: 
  \begin{enumerate}
    \item Denote the reduced form of $z$ by $\tilde{z}$, so $\tilde{z}|_{v} = \tilde{z}|_{v+}$, and hence $\partial^{-}\tilde{z}|_{v} = \partial^{-}\tilde{z}|_{v+}$, namely there is a $d^{\prime}$-dimensional chain rooted at $v$, the chain $\tilde{z}|_{v}$, whose $\partial^{-}$-boundary agrees with the positive syndrome seen by $v$.

      \Cref{claim:findcor} guarantees that there is a (non-empty) valid sweep rule suggestion for $v$. Namely, $\partial^{-} \varphi|_{v} = \partial^{-} z|_{v}$. Thus, adding $\varphi|_{v}$ to $z$ excludes $v$ from the $\partial^{-}$ boundary.
    \item For any other vertex $u \in z$, there is no face $\rotfv{u}{S^{\prime}} \in \partial^{-} z$ such that $v \in \rotfv{u}{S^{\prime}}$. Otherwise, it would contradict $v$ being a maximum of $L_{d+1}$ in $\partial^{-} z$. Thus, $v \notin \textbf{Str}_{0+}(u, \partial^{-} z|_{u})$, so by \Cref{claim:findcor}, $v \notin \varphi|_{u}$.
  \end{enumerate}
Combining the two, we get that $v \notin \partial^{-} (z + \varphi)$. Namely, $v \notin \partial^{-}\xi$. Repeating the above for all the other vertices which get the maximal $L_{d+1}$-value in $\partial^{-} z$ yields the second desired inequality.

\end{proof}

Consider a vertex on the boundary of a chain resulting from an application of the sweep rule. That vertex did not come from nowhere; there must be vertices on the preceding chain, which sit on its boundary, and are the cause for the local view seen by the vertex. The following claim formulates that insight; moreover, it states a relation between the potential walls' values of the vertex and the (part of) vertices who caused it.
\begin{claim} 
  \label{claim:sweepface}
  Let $z$ be a finite $d^{\prime}$-dimensional chain, and let $\xi$ be the result of applying the sweep rule to it. Let $v$ be a vertex in $\partial^{-}\xi$, then there is a vertex $u \in \partial^{-}z$ such $L_{i}(u)\ge L_{i}(v)$ for $i\neq d+1$ and  $L_{d+1}(u) > L_{d+1}(v)$ if $i=d+1$,  and there is a $d^{\prime}$-dimensional face $f$ such $v,u \in f$. 
\end{claim}
\begin{proof}
We start by proving for $i = d + 1$. Now, it splits upon whether $v \in z + \xi$:
  \begin{enumerate}
    \item If $v\in z + \xi$, then $v$ belongs to a face that flipped during the correction, denote its root by $u$ and by definition of the sweep rule $(\partial^{-} z)|_{u+} \neq 0$. Therefore, $u \in \partial^{-}z$, and $u,v$ share a face which $u$ is the root. Since $u$ is the root of the face we get $L_{d+1}(u) > L_{d+1}(v)$. 
    \item Else, if $v \notin z + \xi$, then $v$ has to be in $z$. Combining with the fact that $v \in \xi$, it follows that $v$ is not a local maximum of $L_{d+1}$ in $\partial^{-}z$, otherwise the application of the sweep rule would exclude $v$ (See the proof of \Cref{claim:sweepshrink}), so it wouldn't be in $\xi$. Pick $u$ to be the neighbor of $v$ in $\partial^{-}z$ such that $L_{d+1}(u) > L_{d+1}(v)$.
  \end{enumerate}
  It remains to prove for $i\in [d]$, in this case if $v \in \partial^{-}z$ then take $u = v$. Otherwise, there must be a face $f$ that was flipped during the sweep rule containing $v$. Denote by $g$ the corresponding generator for $i$ and by $u^{\prime}$ the root of $f$. If $L_{g}(u^{\prime}) = L_{g}(v)$ then set $u = u^{\prime}$. Otherwise, $g$ is one of the generators that span $f$, the first condition of the sweep rule (\Cref{def:sweeprule}) requires that for $u^\prime$ flipping a face in the $g$ direction, it has to see the vertex $gu^{\prime}$ in $\partial^{-}z$, in that case, set $u=gu^{\prime}$.
\end{proof}

We finished analyzing the first part of the sweep cycle that acts in the computational basis. In the next subsection, we introduce the transformation into the dual code. Since the dual code is a $(d,d-d^{\prime})$-Toric code, we also find that the sweep cycle shrinks the $\partial^{-}$-boundaries of phase errors.

\subsubsection{Ideal decoder}
 Although the main role of the section is to show that a single sweep-cycle keeps error clusters under control during a noisy run (by admitting a modest shrinking), an immediate corollary of this is that the application of non-noisy multiple iterations of the sweep-cycle has the potential to erase errors, namely a decoder. Here, a classical measurement is given, we act on classical bits, and the sweep-cycle consists only of the first stage that works over the computational basis, without moving to the dual code. We formulate the corollary as follows:
\begin{corollary}
  \label{col:idealcorrection} 
  Consider the $(d,d^{\prime})$-Toric code, induced by a Toric with side length $n$. Given a finite connected reduced error $z \in C_{d^{\prime}}(\mathcal T_\infty^d)$ on that Toric, with $\Size{z} \le n-1$. Then an application of $\Size{z}$ sweep cycles (partial cycles consisting only of the first stage) erases the error.
\end{corollary}
When $z$ is not connected, disjoint components might merge, and it is no longer true that the size of the chain decreases in each iteration. However, we will see later (\Cref{claim:decodestopnoise}, \Cref{claim:replace}, and \Cref{lemma:keylemma}) that performing $\Theta(n^{d+1})$ sweep-cycles erases the error with high probability, if $z$ is assumed to be distributed according to local stochastic noise.

\ifdefined\FULL

\input{shrink_v3.tex}

\section{Investigation}

\ctt{To do: understands how the investigation should look like. Plot an example to such tree.}
The model goes as follow, at each even iteration we toss a noise, then in each odd iteration we perform a single SWEEP rule iteration. That allows to track over the history of the errors.  

We say that a check $e \in $ Error, if there is a face $f$ that adjoin to $e$ and belongs to the Error.

\begin{definition}[Investigation]
  For any unsatisfied check $e$ at time $t$, we define $V$ (the set of checks which participated in 'misleading' of $e$), and two sets of edges on $V$, Arrows and Forks via the following algorithm:

\end{definition}
\tikzset{%
  every neuron/.style={
    circle,
    draw,
    radius=0.1mm
  },
  neuron missing/.style={
    draw=none,
    scale=2,
    text height=0.25cm,
    execute at begin node=\color{black}$\vdots$
  },
}

  \begin{figure}[h]
    \centering
  \begin{algorithm}[H]
    let $V = \{e\}$, Arrows $= \emptyset$, Forks $= \emptyset$\\
    let $Q \leftarrow $ queue $\{ e \}$\\
    \While{ $Q$ is not empty  }{
      $e^{\prime} \leftarrow $ dequeue $Q$ \\
      \If { $e^{\prime} \notin $ Error }{
        $e_{1},..,e_{l} \leftarrow $ Excuse( $e^{\prime}$ ) \\
        Insert $\{e_{i}, e^{\prime} \}$ to Arrows\\
        Insert $\{e_{i}, e_{j} \}$ to Forks
      }
    }
  \end{algorithm}
\end{figure}
\begin{figure}[h]
      \begin{tikzpicture}[scale=1]%
  \draw [dashed, draw=gray] (-1.8,-1.5) -- (7,-1.5);
  \draw [dashed, draw=gray] (-1.8,-3) -- (7,-3);
  \draw [dashed, draw=gray] (-1.8,-4.5) -- (7,-4.5);
  \draw [dashed, draw=gray] (-1.8,-6) -- (7,-6);
  \node (SWEEP1) at (-3, -1.5) {Sweep iteration: };
  \node (NOISE1) at (-3, -3) {Noise iteration: };
  \node (SWEEP2) at (-3, -4.5) {Sweep iteration: };
  \node (NOISE2) at (-3, -6) {Noise iteration: };
  \node (root) at (3.9, -0) { $\leftarrow$ root };
  \node (root) at (6.9, -3) { $\leftarrow$ leaf };
  \node [every  neuron ] (v-1) at ( 3 ,-0) { } ;
  \node [every  neuron ] (v-2) at ( 1 ,-1.5) { } ;
  \node [every  neuron ] (v-3) at ( 5 ,-1.5) { } ;
  \node [every  neuron ] (v-4) at ( 0 ,-3) { } ;
  \node [every  neuron ] (v-5) at ( 2 ,-3) { } ;
  \node [every  neuron ] (v-6) at ( 4 ,-3) { } ;
  \node [every  neuron ] (v-7) at ( 6 ,-3) { } ;
  \node [every  neuron ] (v-8) at ( 0.5 ,-4.5) { } ;
  \node [every  neuron ] (v-9) at ( 1.5 ,-4.5) { } ;
  \node [every  neuron ] (v-10) at ( 2.5 ,-4.5) { } ;
  \node [every  neuron ] (v-11) at ( 1.25 ,-6) { } ;
  \node [every  neuron ] (v-12) at ( 1.75 ,-6) { } ;
\draw [->, draw = blue, thick] (v-1) -- (v-2);
\draw [->] (v-1) -- (v-3);
\draw [->] (v-2) -- (v-4);
\draw [->, draw = blue, thick] (v-2) -- (v-5);
\draw [->] (v-3) -- (v-6);
\draw [->] (v-3) -- (v-7);
\draw [->] (v-4) -- (v-8);
\draw [->, draw = blue, thick] (v-5) -- (v-9);
\draw [->] (v-5) -- (v-10);
\draw [->, draw = blue, thick] (v-9) -- (v-11);
\draw [->] (v-9) -- (v-12);
\end{tikzpicture} 
    \caption{blabla }
    \label{fig:investigation}
  \end{figure}

  \newpage

\fi

\subsection{Logical Gates}

\begin{claim}
  For every generator subset $S^{\prime}$ of size $d^{\prime}$, denote by $x_{S^{\prime}}$ and $z_{S^{\prime}}$ the following $d^{\prime}$-dimensional chains: 
\begin{equation*}
  \begin{split}
    x_{S^{\prime}}  \colonequals \sum_{u \in \expval*{S^{\prime}} }\rotfv{u}{S^{\prime}} \ \ \ z_{S^{\prime}} \colonequals \sum_{u \in \expval*{S/S^{\prime}} }\rotfv{u}{S^{\prime}} 
  \end{split}
\end{equation*}
Then $x_{S^{\prime}}$ and $z_{S^{\prime}}$ are binary representations of anticommuting logical codewords of the $(d,d^{\prime})$-Toric code. In particular, $x_{S^{\prime}} \in \mathcal{C}_{X}/\mathcal{C}_{Z}^{\perp}$, $z_{S^{\prime}} \in \mathcal{C}_{Z}/\mathcal{C}_{X}^{\perp}$, and $\langle x_{S^{\prime}}, z_{S^{\prime}} \rangle = 1$. In addition, let $X_{S^{\prime}}$ and $Z_{S^{\prime}}$ be the Pauli matrices obtained by multiplying each non-trivial coordinate of them by $X$ or $Z$, respectively.
\end{claim}
\begin{proof}
By definition, $x_{S^{\prime}}$ and $z_{S^{\prime}}$ are binary assignments on the $d^{\prime}$-dimensional faces. We start by showing that $x_{S^{\prime}} \in \mathcal{C}_{X}$ and $z_{S^{\prime}} \in \mathcal{C}_{Z}$; namely, that the boundaries $\partial^{-}x_{S^{\prime}}$ and $\partial^{+}z_{S^{\prime}}$ vanish.
\begin{align*}
  \partial^- x_{S'}
  &=\sum_{g\in S'}\sum_{u\in\langle S'\rangle}
    \bigl(\rotfv{u}{S'/g}+\rotfv{gu}{S'/g}\bigr)=0,\\
  \partial^+ z_{S'}
  &=\sum_{g\in S/S'}\sum_{u\in\langle S/S'\rangle}
    \bigl(\rotfv{u}{S'\cup g}+\rotfv{g^{-1}u}{S'\cup g}\bigr)=0.
\end{align*}
In both sums, the two terms cancel after reindexing within the corresponding subgroup.
Thus $x_{S^{\prime}} \in \mathcal{C}_{X}$ and $z_{S^{\prime}}\in \mathcal{C}_{Z}$, now for showing that they are not binary representations of stabilizers, it's enough to show they admit no vanishing product.  
\begin{equation*}
  \begin{split}
    \operatorname{supp}(x_{S'})\cap\operatorname{supp}(z_{S'})
    &=\{\rotf{e}{S'}\},\\
    \langle x_{S'},z_{S'}\rangle&=1\pmod 2.
  \end{split}
\end{equation*}
Therefore, the Paulis $X_{S^{\prime}}$ and $Z_{S^{\prime}}$ are $X$- and $Z$-type anticommuting logical codewords.
\end{proof}

    \begin{figure}[h]

\begin{center}
    \begin{tikzpicture}[scale=1.5]

      \draw[blue!20!black,fill=blue!70!black!70,rounded corners=1,opacity=0.5]%
  (1.9,2.9) rectangle (3.1,3.1);
  \draw[blue!20!black,fill=blue!70!black!70,rounded corners=1,opacity=0.5]%
  (1.9,3.9) rectangle (3.1,4.1);
  \draw[blue!20!black,fill=blue!70!black!70,rounded corners=1,opacity=0.5]%
  (1.9,4.9) rectangle (3.1,5.1);
  \draw[blue!20!black,fill=blue!70!black!70,rounded corners=1,opacity=0.5]%
  (1.9,1.9) rectangle (3.1,2.1);
  \draw[blue!20!black,fill=blue!70!black!70,rounded corners=1,opacity=0.5]%
  (1.9,0.9) rectangle (3.1,1.1);
  \draw[blue!20!black,fill=blue!70!black!70,rounded corners=1,opacity=0.5]%
  (1.9,-0.1) rectangle (3.1,0.1);

  \draw[green!20!black,fill=green!70!black!70,rounded corners=1,opacity=0.5]%
  (-0.1,3.9) rectangle (1.1,4.1);
  \draw[green!20!black,fill=green!70!black!70,rounded corners=1,opacity=0.5]%
  (0.9,3.9) rectangle (2.1,4.1);
  \draw[green!20!black,fill=green!70!black!70,rounded corners=1,opacity=0.5]%
  (1.9,3.9) rectangle (3.1,4.1);
  \draw[green!20!black,fill=green!70!black!70,rounded corners=1,opacity=0.5]%
  (2.9,3.9) rectangle (4.1,4.1);
  \draw[green!20!black,fill=green!70!black!70,rounded corners=1,opacity=0.5]%
  (3.9,3.9) rectangle (5.1,4.1);

  \draw[step=1cm,gray,very thin] (0,0) grid (5,5);
  \foreach \x in {0,1,2,3,4,5}
  \foreach \y in {0,1,2,3,4,5}
  {
  \node[draw,circle,inner sep=1pt,fill] at (\x,\y) {};
}

\end{tikzpicture}
\end{center}
\caption{ 
  In blue and green are the binary presentations $x_{S^{\prime}}$ and $z_{S^{\prime}}$ of logical code words of the $(2,1)$-Toric code. The edge $\{(5,4), (0,4)\}$ is omitted for clarity (it should be a green edge that hovers over the green line). 
}
\label{fig:logicalToric}
\end{figure}
The $(d,d^{\prime})$-Toric code admits a fold-transversal realization of an isomorphism gate into the $(d,d-d^{\prime})$-Toric code, for any dimensions $d,d^{\prime}$. Particularly, the gate exchanges logical Pauli $X_{S^{\prime}}$ (of the first code) with logical Pauli $Z_{S/S^{\prime}}$ (of the second code) and vice versa. It is realized by an application of transversal Hadamards followed by permuting the qubits according to a transformation which we name the \textit{$\phi$-transpose map}. We define the transpose map as follows:
\begin{definition}[The Transpose Map]
  \label{def:trans}
  Let $d, d^{\prime}$ be integers such that $d^{\prime} < d$, and let $\mathcal{T}^{d}_{n}$ be a $(d, d^{\prime})$-Toric complex generated by the generator set $S$.   We define the \textit{transpose map} $\phi : C_{d^{\prime}}(\mathcal{T}_n^d) \rightarrow C_{d - d^{\prime}}(\mathcal{T}_n^d)$ by acting on the basis elements as follows:

  \begin{equation*}
    \begin{split}
      \phi(\rotfv{v}{S^{\prime}}) \colonequals \rotfv{S^{\prime}v}{S/S^{\prime}} 
    \end{split}
  \end{equation*}
\end{definition}
Notice that the action of the transpose map over the qubits is a one-to-one map between the qubits of the $(d,d^{\prime})$-Toric code and the qubits of the $(d,d-d^{\prime})$-Toric code, and therefore can be realized at constant depth using ancillas.
\begin{definition}[ Hadamard-Type Gate $\mathbf{H_{\phi}}$]  
  \label{def:htrans}
We define the logical Hadamard-type gate $ \mathbf{H_{\phi}} \colon L (\mathcal{H}) \rightarrow L (\mathcal{H}) $, to be a code isomorphism $(d,d^{\prime})$-Toric $\rightarrow (d,d-d^{\prime})$-Toric, as follows, for any $ S' \subset S $, of size $ d' $:
  \begin{equation*}
    \begin{split}
      \mathbf{H_{\phi}} ( X_{S^{\prime}} )  \rightarrow Z_{S/S^{\prime}} \ \ \ \  \mathbf{H_{\phi}} ( Z_{S^{\prime}} )   \rightarrow X_{S/S^{\prime}} 
    \end{split}
  \end{equation*}
\end{definition}
\begin{claim}
  The Hadamard-type gate $ \mathbf{H_{\phi}}$ has a fold-transversal implementation obtained by first applying $\mathbf{H}^{\otimes\mathcal{F}_{d^\prime}^{n}}$ over all the physical qubits, and then permuting them by the transpose map $\phi$, namely $\mathbf{\hat{H}}_{\phi} = \phi \circ \mathbf{H}^{\otimes\mathcal{F}_{d^\prime}^{n}}$.
\end{claim}
\begin{proof}
  Clearly, for any subset of generators $S^{\prime} \subset S$, the map $\phi$ sends $x_{S^{\prime}}$ to $z_{S/S^{\prime}}$:
  \begin{equation*}
    \begin{split}
    \phi(x_{S^{\prime}}) = \phi(  \sum_{u \in \expval*{S^{\prime}} }\rotfv{u}{S^{\prime}}   ) =  \sum_{u \in \expval*{S^{\prime}} }\rotfv{ S^{\prime} u}{S/S^{\prime}}  = \sum_{u \in \expval*{S^{\prime}} }\rotfv{u}{S/S^{\prime}} = z_{S/S^{\prime}}
    \end{split}
  \end{equation*}
Moreover, $\phi$ sends binary representations of $Z$-stabilizers of the $(d,d^{\prime})$-Toric code to binary representations of $X$-stabilizers of the $(d,d-d^{\prime})$-Toric code, and vice versa. To see this, consider a $d^{\prime}+1$ dimensional face, and consider the image of the stabilizer it defines.
  \begin{equation*}
    \begin{split}
      \phi \left( \partial^{-}  \rotfv{v}{S^{\prime}} \right) &=  \phi\left( \sum_{g\in S^{\prime}} \rotfv{v}{S^{\prime}/g} +  \rotfv{gv}{S^{\prime}/g}  \right) \\
      & = \sum_{g\in S^{\prime}} \rotfv{ g^{-1}  S^{\prime} v}{(S/S^{\prime})\cup g} +  \rotfv{ g^{-1} S^{\prime}  gv}{(S/S^{\prime})\cup g}  \\ 
      &= \partial^{+} \rotfv{ S^{\prime} v}{S/S^{\prime}}
    \end{split}
  \end{equation*}
  And for the checks set on the $d^{\prime}-1$-dimensional face we have: 
  \begin{equation*}
    \begin{split}
      \phi \left( \partial^{+}  \rotfv{v}{S/S^{\prime}} \right) & =   \phi \left( \sum_{g \in S^{\prime}}\rotfv{ g^{-1} v}{(S/S^{\prime}) \cup g} +   \rotfv{    v}{(S/S^{\prime}) \cup g} \right) \\
      &=  \sum_{g \in S^{\prime}}\rotfv{  (S/S^{\prime})    v}{S^{\prime}/g} +   \rotfv{   g (S/S^{\prime})  v}{S^{\prime} / g} \\
      & =  \partial^{-} \rotfv{ (S/S^{\prime}) v}{S^{\prime}}
    \end{split}
  \end{equation*}
  Thus, $\phi \circ \mathbf{H}^{\otimes\mathcal{F}_{d^\prime}^{n}} $ sends $Z$-stabilizers (at dimension $d^{\prime}$) to $X$-stabilizers ( at dimension $d-d^{\prime}$) and vice versa, thereby sending the satisfied checks to satisfied checks. Combining this with the above, $\phi \circ \mathbf{H}^{\otimes\mathcal{F}_{d^\prime}^{n}}$ realizes the logical isomorphism $\mathbf{H}_{\phi}$ from the $(d,d^{\prime})$-Toric code into the $(d,d-d^{\prime})$-Toric code.
\end{proof}
    \begin{figure}[h]

\begin{center}
    \begin{tikzpicture}[scale=1.5]

  \draw[blue!20!black,fill=blue!70!black!70,rounded corners=1, opacity=0.5]%
  (1.9,2.9) rectangle (2.1,4.1);
  \draw[blue!20!black,fill=blue!70!black!70,rounded corners=1, opacity=0.5]%
  (0.9,2.9) rectangle (2.1,3.1);
  \draw[blue!20!black,fill=blue!70!black!70,rounded corners=1, opacity=0.5]%
  (1.9,1.9) rectangle (2.1,3.1);
  \draw[blue!20!black,fill=blue!70!black!70,rounded corners=1, opacity=0.5]%
  (1.9,2.9) rectangle (3.1,3.1);

  \draw[green!20!black,fill=green!70!black!70,rounded corners=1, opacity=0.5]%
  (2.9,2.9) rectangle (3.1,4.1);
  \draw[green!20!black,fill=green!70!black!70,rounded corners=1, opacity=0.5]%
  (1.9,3.9) rectangle (3.1,4.1);
  \draw[green!20!black,fill=green!70!black!70,rounded corners=1, opacity=0.5]%
  (1.9,2.9) rectangle (2.1,4.1);
  \draw[green!20!black,fill=green!70!black!70,rounded corners=1, opacity=0.5]%
  (1.9,2.9) rectangle (3.1,3.1);

  \draw[black,->]  (2,3) to node[midway, above]{ $\phi$ } (2.5,3.5) ;

  \draw[step=1cm,gray,very thin] (0,0) grid (5,5);
  \foreach \x in {0,1,2,3,4,5}
  \foreach \y in {0,1,2,3,4,5}
  {
  \node[draw,circle,inner sep=1pt,fill] at (\x,\y) {};
}

\end{tikzpicture}
\end{center}
\caption{
  An outline of \Cref{example:stabmap}. The $\phi$-transpose sends $Z$-stabilizers to $X$-stabilizers and vice versa. Here we consider the $(2,1)$ Toric code; namely, the $Z$-stabilizers (parity checks) are set on the vertices, and the $X$-stabilizers (phase checks) are set on the squares. In blue and green, we have the edges corresponding to checks $\partial^{+} \rotfv{(2,3)}{\{ \}}$ and $\partial^{-}\rotfv{(2,3)}{\{ (0,1), (1,0) \} }$. The example shows that indeed $\phi$ maps between the two.
}
\label{fig:transposeToric}
\end{figure}

\begin{example}
  \label{example:stabmap} 
Consider the $(2,1)$-Toric code. Name the generators in $S$ by $g_{1} = (1,0)$ and $g_{2} = (0,1)$. Consider the check corresponding to the $Z$-stabilizer $\partial^{+} \rotfv{(2,3)}{\{ \}}$.
  \begin{equation*}
    \begin{split}
      \partial^{+} \rotfv{(2,3)}{\{ \}} = \{ \rotfv{(2,3)}{ \{ (1,0) \} },  \rotfv{(2,3)}{ \{ (0,1) \} } , \rotfv{(1,3)}{ \{ (1,0) \} } ,\rotfv{(2,2)}{ \{ (0,1) \} }  \} 
    \end{split}
  \end{equation*}
The $\phi$-transpose map acts on it as follows:
  \begin{equation*}
    \begin{split}
      \phi(\rotfv{(2,3)}{ \{ (1,0) \} } ) &= \rotfv{(3,3)}{ \{ (0,1) \} }    \\
      \phi(\rotfv{(2,3)}{ \{ (0,1) \} } ) &=  \rotfv{(2,4)}{ \{ (1,0) \} }\\
      \phi(\rotfv{(1,3)}{ \{ (1,0) \} }) &=  \rotfv{(2,3)}{ \{ (0,1) \} }\\ 
      \phi(\rotfv{(2,2)}{ \{ (0,1) \} } ) &=  \rotfv{(2,3)}{ \{ (1,0) \} }
    \end{split}
  \end{equation*}
  What gives edges on a rectangle that converges with the bits on the support of the $X$-stabilizer $\rotfv{(2,3)}{\{ (0,1), (1,0) \} }$. 
\end{example}

Equipped with $\mathbf{H}_{\phi}$, we gain a mapping into the dual code that both keeps the Toric skeleton and sends the $Z$-checks to $d^{\prime}-1$ dimensional faces. This enables the use of the same local decoder to act on both $\mathcal{C}_{X}$ and its dual. We summarize it in the following immediate corollary:
\begin{corollary}
\label{claim:polesmove}  
Consider a Pauli $P \in \mathcal{P}$, acting on the infinite complex, whose $X$- and $Z$-components are set on the reduced chains $x, z \in C_{d^{\prime}}(\mathcal T_\infty^d)$. Denote by $\tilde{P}$ the Pauli operator resulting from an application of a full sweep-cycle on $P$ and by $\tilde{x}, \tilde{z} \in C_{d^{\prime}}(\mathcal T_\infty^d)$ the chains corresponding to the $X$- and $Z$-components of $\tilde{P}$. If both $x$ and $\phi(z)$ are finite chains, then:
  \begin{enumerate}
    \item For any $i \in [d]$: 
      \begin{equation*}
        \begin{split}
          & \max_{v \in \partial^{-} x} L_{i}(v)  \ge \max_{v \in \partial^{-} \tilde{x}} L_{i}(v) \, \quad \quad \, \max_{v \in \partial^{-} \phi(z)} L_{i}(v)  \ge \max_{v \in \partial^{-} \phi(\tilde{z}) } L_{i}(v) 
      \end{split}
      \end{equation*}
    \item And for $i = d+1$ : 
      \begin{equation*}
        \begin{split}
          & \max_{v \in \partial^{-} x} L_{d+1}(v)  \ge \max_{v \in \partial^{-} \tilde{x}} L_{d+1}(v) + 1 \, \quad \quad \,            \max_{v \in \partial^{-} \phi(z)} L_{d+1}(v)  \ge \max_{v \in \partial^{-} \phi(\tilde{z}) } L_{d+1}(v) + 1
        \end{split}
      \end{equation*}
  \end{enumerate}
  In particular: 
  \begin{equation*}
    \begin{split}
      &\Size{ \partial^{-} \tilde{x}}  \le \Size{ \partial^{-} x}  - 1 \, \quad \quad \,   \Size{ \partial^{-}\phi(\tilde{z} ) }  \le \Size{ \partial^{-}\phi(z) } - 1 
    \end{split}
  \end{equation*}
\end{corollary}

Throughout the construction, we encode each qubit in a code block; namely, we restrict ourselves to using only a single codeword of the $(d,d^{\prime})$-Toric code. From now on, we will think of $S^{\prime} \subset S$ as a fixed $d^{\prime}$-size subset of generators that were fixed in advance. And we will encode the logical $\ket{1}$ state by $x_{S^{\prime}}$. To simplify terminology, we call such usage 'nice encoding'.
\begin{definition}[The Nice Encoding]
  \label{def:niceenc}
  Consider $(d,d^{\prime})$-Toric code and fix a subset $S^{\prime} \subset S$, at size $d^{\prime}$. We define the \textit{nice encoding} to be the subsystem code, which uses $x_{S^{\prime}}$ to encode a single logical qubit. Namely, the logical basis states are: 
  \begin{equation*}
    \begin{split}
      \ket{ \bar{0} } &= \ket{ \mathcal{C}_{Z}^{\perp} } \\ 
      \ket{ \bar{1} } &= \ket{ x_{S^{\prime}} + \mathcal{C}_{Z}^{\perp} } 
    \end{split}
  \end{equation*}
\end{definition}
We are now ready to give a full description of the realization of the logical operators. The logical Paulis $\mathbf{X}$ and $\mathbf{Z}$ are realized by the application of $X_{S^{\prime}}$ and $Z_{S^{\prime}}$. For the gates $\mathbf{S}$ and $\mathbf{T}$, we use single-qubit gate teleportation as presented in \Cref{fig:magic}. In single-qubit gate teleportation, the resource $\ket{\mathbf{S}}$ is initialized as $\mathbf{S}\ket{+}$, similarly for $\ket{\mathbf{T}}$. The gate $\mathbf{1}_{x_{S^{\prime}}}$ stands for the classical procedure that, given the measurement results, checks whether the logical bit $\ket{1}$ is turned on. We will return to this gate with a higher resolution depiction of it after the description of the Hadamard teleportation.
    \begin{figure}[h]
        \centering 
        \begin{quantikz}
          \lstick{$\ket{\psi}$} & \ctrl{1}  &  & \gate{Z_{S^{\prime}}} &  \rstick{$\mathbf{S}\ket{\psi}$} \\
        \lstick{$\ket{\mathbf{S}}$} & \targ{} & \meter{} &  \gate{\mathbf{1}_{x_{S^{\prime}}}} \wire[u][1]{c}   \setwiretype{c}  \\
        \end{quantikz} \qquad \qquad 
        \begin{quantikz}
          \lstick{$\ket{\psi}$} & \ctrl{1}  & & \gate{\mathbf{S}} &  \rstick{$\mathbf{T}\ket{\psi}$} \\
          \lstick{$\ket{\mathbf{T}}$} & \targ{} & \meter{} & \gate{\mathbf{1}_{x_{S^{\prime}}}} \wire[u][1]{c}   \setwiretype{c}  \\
        \end{quantikz}
        \caption{ Simulating the $\mathbf{S}$ and $\mathbf{T}$ gates, using the $\ket{\mathbf{S}}$, $\ket{\mathbf{T}}$ magic states. }   
\label{fig:magic}
\end{figure}

For the Hadamard gate, we use a modification of the standard two-qubit gate teleportation as shown in \Cref{fig:magich}. The realization differs from the standard realization as the measurement in the phase basis is done by first moving to the dual space via the application of physical $\mathbf{H}^{\otimes\mathcal{F}_{d^\prime}^{n}}$, measuring, reorganizing the qubits through $\phi$, and then checking for the $x_{S/S^{\prime}}$ outcome (thus, it can be thought of as the application of $\mathbf{H}_{\phi}$ and then measurement). In addition, unlike the single-qubit gate teleportation, here $\ket{\mathbf{H}}$ is initialized as $I\otimes \mathbf{H} (\ket{00} + \ket{11})$ (up to normalization).
    \begin{figure}[h]
        \centering 
        \begin{quantikz}
          \lstick{$\ket{\psi}$} & \ctrl{1}  & \gate{\mathbf{H}^{\otimes\mathcal{F}_{d^\prime}^{n}}} & \meter{} &  \gate{ \phi }  \setwiretype{c} &  \gate{\mathbf{1}_{x_{S/S^{\prime}}}} \wire[d][2]{c} \\
          \lstick[2]{$\ket{\mathbf{H}}$} & \targ{} & & \meter{}   &  \gate{\mathbf{1}_{x_{S^{\prime}}}} \wire[d][1]{c}   \setwiretype{c} \\
         & &  & & \gate{Z_{S^{\prime}}}   & \gate{X_{S^{\prime}}}  & \rstick{$\mathbf{H}\ket{\psi}$} \\
        \end{quantikz} 
        \caption{ Gate teleportation of the Hadamard gate, using the resource $\ket{\mathbf{H}}$. We emphasize that although the Hadamard (at least Hadamard-type) has an $O(1)$-depth gate realization (at least when $d^{\prime}=d/2$), we still need all the operations on the quantum register to be strictly transversal for the sake of the proof. Thus, we realize the Hadamard via gate teleportation. }   
\label{fig:magich}
\end{figure}

The classical procedure of logical codeword checking, namely $ \mathbf{1}_{x_{S^{\prime}}}$ / $ \mathbf{1}_{x_{S/S^{\prime}}}$, is divided into five stages. In the first, we find all the connected components of the $\partial^{-}$-boundary. Finding connected components can be done in $O(\log^{2}(n))$ depth (see \cite{SHILOACH198257} and \cite{1676869} for logarithmic time connected components in the PRAM model and \cite{Lev1980} for the simulation of PRAM machines using bounded-fan circuits). Then, we broadcast the connected components. Conceptually, each connected component is sent to its own Toric; technically, we copy components of the syndromes into clean registers. The broadcast also requires at most $O(\log(n))$ depth since there can be at most \#checks $=O(n^{d})$ connected components and therefore at most $O(n^{d})$ copies. Then, we decode each of the copies in parallel. The separation of the components guarantees that during decoding, components will not merge with each other. For the decoding, we iterate sweep rule cycles (defined in \Cref{def:sweeprule}) until erasing the syndrome. Later, we will see that $n$ cycles succeed to erase the syndrome with high probability\footnote{ In the proof of the \Cref{theorem:main} we show that $\Theta(n)$-size connected component of the syndrome is not likely to occur.}. We erase the syndrome by combining the results of all the decoding sessions together by Xoring their correction suggestions. Each of the Xor operations is between at most $O(n^{d})$ bits and therefore requires at most $O(\log(n))$ depth. At this point, we assume that the combined register stores the measurement without errors. We finish by getting back into the logical space, namely a multiplication of binary matrix, which can be done at logarithmic depth $O(\log(n))$ (at the cost of broadcasting into $O(n^{2d})$-wide again). So the overall depth for performing a logical measurement is $O(n)$. The additional classical width cost is due to the broadcasting which introduce at most $O(n^{2d})$ additional classical bits.

\newpage

\subsection{Propagation Through Gate-Teleportation}
\label{subsection:logical_gates}
Having an explicit description of the gate's realizations enables tracking how faults propagate through the computation. The next claim formulates the transitions of faults via the gate-teleportation gadgets. 

\begin{claim}
\label{claim:paulipropcon}  
The propagation of a Pauli through the (conditioned) gate teleportation gadgets has a well-defined structure. 
\end{claim}
\begin{proof}
For chains $x,z \in C_{d^{\prime}}(\mathcal{T}_n^d)$, we denote by $P_{x,z}$ the Pauli consisting of $X$'s and $Z$'s on the faces of $x$ and $z$. We prove the claim by propagating Paulis manually. We start by propagating Pauli through the $\mathbf{S}$-teleportation, let $P_{x_{1},z_{1}}$ and $P_{x_{2},z_{2}}$ be the Pauli operators that act on the first and the second registers. Then: 
\begin{figure}[H]
        \centering 
        \begin{equation*}
        \begin{split}
        & \begin{quantikz}
          \gate{\inr{ P_{x_{1},z_{1}} }  } & \ctrl{1}  &  & \gate{Z_{S^{\prime}}} &   \\
        \gate{\inr{ P_{x_{2},z_{2}} }  } & \targ{} & \meter{} &  \gate{\mathbf{1}_{x_{S^{\prime}}}} \wire[u][1]{c}   \setwiretype{c}  \\
        \end{quantikz}  \ = \  
        \begin{quantikz}
          & \ctrl{1}  & \gate{\inr{ P_{x_{1},z_{1} + z_{2} } }  }  &  & \gate{Z_{S^{\prime}}} &   \\
           & \targ{} &  \gate{\inr{ P_{x_{1} + x_{2},z_{2}} }  } & \meter{} &  \gate{\mathbf{1}_{x_{S^{\prime}}}} \wire[u][1]{c}   \setwiretype{c}  \\
        \end{quantikz} \\
        = \, & \begin{quantikz}
          & \ctrl{1}  &  &  \gate{\inr{ P_{x_{1},z_{1} + z_{2} } }  } & \gate{Z_{S^{\prime}}} &   \\
          & \targ{} &  \meter{} &  \gate{\inr{ P_{x_{1} + x_{2}, }}  } \setwiretype{c} & \gate{\mathbf{1}_{x_{S^{\prime}}}} \wire[u][1]{c}     \\
        \end{quantikz} \ = \  
        \begin{quantikz}
          & \ctrl{1}  &  & & \gate{Z_{S^{\prime}}} &   \gate{\inr{ \pm P_{x_{1},z_{1} + z_{2} } }  } & \\
           & \targ{} &  \meter{}  & \gate{\inr{ P_{x_{1} + x_{2}, }}  } \setwiretype{c} & \gate{\mathbf{1}_{x_{S^{\prime}}}} \wire[u][1]{c}     \\
        \end{quantikz} 
         \end{split}
        \end{equation*}
        \caption{ Pauli propagation through $\mathbf{S}$ teleportation. }   
\end{figure}
For the $\mathbf{T}$-teleportation, we denote by $P_{x_{3},z_{3}}$ the Pauli acting on the third register containing the resource state $\ket{ \mathbf{S} }$, which might be used through the gadget to `correct` the teleportation. Let $\mathbf{1}^{2}_{x_{S^{\prime}}}$ be the indicator for the event in which measuring the resource register (that prior to the entangling, holds the state $ P_{x_{2},z_{2}} \ket{\mathbf{T}}$) results in $\ket{\bar{1}}$. Then:
    \begin{figure}[H]
        \centering 
        \begin{equation*}
        \begin{split}
        & \begin{quantikz}
         & \gate{\inr{ P_{x_{1},z_{1}} }  } & \ctrl{2}  & & \gate[2]{\mathbf{S}} &   \\
          \lstick{$ \ket{\mathbf{S}}$} & \gate{\inr{ P_{x_{3},z_{3}}  }} & &  &    \\
         &\gate{\inr{ P_{x_{2},z_{2}} }  }    & \targ{} & \meter{} & \gate{\mathbf{1}_{x_{S^{\prime}}}} \wire[u][1]{c}   \setwiretype{c}  \\
        \end{quantikz} \ = \  
        \begin{quantikz}
          & \ctrl{2}  & & \gate{\inr{ P_{x_{1},z_{1}+z_{2}} }  } & \gate[2]{\mathbf{S}} &   \\
          \lstick{$ \ket{\mathbf{S}}$} & & &\gate{\inr{ P_{x_{3},z_{3}}  }}   &    \\
          &\targ{} & \meter{} & \gate{\inr{ P_{x_{2}+x_{1},} }  }   \setwiretype{c}   & \gate{\mathbf{1}_{x_{S^{\prime}}}} \wire[u][1]{c}   \\
        \end{quantikz} \\
        = \, & \begin{quantikz}
          & \ctrl{2}  & & &\gate[2]{\mathbf{S}} &   \gate{ \inr{ \pm P_{x_{1},z_{1}+z_{2}  +  \mathbf{1}^{2}_{x_{S^{\prime}}}  z_{3}} }  } &  \\
          \lstick{$ \ket{\mathbf{S}}$} & & &\gate{\inr{ P_{x_{3}+x_{1},z_{3}}  }}   &    \\
          &\targ{} & \meter{} & \gate{\inr{ P_{x_{2}+x_{1},} }  }  \setwiretype{c}   & \gate{\mathbf{1}_{x_{S^{\prime}}}} \wire[u][1]{c}    \\
        \end{quantikz} 
         \end{split}
        \end{equation*}
        \caption{ Pauli propagation through $\mathbf{T}$ teleportation. }   
\end{figure}
Where to infer the last passage we used the Pauli-propagation through $\mathbf{S}$-teleportation. Finally, for the $\mathbf{H}$-teleportation we have: 
    \begin{figure}[H]
        \centering 
        \begin{equation*}
          \begin{split}
        & \begin{quantikz}
          \gate{\inr{ P_{x_{1},z_{1}} }} & \ctrl{1}  & \gate{\mathbf{H}^{\otimes\mathcal{F}_{d^\prime}^{n}}} & \meter{} &  \gate{ \phi }  \setwiretype{c} &  \gate{\mathbf{1}_{x_{S/S^{\prime}}}} \wire[d][2]{c} \\
          \gate{\inr{ P_{x_{2},z_{2}} }}  & \targ{} & & \meter{}   &  \gate{\mathbf{1}_{x_{S^{\prime}}}} \wire[d][1]{c}   \setwiretype{c} \\
        \gate{\inr{ P_{x_{3},z_{3}} }}  & &  & & \gate{Z_{S^{\prime}}}   & \gate{X_{S^{\prime}}}  &  \\
        \end{quantikz}  \\
        = \, & \begin{quantikz}
          & \ctrl{1}  & \gate{\mathbf{H}^{\otimes\mathcal{F}_{d^\prime}^{n}}} & \meter{} &  \gate{ \phi }  \setwiretype{c} & \gate{\inr{ P_{\phi(z_{1} + z_{2}), }}}   & \gate{\mathbf{1}_{x_{S/S^{\prime}}}} \wire[d][2]{c} \\
            & \targ{} & & \meter{}   & \gate{\inr{ P_{x_{1}+x_{2},} }}    \setwiretype{c}   &\gate{\mathbf{1}_{x_{S^{\prime}}}} \wire[d][1]{c}  \\
            & & &  & & \gate{Z_{S^{\prime}}}   & \gate{X_{S^{\prime}}}  & \gate{\inr{ \pm P_{x_{3},z_{3}} }}   \\
        \end{quantikz} 
          \end{split}
        \end{equation*}
        \caption{ Pauli propagation through $\mathbf{H}$ teleportation. }   
\end{figure}
Notice that the Pauli $P_{\phi(z_{1}+z_{2}),}$ consists of applying $X$'s on the $d-d^{\prime}$-dimensional faces of the chain $\phi(z_{1}+z_{2})$.
\end{proof}
We summarize by restating the claim above for Clifford + measurement gates, which also use sweep-cycles.
\begin{corollary}
\label{claim:pauliprop}  
Let $C$ be a quantum circuit, such that all its registers are encoded via the nice encoding, and the gates of $C$ are either logical Pauli, transversal $\mathbf{CX}$, sweep-cycles, or $\mathbf{H}, \mathbf{S}$, and $\mathbf{T}$ realized via gate-teleportation. Then the propagation of Pauli through a (conditioned) subjection of $C$ has a well defined structure. In particular, the logical operators propagate Paulis in transversal manner. 
\end{corollary}

\section{Toom's Contours}

\subsection{The Barrier Argument}
We pass repeatedly between the finite Toric complex and its infinite cubical cover. We first extend the size function and the notion of a reduced chain to the finite complex.

\begin{definition}
  \label{def:mincover}
Let $x \in C_{i}(\mathcal{T}_n^d)$ be an $i$-dimensional chain on the finite Toric complex. We extend the size function as follows. Let $x^{\prime}$ be the $\infty$-lift of $x$, so $x^{\prime} \in C_{i}(\mathcal T_\infty^d)$. Let $y$ be a finite-support subchain of $x^{\prime}$ satisfying:
  \begin{enumerate}
    \item The projection of $y$ back to the finite complex covers $x$.
    \item The size of $y$ is minimal. 
  \end{enumerate}
  Then, we set: 
  \begin{equation*}
    \begin{split}
      \Size{x} \colonequals \Size{y}
    \end{split}
  \end{equation*}
We call $y$ a \emph{minimizing cover} of $x$. We extend the notation of reduced form to a $d^{\prime}$-dimensional chain on the finite complex $x \in C_{d^{\prime}}(\mathcal{T}_n^d)$, to be its translation by stabilizers that gets the minimal size. Namely, for $z \in \image \partial^{-} : C_{d^{\prime}+1}(\mathcal{T}_n^d) \rightarrow C_{d^{\prime}}(\mathcal{T}_n^d)$, the chain $x+z$ is a reduced form of $x$ if $\Size{x+z}$ is minimal among the choices of $z$.
\end{definition}
In the following, we prove that the definition of the size function on intersecting finite chains satisfies a subadditivity bound.
\begin{claim}
  \label{clima:finitetrig}
  Let $x,y\in C_{i}(\mathcal{T}_n^d)$ be chains such that $V(x) \cap V(y) \neq \emptyset$, then: 
  \begin{equation*}
    \begin{split}
      \Size{x + y} \le \Size{x} + \Size{y}
    \end{split}
  \end{equation*}
\end{claim}
\begin{proof}
  First, notice that if the finite-support covering chains $x^{\prime}, y^{\prime} \in C_{i}(\mathcal T_\infty^d)$ satisfy that the projection of $x^{\prime}$ covers $x$ and the projection of $y^{\prime}$ covers $y$, then the projection of $x^{\prime} \cup y^{\prime}$ covers $x+y$. Denote the $i$-dimensional faces in $x, y$ that share a vertex by $f,g$, and pick $x^{\prime}, y^{\prime}$ such that $f, g$ are also in $x^{\prime}, y^{\prime}$\footnote{If $f$ and $g$ meet across the periodic boundary, then reprove for translations of $x,y$. }. Thus, $V(x^{\prime})\cap V(y^{\prime}) \neq \emptyset$, and by \Cref{claim:tringinq}, we have:
  \begin{equation*}
    \begin{split}
      \Size{x + y} \le \Size{x^{\prime} \cup y^{\prime}} \le \Size{x^{\prime}} + \Size{y^{\prime}}
    \end{split}
  \end{equation*}
To finish, we argue that we can choose $x^{\prime}, y^{\prime}$ such that $\Size{x} = \Size{x^{\prime}}$ and $\Size{y} = \Size{y^{\prime}}$. Let $\tilde{x} \in C_{i}(\mathcal T_\infty^d)$ be the $i$-dimensional chain on the infinite skeleton such that $\Size{x} = \Size{\tilde{x}}$. Split into cases depending on whether $f \in \tilde{x}$.
  \begin{itemize}
    \item If $f \in \tilde{x}$, then we can just choose $x^{\prime} = \tilde{x}$.
    \item If not, then since the projection of $\tilde{x}$ covers $x$, there must be a vertex $v \in \mathcal{G}_{d}$, such that $v$ is a product of the integer powers of $g_{1}^{n},\ldots, g_{d}^{n}$, and a $i$-dimensional face $\tilde{f}$ in $\tilde{x}$ such that $\tilde{f}$ is a translation of $f$ by $v$. Then, construct $x^{\prime}$ by taking the translations of $\tilde{x}$ faces by $v^{-1}$. The size function is invariant to translations, so:
      \begin{equation*}
        \begin{split}
          \Size{x} = \Size{\tilde{x}} = \Size{v^{-1}\tilde{x}} = \Size{x^{\prime}}
        \end{split}
      \end{equation*}
      and in addition, $f = v^{-1}\tilde{f} \in x^{\prime}$. Moreover, since the projection is invariant to translation by integer powers of $g_{1}^{n},\ldots, g_{d}^{n}$, the projection of $\tilde{x}$ equals the projection of $x^{\prime}$, namely, covers $x$.
  \end{itemize}
  So, 
  \begin{equation*}
    \begin{split}
      \Size{x + y} \le \Size{x^{\prime}} + \Size{y^{\prime}} = \Size{x} + \Size{y}
    \end{split}
  \end{equation*}
\end{proof}

We will use the following two elementary lifting facts and omit their proofs. They are discrete instances of the standard local relation between a Torus and its universal cover: below the periodicity scale, the covering map is injective on the relevant region, so chains may be treated as chains in the infinite lattice. In the Toric-code literature, the same principle appears in the identification of non-trivial logical operators with homologically non-trivial cycles of linear extent \cite{Dennis_2002}, and in the restriction of Toric arguments to local regions whose diameter is smaller than the system size \cite{Kubica_2019}. The claims below merely formulate this observation for the size function and minimizing covers used here.

\begin{claim}
  \label{claim:liftb}
  Let $x \in C_{d^{\prime}-1}(\mathcal{T}_n^d)$ be a $d^{\prime}-1$ dimensional chain of the finite complex, and let $y \in C_{d^{\prime}-1}(\mathcal T_\infty^d)$ be the $d^{\prime}-1$ dimensional chain on the infinite complex such that $y$ is the minimizing cover of $x$. If $\Size{x} < n-1$, and:
\begin{equation*}
  \begin{split}
    x \in \image \partial^{-} : C_{d^{\prime}}(\mathcal{T}_n^d) \rightarrow C_{d^{\prime}-1}(\mathcal{T}_n^d)
  \end{split}
\end{equation*}
Then, also:
\begin{equation*}
  \begin{split}
    y \in \image \partial^{-} : C_{d^{\prime}}(\mathcal T_\infty^d) \rightarrow C_{d^{\prime}-1}(\mathcal T_\infty^d)
  \end{split}
\end{equation*}
\end{claim}

\begin{claim}
  \label{claim:liftc}
Let $x \in C_{d^{\prime}}(\mathcal{T}_n^d)$ be a $d^{\prime}$-dimensional chain of the finite complex. Suppose that $\Size{x} < n-1$. Then there is a translation $x^{\prime}$ of $x$ (on the finite complex), such that $x^{\prime}$ has a minimizing cover $y \in C_{d^{\prime}}(\mathcal T_\infty^d)$, such that the faces of $y$ align with the finite complex. In other words, the chosen representatives coincide.
\end{claim}

Next, we translate the fact that a chain enclosing a non-trivial loop has a size which is at least the side of the Toric.
\begin{claim}
  \label{claim:sizenont}
Suppose that the chain $x \in C_{i}(\mathcal{T}_n^d)$ closes a non-trivial loop on the Toric, then $\Size{x} \ge n - 1$.
\end{claim}
\begin{proof}
  Let $x^{\prime} \in C_{i}(\mathcal T_\infty^d)$ be a minimizing cover for $x$. In particular, there are vertices $v_{1}, \ldots, v_{l}$ in $x^{\prime}$ such that their projection corresponds to the non-trivial loop in $x$. Assume the non-trivial cycle is in the $g$-direction, thus the $L_{g}$ values of the projections of $v_{1}, \ldots, v_{l}$ cover all the integer range $\{0,1,\ldots, n-1\}$. Thus, there are two vertices $v,u \in v_{1},\ldots,v_{l}$ such that their $L_{g}$-difference is at least $L_{g}(u) - L_{g}(v) \ge n-1$. Thus,
  \begin{equation*}
    \begin{split}
      \Size{x} = \Size{x^{\prime}} \ge \Size{\{u,v\}} \ge n-1
    \end{split}
  \end{equation*}
\end{proof}

Using the claim above, we are able to introduce the notion of filling boundaries. The following claim states that if the connected components of the boundary are all of size lower than $n$, then we can match each of them to a chain such that its boundary agrees with the component and its size is lower than the component size. That will lead us to define the filling operator (\Cref{def:connsyn}).
\begin{claim}
  \label{claim:fillchain}
  Let $x \in C_{d^{\prime}}(\mathcal{T}_n^d)$ be a $d^{\prime}$-dimensional chain. Suppose that there is an integer $l \le n-1$ such each connected component of $\partial^{-}x$ has size $ < l$. Then there are $d^{\prime}$-dimensional chains $y_{1}, y_{2}, \ldots$ such that each $\partial^{-}y_{i}$ is a connected component of $\partial^{-}x$ and $\Size{y_{i}} \le \Size{\partial^{-}y_{i}}$.
\end{claim}

\begin{proof}
  Consider a $d^{\prime}$-dimensional chain $x \in C_{d^{\prime}}(\mathcal{T}_n^d)$ such that $\partial^{-} x$ has a decomposition $\partial^{-}x = \tilde{x}_{1} + \tilde{x}_{2}$, where $\tilde{x}_{1}, \tilde{x}_{2}$ are $d^{\prime}-1$-dimensional chains, and $\tilde{x}_{1}$ is a connected component of $\partial^{-}x$. In particular $\tilde{x}_{1}$ is not connected to $\tilde{x}_{2}$. By disconnectivity, $\tilde{x}_{1}$ does not share a $d^{\prime}-2$ dimensional face with $\tilde{x}_{2}$. Thus, $\partial^{-} \tilde{x}_{1}$ and $\partial^{-} \tilde{x}_{2}$ are disjoint. On the other hand, since $\partial^{- \circ 2} x = 0$, then combining with disjointness, it follows that:
  \begin{equation*}
    \begin{split}
      \partial^{-}( \tilde{x}_{1} + \tilde{x}_{2} ) = 0 \Rightarrow \partial^{-} \tilde{x}_{1} = \partial^{-}\tilde{x}_{2} = 0 
    \end{split}
  \end{equation*}  Namely, $\tilde{x}_{1} \in \ker \partial^{-} : C_{d^{\prime}-1}(\mathcal{T}_n^d) \rightarrow C_{d^{\prime}-2}(\mathcal{T}_n^d)$.
Suppose $\tilde{x}_{1} \notin \image \partial^{-} : C_{d^{\prime}}(\mathcal{T}_n^d) \rightarrow C_{d^{\prime}-1}(\mathcal{T}_n^d)$, then we have that $\tilde{x}_{1}$ is a non-trivial codeword of the $(d,d^{\prime}-1)$-Toric code. 
Since $d^{\prime} \ge 2$, and $\tilde{x}_{1}$ has to close a non-trivial loop, so by \Cref{claim:sizenont}, the size of $\tilde{x}_{1}$ is at least $n-1$, which is a contradiction. So we get that $\tilde{x}_{1}\in \image \partial^{-} : C_{d^{\prime}}(\mathcal{T}_n^d) \rightarrow C_{d^{\prime}-1}(\mathcal{T}_n^d)$. In particular, there is a $d^{\prime}$-dimensional chain $x_{1}$ such that $\partial^{-}x_{1} = \tilde{x}_{1}$. We argue that $x_{1}$ can be chosen such $\Size{x_{1}} \le \Size{\tilde{x}_{1}}$.

By \Cref{claim:liftc}, since $\Size{\tilde{x}_{1}} < n-1$, there is a translation of the finite complex for which the minimizing cover of the translation of $\tilde{x}_{1}$ aligns with the finite complex. Since the size function is invariant to translations, we can assume, without loss of generality, that the given finite complex is already translated. Consider the minimizing cover of $\tilde{x}_{1}$, and denote it by $\tilde{y_1}$. Since $\Size{\tilde{x}_{1}} < n-1$, then by \Cref{claim:liftb} we have that $\tilde{y}_{1}$ can be chosen such:
\begin{equation*}
  \begin{split}
    \tilde{y}_{1} \in \image \partial^{-} : C_{d^{\prime}}(\mathcal T_\infty^d) \rightarrow C_{d^{\prime}-1}(\mathcal T_\infty^d)
  \end{split}
\end{equation*}
Let $y_{1} \in C_{d^{\prime}}(\mathcal T_\infty^d) $ be a $d^{\prime}$-dimensional reduced chain such that $\partial^{-}y_{1} = \tilde{y}_{1}$. By the definition of the algorithm in \Cref{claim:reducederr}, from which $y_{1}$ is obtained, it follows that the potential walls that bound $\tilde{y}_{1}$ also bound $y_{1}$, and therefore $y_{1}$ is also aligned with the finite complex. Thus, we can take $x_{1}$ to be equal to $y_{1}$. By \Cref{claim:reducederr}:
\begin{equation*}
  \begin{split}
    \Size{x_{1}} = \Size{y_{1}} = \Size{\partial^{-}y_{1}} = \Size{\tilde{y}_{1}} = \Size{\tilde{x}_{1}}
  \end{split}
\end{equation*}
So $x_{1}$ satisfies the desired condition for the connected component $\tilde{x}_{1}$. To complete, we repeat recursively on the connected components of $\tilde{x}_{2}$. 
\end{proof}

\begin{definition}
  \label{def:connsyn}
Let $x \in C_{d^{\prime}}(\mathcal{T}_n^d)$ be a $d^{\prime}$-dimensional chain. $x$ will be said to be a fillable chain if each of the connected components of its $\partial^{-}$-boundary, namely $\partial^{-} x$, has size $< n - 1$. By \Cref{claim:fillchain}, there are $d^{\prime}$-dimensional reduced chains $x_{1}, \ldots,$ such that each $\partial^{-}x_{i}$ is a connected component of $\partial^{-}x$. Thus:
    \begin{equation*}
      \begin{split}
      \partial^{-} (x + \sum_{i}x_{i}) = 0  \quad \Rightarrow x + \sum_{i}x_{i} \in \ker \partial^{-} \colon C_{d^{\prime}}(\mathcal{T}_n^d) \rightarrow C_{d^{\prime}-1}(\mathcal{T}_n^d)
    \end{split}
  \end{equation*}
  Namely, $x + \sum_{i}x_{i}$ is a codeword in $\mathcal{C}_{X}/ \mathcal{C}_{Z}^{\perp}$. We define the filling operator, denoted by $[x]$, which for a given fillable chain $x$, returns the homology class (logical operator) of $x + \sum_{i}x_{i}$. Lastly, to ensure that the filling operator is canonical, fix for each $d^{\prime}-1$-dimensional chain, which is a connected component, a unique reduced $d^{\prime}$-dimensional chain. With the notations above, for each $\partial^{-} x_{i}$ fix a unique $x_{i}$.
\end{definition}

The following claim tracks the evolution of the connected component of the boundary (in terms of the maximal size), where the process is restricted such that in each step the chain is allowed to differ by at most a single face.
\begin{claim}
  \label{claim:mergebyf}
  Let $x_{1},x_{2} \in C_{d^{\prime}}(\mathcal{T}_n^d)$ be $d^{\prime}$-dimensional chains, and let $l$ be an integer such the Hamming distance between $x_{1}$ and $x_{2}$ is at most $l$. Denote by $r$ the size of the maximal connected component of $\partial^{-}x_{1}$. Then the maximal size of a connected component of $\partial^{-}x_{2}$ is at most $4{ d^{\prime} \choose 2 }l(r+d^{\prime}) $
\end{claim}
\begin{proof}
  Consider the adjacency graph (\Cref{def:asontoric}) with respect to the $d^{\prime}-1$-dimensional chain $\partial^{-}x_{1}$, where each vertex is a $d^{\prime}-1$-dimensional face of $\partial^{-}x_{1}$, and each edge in the graph is a $d^{\prime}-2$ face in the intersection of two $d^{\prime}-1$-dimensional faces in $\partial^{-}x_{1}$. A $d^{\prime}$-dimensional face contains $4{ d^{\prime} \choose 2 }$ faces of dimension $d^{\prime}-2$ . Hence, introducing a new $d^{\prime}$-dimensional face to $x_{1}$ adds at most $4{ d^{\prime} \choose 2 }$ edges to the adjacency graph, each corresponding to a $d^{\prime}-2$ face contained in the new $d^{\prime}$-dimensional face. So introducing $l$ faces into $x_{1}$ adds at most $4 { d^{\prime} \choose 2 } l$ new edges, and therefore connects at most the same number of connected components in $\partial^{-}x_{1}$, or a connected component induced by the additional $l$ faces. Using the triangle inequality for intersecting sets (\Cref{clima:finitetrig}), we have that any connected component of $\partial^{-}x_{2}$ is at most $4{ d^{\prime} \choose 2 } l$ times the maximal size of a connected component of $\partial^{-}x_{1}$, namely $4 { d^{\prime} \choose 2 }l (r+d^{\prime})$ (Where the $d^{\prime}$ in the closure stands for the connected components of a single face, each corresponding to one of the additional $l$ faces.)
\end{proof}

Next, we present the main claim of the section, which argues that for a sequence of chains $x_{1}, \ldots, x_{t}$, such that any two sequential chains differ by at most a single face, if $[x_{1}] \neq [x_{t}]$, then there must be an intermediate step $i$ such that $\partial^{-}x_{i}$ has a connected component with sufficiently large size. After that, we present \Cref{claim:barrierconstant}, which generalizes the claim to allow sequential elements to be at a distance greater than one.

\begin{claim}
  \label{claim:barrier}
Let $x_{1}, \ldots, x_{t} \in C_{d^{\prime}}(\mathcal{T}_n^d)$ be $d^{\prime}$-dimensional chains, such that $x_{1}$ is a trivial chain and for any $i \in [t-1]$, the Hamming distance between $x_{i}$ and $x_{i+1}$ is one. There is a constant $c \in (0,1)$, depending only on $d, d^{\prime}$, such that if $n$ is sufficiently large, and for any $x_{i} \in x_{1}, \ldots, x_{t}$, the size of each of the connected components of $\partial^{-} x_{i}$ is bounded by $c n$, then:
  \begin{equation*}
    \begin{split}
      [0]= [x_{1}] = [x_{2}] = ... = [x_{t}] 
    \end{split}
  \end{equation*}
\end{claim}
\begin{proof}
  Assume through contradiction that there is $i$ such that $0 = [x_{i}] \neq [x_{i+1}]$. Denote by $x_{i}^{1}, x_{i}^{2}, \ldots$ and $x_{i+1}^{1}, x_{i+1}^{2}, \ldots$ the reduced $d^{\prime}$-dimensional chains such that $\partial^{-}x_{i}^{j}$ is a connected component of $\partial^{-}x_{i}$, as defined in \Cref{def:connsyn}. Hence:
  \begin{equation*}
    \begin{split}
      [x_{i+1}] \sim_{\mathcal{C}_{X}} [x_{i+1}] + [x_{i}] \sim_{\mathcal{C}_{X}}  x_{i} + x_{i+1} + \sum_{j}x_{i}^{j} + \sum_{j}x_{i+1}^{j}
    \end{split}
  \end{equation*}
Denote by $w$ the chain:
  \begin{equation*}
    \begin{split}
      w \colonequals x_{i} + x_{i+1} + \sum_{j}x_{i}^{j} + \sum_{j}x_{i+1}^{j}
    \end{split}
  \end{equation*}
  Since the Hamming distance between $x_{i}$ and $x_{i+1}$ is at most one, and since each $d^{\prime}$-dimensional face contributes a constant support over $d^{\prime}-1$-dimensional chains, the $\partial^{-}$-boundaries of $x_{i}$ and $x_{i+1}$ are at a Hamming distance that is at most constant. Therefore, there are only constant connected components of $\partial^{-}x_{i}$ which are not connected components in $\partial^{-}x_{i+1}$ and vice versa. Denote by $y_{1}, \ldots, y_{m}$ the chains in $x_{i}^{1}, \ldots$ and $x_{i+1}^{1}, \ldots$ such that they belong to only one of the collections. Where $m$ is bounded by a function of $d,d^{\prime}$, namely $m \le M(d,d^{\prime})$. So:
  \begin{equation*}
    \begin{split}
      \sum_{j}x_{i}^{j} + \sum_{j}x_{i+1}^{j} = \sum_{j}^{m}y_{j}
    \end{split}
  \end{equation*}
Moreover, the vertices of any of $y_{1}, \ldots, y_{m}$ and $x_{i+1}+x_{i}$ have a nontrivial intersection. To see this, suppose that the vertices of $y_{j}$ do not intersect the face $x_{i+1}+x_{i}$, and assume without loss of generality that $\partial^{-}y_{j}$ is a connected component of $\partial^{-}x_{i}$. Then, the vertices of $\partial^{-}y_{j}$ also do not intersect the vertices of $\partial^{-}(x_{i} + x_{i+1})$. So $\partial^{-}y_{j}$ is also a connected component of
  \begin{equation*}
    \begin{split}
    \partial^{-}x_{i} + \partial^{-}(x_{i+1} + x_{i})  = \partial^{-}x_{i+1}
    \end{split}
  \end{equation*}
  which is contradiction to the choices of the $y_{j}$'s.  Therefore:
\begin{equation*}
  \begin{split}
    \Size{ w } \le \Size{  x_{i} + x_{i+1} } + \sum_{j}^{m} \Size{y_{j}} \le d + M(d,d^{\prime}) c n  
  \end{split}
\end{equation*}
Thus, for sufficiently large $n$ ,if $c < \frac{1}{M(d,d^{\prime})+1}$, we have that $\Size{w} < n-1$, a contradiction for $w$ being a non-trivial codeword.
\end{proof}

\begin{claim}
  \label{claim:barrierconstant}
  Let $x_{1}, \ldots, x_{t} \in C_{d^{\prime}}(\mathcal{T}_n^d)$ be $d^{\prime}$-dimensional chains defined as in \Cref{claim:barrier}, with the change that any two sequential chains $x_{i}, x_{i+1}$ are at a Hamming distance of at most $\Delta$, and setting the constant $c^{\prime} \leftarrow c/(8{ d^{\prime} \choose 2 }  \Delta )$. Then the claim still holds.
\end{claim}

\begin{proof}
  Let $x_{1}^{\prime}, \ldots, x^{\prime}_{\Delta t + 1}$ be the chains obtained from $x_{1}, \ldots, x_{t}$ by stacking $\Delta-1$ intermediate chains between any two sequential elements, such that any two sequential chains of the resulted set are at Hamming distance at most one. Namely, for any $i \in \{0, 1, \ldots t-1 \} $, we have that $x_{1 + i\Delta}^{\prime} = x_{i + 1}$ and $x^{\prime}_{i+1}$ differs from $x^{\prime}_{i}$ by at most one coordinate. By \Cref{claim:mergebyf}, each connected component of $\partial^{-}x^{\prime}_{i}$ has a size of at most $4{ d^{\prime} \choose 2 }\Delta (c^{\prime} n+d^{\prime}) \le c n$. 

  \paragraph{}
  Thus, by \Cref{claim:barrier},we have that:
  \begin{equation*}
    \begin{split}
      [0] = [x_{1}^{\prime}] =[x_{2}^{\prime}] = \ldots = [x_{\Delta t + 1}^{\prime}] 
    \end{split}
  \end{equation*}
  In particular: 
  \begin{equation*}
    \begin{split}
      & [0] = [x_{1}^{\prime}] =[x_{1 + \Delta}^{\prime}] = \ldots  =  [x_{1 + i \Delta}^{\prime} ]  =  \ldots = [x_{\Delta t + 1}^{\prime}] \\
      \Rightarrow  & [0] = [x_{1}] =[x_{2}] =   \ldots = [x_{t}] \\
    \end{split}
  \end{equation*}
\end{proof}

We summarize the section with the following corollary that incorporates the above insight into an ongoing computation. Namely, changing the logical codeword implies an intermediate step with a sufficiently large connected component of the syndrome. In the computation context, the sequence of the chains is derived from the Paulis in the step-propagation of a conditioned subjected circuit.

We are stressing that the Paulis in the step-propagation record the deviation from an ideal running. In particular, a flattened into steps running, which introduces large syndrome clusters, does not imply a large connected component of the Paulis in the step-propagation. For example, consider CNOTing a register containing a large stabilizer into a second empty register. The ideal running exhibits an intermediate step in which the syndrome of the second register has a large connected component, yet the Paulis in the 'step-propagation of the ideal running' (namely where $\mathcal{E} = \emptyset$) are the identity.
\begin{corollary}[Barrier argument]
  \label{coro:barrier}
  Let $c^{\prime} \in (0,1)$ be the constant as defined in \Cref{claim:barrierconstant}. Let $P_{1},\ldots,P_{\tau_{f}}$ be a step-propagation of a conditioned subjected circuit $C[\mathcal{E},\alpha]$ (\Cref{def:propagation}). Assume that the arity of the gates of $C$ is bounded by $\Delta$, and that each of the registers in $C$ is encoded in the $(d,d^{\prime})$-Toric code. Denote by $R$ the number of quantum registers in $C$. For a Pauli $P_{\tau}$ in the step-propagation, denote by $x_{\tau} \in C_{d^{\prime}}(\mathcal{T}_n^d)^{\otimes R}$ the $X$-component of $P_{\tau}$ and by $x_{\tau}|_{R_{j}}$ the restriction of $x_{\tau}$ to register $R_{j}$. Then, if for any step $\tau$ and for any register $R_{j}$, there is not a connected component of $\partial^{-} x_{\tau}|_{R_{j}}$ of size greater than $c^{\prime}n$, then:
  \begin{equation*}
    \begin{split}
      [x_{\tau_{f}} |_{R_{j}}] = [0] 
    \end{split}
  \end{equation*}
A similar result holds for the $Z$-components, where we define $z_{\tau}|_{R_{j}}$ to be the $d-d^{\prime}$-dimensional chain obtained by applying the $\phi$-transpose on the chain supporting the $Z$-component of $P_{\tau}$ restricted to the register $R_{j}$.
\paragraph{}
Equivalently, if the subjected computation introduces a logical error, then at some intermediate step one of the $X$- or $Z$-syndromes has a connected component of size at least $c^{\prime}n$. 
\end{corollary}
\begin{proof}
The proof follows immediately from \Cref{claim:barrierconstant}, where we use the fact that the arity is bounded by $\Delta$, and therefore sequential Paulis in the step-propagation are at a Hamming distance of at most $\Delta$.
\end{proof}

\begin{remark}
Consider the application of an iterative sweep rule over a chain whose boundary is at a size lower than $c^{\prime} n$. Since the sweep rule only decreases the size of the boundary, it is easy to see that along the application, the size of the connected component of the boundary is kept below $c^{\prime}n$. Thus, each time, the filling operator will complete the chain to the same codeword. In particular, we get that decoding a measurement as presented in \Cref{def:niceenc} doesn't result in a logical error, provided that each of the connected components of the boundary, prior to the broadcast, does not have a size exceeding $c^{\prime}n$.
\end{remark}

\subsection{Toom's Contours}
\label{subsection:toricencoded}
We are about to zoom out from a single register encoded in the Toric code to describe a fully fault-tolerant computation. For a given quantum circuit, assumed to consist of gates $\mathbf{X}, \mathbf{Z}, \mathbf{CX}, \mathbf{H}, \mathbf{S}, \mathbf{T}$ and measurements, we define its Toric encoded circuit as its version where each qubit is encoded by the Toric code. Formally:
    \begin{definition}[Toric Encoded Circuit]
      Let $C$ be a quantum circuit. We say that $C$ is a \textit{Toric encoded circuit} if the following holds:
      \begin{enumerate}
        \item Any layer of the circuit consists of classical and quantum registers, and there is a Toric code $\mathcal{C}$ such that the quantum registers are encoded by the \textit{nice encoding of} $\mathcal{C}$. At time $t=0$, the quantum registers are initialized to be either $\ket{\bar{0}}$, encoded $\ket{\mathbf{H}}$, encoded $\ket{\mathbf{S}}$, or encoded $\ket{\mathbf{T}}$. The classical registers are initialized to be zeros.
        \item The layers of $C$ on the quantum registers at even times are sweep-cycles, and the layers at odd times are realizations of either logical Paulis, $\mathbf{CX}$, or measurements in the computational and phase bases. On the classical registers, including the measured register, we allow any two-bit gates.
        \end{enumerate} 
    \end{definition}

For Toric encoded circuits, we define their \textit{base graph}, which can be thought of as a recording of a running from the viewpoint of the Toric complex. The \textit{base graph} is obtained by taking the Toric complex's vertices at different times as the vertices and connecting any two sequential in time if they support faces — qubits — which are either the same or are entangled by a quantum gate.

  \begin{definition}[Base graph $G_{o}$]
    Let $C$ be a Toric encoded circuit at depth $\mu$. Notice that measurements take quantum registers into classical data; thus each quantum register has a final time coordinate, which we denote by $\{\mu_i\}_i$. Denote by $\{\mathcal{G}_{i}\}_i$ the skeletons of the $\infty$-lift of the Toric complexes that induce the codes. We define the base graph of $C$, denoted by $G_{o} = (V_{o}, E_{o})$, as follows:
    \begin{enumerate}
      \item The vertices $V_{o}$ are pairs, where the first coordinates correspond to a vertex in the disjoint union over the vertices in $\{\mathcal{G}_{i}\}$, and the second coordinate is associated with a time. Namely:
        \begin{equation*}
          \begin{split}
            V_{o} \colonequals \bigcup_{i} \bigl( V(\mathcal{G}_{i}) \times [\mu_{i}+1] \bigr)
          \end{split}
        \end{equation*}
      \item The edges $E_{o}$ are derived from the original edges in $\mathcal{G}_{1}, \ldots, \mathcal{G}_{k}$, with the addition that any two vertices at consecutive times $t$ and $t+1$ that support qubits (faces) such that $C^{\infty}$ acts non-trivially on those qubits at time $t$ are also connected by an edge. Namely:
        \begin{equation*}
          \begin{split}
            E_{o} & \colonequals \big\{ \{(v,t), (u,t^{\prime}) \} :  \text{ if }     \\ 
              & \text{either there exists } i \text{ such that }  \{v,u\} \in \mathcal{G}_{i} \text{ and } t^{\prime} = t+1 \\
            & \text{or there exists } u_i = (g_{i},l_{i}) \in C^{\infty} \text{ such  that }  v,u \in l_{i,2} \text{ and } t = l_{i,1}, t^{\prime} = l_{i,1}+1  \big\} 
          \end{split}
        \end{equation*}
        Where we define a vertex $v$ to belong to location $l$ if there is a qubit-face $f$ in $l$ such that $v$ belongs to its associated face. %
    \end{enumerate} 
  \end{definition}
Notice that since the gates' locations are invariant to the conditioning of measurement outcomes (\Cref{remark:condmeas}), we have that for any outcomes vector $\alpha \in \mathbb{F}_{2}^{(\cdot)}$, the base graph of the conditioned circuit $C[\mathcal{E},\alpha]$ is the same as for $C[\mathcal{E}]$.

\paragraph{}

As a concept, if the base graph is a recording of a run from the Toric complexes' point of view, then a \textit{Toom's contour} is a recording of the deviation's evolution in the eyes of the Toric complexes' vertices. We use the notion of $x,z$-component of a lifted Pauli and the Pauli-path to describe a deviation:
\begin{definition}
Let $P \in \mathcal{P}^{(\cdot)}$ be a Pauli, and let $Q \in \mathcal{P}^{(\cdot)}$ be its $\infty$-lift. Assume that $P$ acts on a Toric encoded circuit, namely it has a decomposition such that each element of it acts on a register encoded in the $(d,d^{\prime})$-Toric code. We denote by $x(Q)$ the $d^{\prime}$-dimensional chain in the infinite complex that supports the $X$-component of $Q$, and by $z(Q)$ the $d-d^{\prime}$-dimensional chain, such that the inverse $\phi$-transpose sends it to the $d^{\prime}$-dimensional chain that supports the $Z$-component of $Q$, namely $\phi^{-1}(z(Q))$ is the chain that supports the $Z$-component.

\end{definition}

\begin{definition}
  \label{def:paulipath}
  Consider the propagation $(\mathbf{P},\mathbf{a})$ of a conditioned subjected circuit $C[\mathcal{E},\alpha]$ (\Cref{def:propagation}). A Pauli-path of $C[\mathcal{E},\alpha]$ is the $\infty$-lift of the Paulis in $\mathbf{P}$.

\end{definition}

For a Pauli-path $P_{1}, \ldots ,P_{\mu+1}$, the Toom's contour is the restriction of the base graph to vertices that witness a deviation, namely, those that belong to qubits supporting $P_{1},P_{2},..,P_{\mu+1}$.

\begin{definition}[Toom's Contour Graph (Standard Form)]
    \label{def:contour}
    For a Toric encoded quantum circuit $C$, set of faults $\mathcal{E}$, measurement outcomes $\alpha$, and a Pauli-path $P_{1}, \ldots, P_{\mu +1}$ of $C[\mathcal{E}, \alpha]$, We define the \textit{Toom's contour graph} $\mathcal{H} = (V_{\mathcal{H}}, E_{\mathcal{H}})$ as follows: 
    \begin{enumerate}
      \item The vertices $V_{\mathcal{H}} = V_{\mathcal{H}, X} \cup V_{\mathcal{H}, Z}$ correspond to vertices of faces supporting the $\partial^{-}$-boundary of the Pauli-path. They consist of the triple $(v,t,\cdot)$ where $(v,t)$ is a vertex of the base graph, and the third is a symbolic coordinate that is either $\{X\}$ or $\{Z\}$ corresponding to the component of the Pauli that $v$ supports at time $t$.
        \begin{equation*}
          \begin{split}
            V_{\mathcal{H}, X} \colonequals & \big\{ v_{t} \times \{X\} : v_{t}=(v,t) \in V_{o} \\
            & \, \, \, \text{ and there exists a face } f \in \partial^{-} x(P_{t}) \text{ such that } v \in f \big\} \\
             V_{\mathcal{H}, Z} \colonequals & \big\{ v_{t} \times \{Z\} : v_{t}=(v,t) \in V_{o} \\
             & \, \, \, \text{ and there exists a face } f \in \partial^{-} z(P_{t}) \text{ such that } v \in f \big\} 
          \end{split}
        \end{equation*}

      \item The edge set $E_{\mathcal{H}}$ is partitioned into two types $A_{\mathcal{H}}$ and $F_{\mathcal{H}}$. For any realization of logical gate $g$ which is applied at time $t$ in $C[\mathcal{E}]^{\infty} $, we denote by $E_{o}|_{g}$ the edges in $E_{o}$ which are induced by $g$. Then, we add edges to $A_{\mathcal{H}}$ as follows: 

        \begin{enumerate}
          \item If $g$ does not belong to the faults, namely $g \notin \mathcal{E}$, and $g$ is not a measurement in the phase-basis, then we add the arrows: 
            \begin{equation*}
              \begin{split}
                 \bigl\{  \{ \tilde{u} = (u,t, \cdot),  \tilde{v} = (v,t+1, \cdot)\} : \{\{u,t\},\{v,t+1\}\} \in E_{o}|_{g} \text{ and } \tilde{v},\tilde{u} \in V_{\mathcal{H}, \cdot} \bigr\}
              \end{split}
            \end{equation*}
            Where if $g$ is a measurement in the phase-basis, then we add type-alternating arrows as follows:
            \begin{equation*}
              \begin{split}
                & \bigl\{  \{ \tilde{u} = (u,t, \{Z\}),  \tilde{v} = (v,t+1, \{X\})\} : \{\{u,t\},\{v,t+1\}\} \in E_{o}|_{g} \\
              &  \, \, \, \text{ and } \tilde{v}\in V_{\mathcal{H},X},\tilde{u} \in V_{\mathcal{H},Z} \bigr\} 
              \end{split}
            \end{equation*}
          \item Else, if $g \in \mathcal{E}$ then, we add:   
            \begin{equation*}
              \begin{split}
                & \bigl\{  \{ \tilde{u} = (v,t, \cdot),  \tilde{v} = (v,t+1, \cdot)\} : \text{ such that either } \tilde{u} \in V_{\mathcal{H},X} \text { or } \tilde{u} \in V_{\mathcal{H},Z} \bigr\} 
              \end{split}
            \end{equation*}
        \end{enumerate}
            To construct $F_{\mathcal{H}}$, we connect any pair of vertices $(v,t)$, $(u,t)$ if there exists a vertex $(w,t+1)$ such that both of them are connected to it through $A_{\mathcal{H}}$. Namely: 
        \begin{equation*}
          \begin{split}
            F_{\mathcal{H}} &= \big\{ e = \{u,v\} : \text{ exists } w \in V_{\mathcal{H}} \text{ such that } \{v,w\},\{u,w\} \in A_{\mathcal{H}} \big\} 
          \end{split}
        \end{equation*}
    \end{enumerate}
    We call $A_{\mathcal{H}}$ and $F_{\mathcal{H}}$ the \textit{arrows} and \textit{forks} edges respectively. 
\end{definition}

For the analyses, we would like to extend Toom's contour to record in their final layer a partial time-step, namely a computation up to a certain step. For this, we first consider Toom's contour of the partial computation circuit up to a step (\Cref{def:steps}). Then, since we would like to refer to the connected chain at the final layer as the 'final error clusters', we add at the end a layer whose purpose is to connect in time (by arrows) the connected components of Toom's contour in the prior layer, which are connected by checks (their union is identified with a connected chain).

\begin{definition}[Analysis Toom's Contour Up to Step $\tau$]
  For a Toric encoded quantum circuit $C$, a set of faults $\mathcal{E}$, outcomes $\alpha$, and a step $\tau$, let $C[\mathcal{E},\alpha]_{\tau}$ be the partial computation circuit of $C[\mathcal{E},\alpha]$ up to step $\tau$. We define the \textit{analysis Toom's contour at step $\tau$}, $\mathcal{H}_{\tau} = (V_{\mathcal{H}}, E_{\mathcal{H}})$ to be the graph obtained by taking the Toom's contour of $C[\mathcal{E},\alpha]_{\tau}$ and adding vertices and edges as follows:
  \begin{enumerate}
    \item Let $t$ be the depth of $C[\mathcal{E},\alpha]_{\tau}$. We add a layer corresponding to the time $t+1$, which duplicates the vertices of time $t$. Namely, we add to $V_{\mathcal{H}}$:
\begin{equation*}
  \begin{split}
    \big\{ (v,t+1, \cdot ) : \text{ if }  (v,t, \cdot ) \in V_{\mathcal{H}}  \big\}
  \end{split}
\end{equation*}
    \item Then, we add the following additional arrows:
        \begin{equation*}
          \begin{split}
            \bigl\{ \bigl\{ \tilde{u} & = (u,t, \cdot ),  \tilde{v}= (v,t+1,\cdot)  \bigr\}  \text{ if  either there exists }\\ 
           & \text{ a } d^{\prime}-1  \text{ dimensional face } f  \text{ such that } \tilde{u},\tilde{v} \in V_{\mathcal{H},X} \text{ and } u,v \in f  \\
          & \text{or there exists a } d - d^{\prime}-1  \text{ dimensional face } f \text{ such that } \tilde{u},\tilde{v} \in V_{\mathcal{H},Z} \text{ and } v,u \in f   \bigr\}  
          \end{split}
        \end{equation*}
      \item Finally, we add forks to preserve the property that any two vertices with the same time coordinate, which are connected to a third (with a $+1$ higher time coordinate) via arrows, are also connected by a fork.
  \end{enumerate}
\end{definition}

 \subsection{Space-Time Graph}
  \label{subsection:spacetimegraph}  
  Several properties of Toom's contours follow from their directed space--time structure, independently of the underlying Toric complex. We isolate these properties using \emph{space--time graphs}, whose edges are partitioned into arrows and forks. 
  
  Here, we present the results and proofs first given by Berman and Simon \cite{Berman} and later by Gacs \cite{gacs2021newversiontoomsproof} about \textit{space-time graphs}. Although the proofs we give are almost identical, we choose to present them rather than referring to the sources. The reason for this is that in the formulations in \cite{Berman} and \cite{gacs2021newversiontoomsproof}, vertices are bits\footnote{It is also true that the gates are different, yet that difference alone does not constitute a conceptual change.}, so conceptually their Toom's contours track different objects. Thus, jumping into their text while keeping our image in mind might cause confusion, which we would like to avoid.

  \begin{definition}
    Consider a graph $G = (V, A, F)$, where $V$ is a set of vertices, and $A$ and $F$ are sets of undirected edges. 
    Let $ T \colon V \rightarrow \mathbb{N}$. We say that $(G,T)$ is a space-time graph if and only if:
\begin{enumerate}
  \item $\{x,y\} \in A \Rightarrow \Time{y} = \Time{x} \pm 1$  
  \item For $x,y \in V $  $\{x,y\} \in F $ if and only if there exists $z \in V $ such both $\{x,z\},\{y,z\} \in A$, and $\Time{x} = \Time{y} = \Time{z}-1$. 
\end{enumerate} 
Additionally, for any vertex $v \in V$ of the space-time graph, denote by $\pre{v}$ the set of vertices defined as follows: 
    \begin{equation*}
      \begin{split}
\pre{v} \colonequals \{ u \in V \colon \{u,v\} \in A \text{ and } \Time{v} = \Time{u} + 1 \}
      \end{split}
    \end{equation*}
We extend the notion for subsets of vertices $X \subset V$ by $\pre{X} = \bigcup_{v \in X} \pre{v}$. We will call $T$ the time coordinate, $A$ arrows, and $F$ forks. Observe that from the second requirement, it follows that forks can connect only vertices that have the same time value (set by $T$).
  \end{definition}
  \begin{remark}
    The Toom's contour graph defined in \Cref{def:contour} is, by definition of its construction, a space-time graph.
\end{remark}
The history of a vertex consists of all the vertices that are connected to it by arrows. Here we allow passing through arrows in both directions, so it might be possible that the history of vertex $u$ contains vertices with the same time coordinate as $u$. That motivates to define \textit{slices}. A \textit{slice} is a (maximal) subset of vertices with the same time coordinate such they share the same history. In \cite{Berman}, slices take the role of error droplets.
  \begin{definition}[History, Slice]
    \label{def:slice}
For a time graph $(G,T)$ and a vertex $u \in V$, let the \textit{history} of $u$, denoted by $H_{u}$, be the restriction of $G$ to vertices that both precede $u$ in time (namely, vertices whose time $\le T(u)$) and connect to $u$ via a path of arrows. A \textit{slice} is a maximal subset of vertices $K \subset V$ such that for any two vertices $u,v \in K$, $\History{u} = \History{v}$ is satisfied. As a consequence, the time coordinate in $K$ has to be constant. Thus, we can extend the time coordinate and the history to slices and write for a slice $K$, $\Time{K}$ and $\History{K}$.
  \end{definition}
The following claim ensures us an 'independent' argument about slices. Here we state that two slices with the same time coordinate have to hold disjoint histories. Another way to think about this is that if the first and the second slices are 'induced by $n_{1}$ and $n_{2}$ faults,' then the slice obtained by merging them 'is induced by $n_{1}+n_{2}$ faults.'
  \begin{claim} 
    \label{claim:disj}
    Let $K_{1}, K_{2}$ be two slices such that $\Time{K_1} = \Time{K_2}$, then either $K_{1}=K_{2}$ or $\History{K_1}, \History{K_2}$ are disjoint.
  \end{claim}
  \begin{proof}
Assume a non-empty intersection, so there exists $z \in \History{K_{1}}\cap \History{K_{2}}$ such that for any $x \in K_{1}$ and $y \in K_{2}$, there exist arrow-paths $z \rightarrow x$ and $z \rightarrow y$. Thus, for any $w \in \History{K_{1}}$, one can concatenate a path $w \rightarrow x \rightarrow z \rightarrow y$ and conclude that $w \in \History{K_{2}}$.
  \end{proof}

Next, we would like to study how slices look a moment before they merge. For this, we define the \textit{graph of a slice}. We construct this graph by taking a slice and considering all the slices that precede it by a single time unit. We take those slices as vertices and connect any two of them that have vertices sharing a fork.
  \begin{definition}[The Graph of Slice]
    \label{def:gslice}
    Let $K$ be a slice. The graph of $K$, denoted by $G_{K} = (V_{K}, E_{K})$, is the graph whose vertices are slices $R \subset \History{K}$ with $\Time{R} = \Time{K} - 1$. Two vertices are connected by an edge if and only if there is a fork whose ends belong to the slices associated with them. Namely, $S, R \in V_{K}$ are connected in $G_{K}$ if and only if there is $\{s, r\} = f \in F$ such that $s \in S$ and $r \in R$.
  \end{definition}

  The graph of slices is connected. That fact means that a spanning tree of a (subgraph of) a slice's history can be found by combining the spanning trees of its predecessors. The following gives the proof:
  \begin{claim}
    \label{claim:connected}
    $G_{K}$ is connected. 
  \end{claim}

  \begin{proof}
    Let $R, S \in V_{K}$ and consider $r,s \in V$ such that $r \in R, s \in S$. Recall that $R,S \subset \History{K}$ and therefore $r,s \in \History{K}$. Thus, by definition of the History there exists a path, passing through arrows only, from $r$ to $s$ in $\History{K}$.
    Consider a path that is minimal in the number of vertices with a time coordinate equal to $\Time{K}$; denote that number by $k$. We prove by induction on $k$ that $R$ and $S$ are connected in $G_{K}$.

      \begin{enumerate}
      \item Base. $k=0$, In this case, any vertex on the path between $r$ and $s$ has an equal or lower time coordinate than $r$ and is also connected to $r$ via a path of arrows, namely belongs to $H_{r}$. In particular, $s$ is also a vertex on the path, and therefore $s \in \History{r}$ $\Rightarrow \History{r} = \History{s}$, so $R = S$.
      \item Induction step. Since $k > 0 $, there exists a vertex $y$ on the path from $r$ to $s$ such that $\Time{y} = \Time{K}$. Let $x$ be the vertex which precedes $y$ on the path, and $z$ be the vertex that follows $y$. Since the values of $T$ at the two ends of an arrow differ by $1$, and since $y$ gets the maximal $T$ value in $\History{K}$ it follows that: 
        \begin{enumerate}
          \item $\{x,z\} \subset \pre{y} \Rightarrow \{x,z\} \in F$. 
          \item The slices of $x$ and $z$, say $K_{x}$ and $K_{z}$ have time value $\Time{K_{x}} = \Time{K_{z}} = \Time{K} -1$, Namely they are both members of $V_{K}$. 
        \end{enumerate}
        Since the path connecting $r,x$ in $\History{K}$ has $k-1$ vertices with $\Time{\cdot} = \Time{K}$, then by the induction hypothesis, $R$ and $K_{x}$ are connected in $G_{K}$, and by similar argument so are $K_{z}$ and $S$. On the other hand, $ \{x,z\} \Rightarrow \{K_{x},K_{z}\} \in E_{K}$. Thus, $R$ and $S$ are connected in $G_{K}$.
    \end{enumerate}
  \end{proof}
  Now, we would like to plug in the fact that the vertices originate from the infinite plane. To do so, we equip slices with poles. The poles are the vertices in the slice, which are the maximal points of potential walls.
  \begin{definition}[Poles Selection]
    Let $(G,T)$ be a space--time graph, whose vertices are associated with points in $\mathbb{R}^{d}$. Let $A$ be either a slice, arrow, or a fork in $(G,T)$. We define a poles selection for $A$ to be a selection of $d+1$ vertices, $v_{1},..,v_{d+1}$ , not necessarily distinct, from $A$. The size of $A$ is defined to be:
    \begin{equation*}
      \begin{split}
        \Size{A} \colonequals  \sum_{i=1}^{d+1}  \max_{v\in A} L_{i}(v) 
      \end{split}
    \end{equation*} 
    Thus, for any selection of poles $v_{1},..,v_{d+1}$ we have that: 
    \begin{equation*}
      \begin{split}
        \Size{A} \ge \sum_{i=1}^{d+1} L_i(v_i) 
      \end{split}
    \end{equation*} 
    For a vertex $u$, we define the $i$th predecessor of $u$, denoted by $\prei{u}$, to be its predecessor which achieves the maximal value of $L_{i}$,  namely:
    \begin{equation*}
      \begin{split}
        \prei{u} \colonequals  \arg \max_{v\in \pre{u}} L_{i}(v) 
      \end{split}
    \end{equation*}
  \end{definition}
    
 \paragraph{}  The following claim relates the size of a given slice to the sizes of its predecessor slices.
  \begin{claim}
    \label{claim:prevslices}
Consider a time graph $(G,T)$ whose vertices are associated with points in $\mathbb{R}^{d}$. Let $K$ be a slice in $G$, such that $K \neq \History{K}$. Let $v_{1}, \ldots, v_{d+1}$ be a pole selection for $K$. Then for some integer $k$, there exist $k+1$ slices $K_{0}, K_{1}, \ldots, K_{k}$ and $k$ forks $F_{1}, \ldots, F_{k}$ with pole selections for them that satisfy:
    \begin{enumerate}
      \item For any $i \in [d+1]$,  the $i$th predecessor $\prei{v_{i}} $ belongs to the selected poles of $K_{0}, \ldots, K_{k}$.
      \item Consider the graph whose vertices are associated with $K_{0}, \ldots, K_{k}$, and there is an edge between $K_{i}$ and $K_{j}$ if and only if there is a fork in $F_{1}, \ldots, F_{k}$ that intersects both $K_{i}$'s and $K_{j}$'s selected poles. Then the graph formed from $K_{0}, \ldots, K_{k}$ is connected.
      \item  For any $j\in [0,k]$, denote by $v_{j}^{1},..,v_{j}^{d+1}$ the selected poles for $K_{j}$, then:  
        \begin{equation*}
          \begin{split}
            \sum_{j=0}^{k}\sum_{i=1}^{d+1}L_{i}(v_{j}^{i}) +  \sum^{k}_{i=1}\Size{F_{i}} \ge \sum_{i=1}^{d+1}L_{i}\left( \prei{v_{i}} \right) 
          \end{split}
        \end{equation*}
        
    \end{enumerate}
  \end{claim}
  \begin{proof}
First, notice that $K$ doesn't contain leaves in $(V,A)$, since a leaf is connected only to itself via arrows, and therefore its 'slice' equals its 'History'. Thus, $\prei{v_{i}}$ is not empty for any $i \in [d+1]$. Next, consider the graph $G_{K} = (V_{K},E_{K})$ from \Cref{def:gslice}. Since $\{ v_{i}, \prei{v_{i}} \}$ is an arrow, $\prei{v_{i}}$ belongs to a slice in $V_{K}$. For $i=1,\ldots,d+1$, denote by $R_{i}$ the slice in $V_{K}$ that contains $\prei{v_i}$.

    By \Cref{claim:connected} $G_{K}$ is connected and therefore there exist a minimal collection $K_{0},..,K_{k}$ of slices from $V_{K}$ which contains $R_{1},R_{2},..R_{d+1}$, and which is connected in $G_{K}$. We pick a set of forks $\mathbf{F} = \{F_{1},..,F_{k} \}$ which realizes a spanning tree for this collection of slices. %

    Let us denote by $J_{0},..,J_{2k}$ the enumerated union $K_{0},..K_{k},F_{1},..,F_{k}$. Observe that for any choice of $(d+1)(2k+1)$ elements $z_{0}^{1},z_{0}^{2}, z_{0}^{3},..,z_{0}^{d+1}..,z_{2k}^{d+1}$ in $\mathbb{R}^{d}$, such that $z_{j}^{i} \in J_{j}$ it holds that:
    \begin{equation}
      \label{equ:spanlow}
      \begin{split}
         \sum_{j=0}^{k}\sum_{i=1}^{d+1}L_{i}(z_{j}^{i}) + \sum^{k}_{i=1}\Size{F_{i}}  \ge \sum_{j,i}L_{i}(z_{j}^{i})
      \end{split}
    \end{equation}

Later, we are going to take the points $\{z_{j}^{i}\}$ from the intersection of $J_{j}$'s. For convenience, let us assume that for the index subset $X \subset [2k]\cup\{0\}$, the intersection $\bigcap_{j \in X}J_{j}$ returns a single point. If there is more than one, then decide on the point in a way that all the chosen points simultaneously satisfy that for any index subsets $Y, X$, such that $X \cap Y \neq \emptyset$ and they both introduce a non-trivial intersection $\bigcap_{j \in Z} J_{j} : Z\in \{X,Y\}$, then:
    \begin{equation*}
      \begin{split}
        \bigcap_{j \in X} J_{j} = \bigcap_{j \in Y } J_{j} 
      \end{split}
    \end{equation*}

    Now consider a non-trivial maximal intersection $\bigcap_{j \in X} J_{j}$. The intersection is maximal in the sense that for any $j^{\prime} \notin X$, it holds that $\bigcap_{j \in X\cup \{j^{\prime} \}} J_{j} = \emptyset$. We claim that for any $i$ and for any such $X$, there is $t \in X$ such that $J_{t}$ is the successor of $J_{j}$ on the path to $R_{i}$ for any $j \in X/\{t\}$. Assume not. Let $x, y \in X$, and denote the paths from $J_{x}, J_{y}$ to the $R_{i}$ by $J_{x}, J_{x_2}, \ldots, R_{i}$, and $J_{y}, J_{y_2}, \ldots, R_{i}$, such that $J_{y} \neq J_{x_{2}}$. If $y = x_{2}$, $y_{2} = x$, or $x_{2} = y_{2} \in X$, then pick other $x, y$ (if there is no such, we get an immediate contradiction). In that case, $J_{x}, J_{y}, J_{y_{2}}, \ldots, R_{i}$ is a path from $J_{x}$ to $R_{i}$ which is different from $J_{x}, J_{x_{2}}, \ldots, R_{i}$. Therefore, there are at least two paths from $J_{x}$ to the $R_{i}$, namely, there is a cycle in contradiction to the fact that the collection $K_{0}, \ldots, K_{k}, F_{1},..,F_{k}$ is a tree.

    We will choose the $z_{j}^{i}$ by the following process. Pick a $z_{j}^{i}$ that has not been set yet. If $J_{j} = R_{i}$, then set $z_{j}^{i} \leftarrow \prei{v_{i}}$ - we call this a type-1 assignment. Otherwise, consider the path from $J_{j}$ to $R_{i}$ (by connectivity, it has to be such a one, and by minimality, there is exactly one). Let $J_{t}$ be the successor on the path, then pick $z_{j}^{i} \in J_{j} \cap J_{t}$ - similarly, we call this assignment a type-2 assignment. Since any type-2 assignment is associated with a non-trivial intersection, and any non-trivial intersection induces a type-2 assignment for all $i$'s (since for all $i$'s, one of the $J_{j}$ on its support is a successor of the rest on their path to the $R_{i}$), we have that:
    \begin{equation}
      \label{equ:span}
      \begin{split}
        \sum_{j,i}L_{i}(z_{j}^{i}) &= \sum_{j,i, z_{j}^{i} \in \text{type-1}}L_{i}(z_{j}^{i}) + \sum_{j,i, z_{j}^{i} \in \text{type-2}}L_{i}(z_{j}^{i}) \\
        &= \sum_{i=1}^{d+1}L_{i}\left( \prei{v_{i}} \right) + \sum_{X \colon \bigcap_{j\in X}J_{j} \neq \emptyset }\left(|X|-1\right) \sum_{i=1}^{d+1} L_{i}\left(\bigcap_{j\in X}J_{j} \right) \\ 
        &=\sum_{i=1}^{d+1}L_{i}\left( \prei{v_{i}} \right) 
      \end{split}
    \end{equation}
To complete, we select the $i$th pole of $K_{j}$ to be $z^{i}_{j}$. Then, for $i=1,\ldots,d+1$, the $i$th predecessor $\prei{v_{i}}$ is the $i$th pole of $R_{i}$. This assures (1). Since for any intersection $K_{i}\cap F_{j}$ a type-2 assignment is picked from the intersection to $K_{i}$, we have that the type-2 selected poles of $K_{0},K_{1},\ldots,K_{k}$ are covered by the vertices of $F_{1},F_{2},\ldots,F_{k}$. This assures (2). Plugging \Cref{equ:span} into \Cref{equ:spanlow} assures (3).
  \end{proof}

Until now, the claims stated were true for any kind of space-time graph (where their vertices can be associated with coordinates in $\mathbb{R}^{d}$). The next claims assume that the given time graph is a Toom's contour induced by the noisy running of a Toric encoded circuit.

\begin{claim}
  \label{claim:maxfork}
  For Toom's contours, the maximal size of a fork is at most $2d$.
\end{claim}
\begin{proof}
Consider the rules of forks connecting in \Cref{def:contour}. For forks obtained by rules (a), (b), or as a result of a fault, it's clear that their vertices at their ends are on intersecting faces (when projecting the vertices on $\mathcal{G}$). The maximal fork, in those cases, has the form of:
  \begin{equation*}
    \begin{split}
   F =  \bigl\{ v , \bigl( \prod_{g \in S^{\prime}_{1}} g\bigr) \bigl(  \prod_{g \in S^{\prime}_{2}} g \bigr) v    \bigr\}
    \end{split}
  \end{equation*}
Where $S^{\prime}_{1}$ and $S^{\prime}_{2}$ are subsets of $S$ of size either $d^{\prime}$ or $d - d^{\prime}$, and therefore the size of such forks is at most $\Size{F} = 2 \max \{ d^{\prime}, d - d^{\prime} \} \le 2d$.
\end{proof}

While in the construction presented in \cite{Berman}, all the layers were the same (consisting of the same gates up to wiring), here different layers might differ both in realization and role. Thus, we would like to distinguish between them. We will define the \textit{predecessor decomposition} of a slice to be the collection of slices and forks which satisfy the condition in \Cref{claim:prevslices}. The order of a slice is a number in $\{0,1,2,3\}$ that matches the time modulo four of the layer that results in the slice. Therefore, it encodes the layer type.

  \begin{definition}
    Let $K$ be a slice, and let $v_{1}, \ldots, v_{d+1}$ be a selection of poles for it. Then, let $K_{0}, \ldots, K_{k}$ be slices in $H_{K}$, and let $F_{1}, \ldots, F_{k}$ be forks, such that $K_{0}, \ldots, K_{k},$ $F_{1}, \ldots, F_{k}$ satisfy the conditions in \Cref{claim:prevslices} with respect to the poles selection. We call $K_{0}, \ldots, K_{k},$ $F_{1}, \ldots, F_{k}$ the \textit{predecessor decomposition} of $K$ \textit{with respect to the pole selection} $v_{1}, \ldots, v_{d+1}$. When the pole selection is either clear from the context or can be chosen arbitrarily, we will omit it.
  
    Furthermore, we say that $K$ is a result of a \textit{sweep-cycle}, \textit{faults}, or a \textit{logical operation} if the layer at time $T(K) - 1$ consists of sweep-cycle gates, fault gates (whose source is $\mathcal{E}$), or either measurement or logical Pauli. The \textit{order} of the slice is its time plus $3$ modulo $4$, namely $\text{order}(K) = T(K) + 3 \mod 4$. So the orders of slices that result from a sweep-cycle, faults, logical operation, and again faults are $0, 1, 2, 3$ respectively.
  \end{definition}
The distinction between slices resulting from different layers allows us to adjust \Cref{claim:prevslices} by conditioning on the slice type. Yet, before doing so, we will have to extend the infinity characterization to slices.

  \begin{corollary}
    \label{claim:shrinkslice}
    Consider a slice $K$ in $\mathcal{H}$, and a selection of poles $v_{1}, \ldots, v_{d+1}$, such that $K \neq H_{K}$. Since $K \neq H_{K}$, it follows that $K$ has a predecessor decomposition with respect to $v_{1}, \ldots, v_{d+1}$, denoted by $K_{0}, \ldots, K_{k}, F_{1}, \ldots, F_{k}$. For each slice $K_{j}$, denote its selected poles by $z_{j}^{1}, \ldots, z_{j}^{d+1}$. The inequality in \Cref{claim:prevslices} becomes:
    \begin{enumerate}
      \item If $K$ is a result of a sweep-cycle, then:
        \begin{equation*}
          \begin{split}
            \sum_{j=0}^{k}\sum_{i=1}^{d+1} L_{i}(z_{j}^{i}) +  \sum^{k}_{i=1}\Size{F_{i}} \ge \sum_{i=1}^{d+1}L_{i}(v_i) + 1
          \end{split}
        \end{equation*}
          \item If $K$ is a result of either faults or a logical operation, then: 
        \begin{equation*}
          \begin{split}
            \sum_{j=0}^{k}\sum_{i=1}^{d+1} L_{i}(z_{j}^{i}) +  \sum^{k}_{i=1}\Size{F_{i}} \ge \sum_{i=1}^{d+1}L_{i}(v_i)   
          \end{split}
        \end{equation*}
      \end{enumerate}
  \end{corollary}
  \begin{proof}
 Let us split upon what $K$ results from:
    \begin{enumerate}
      \item For the sweep-cycle part, we use \Cref{claim:polesmove} and \Cref{claim:sweepface}. The first states that within an application of the sweep-cycle over a chain, the maximal $L_{d+1}$ value of the $\partial^{-}$-boundary of the resulting chain decreases by at least one, while the values of each $L_{g}$ do not increase. The second assures that for any local maximum of $L_{i}$ in the resulting $\partial^{-}$-boundary, there is a vertex in the $\partial^{-}$-boundary of the original such that they (1) share a $d^{\prime}$-dimensional face and (2) the $L_{i}$-value of the vertex in the result is at most the $L_{i}$-value of the vertex in the original, where if $i=d+1$, the relation becomes strictly less. Thus, prior to forming the spanning tree of the predecessor slices, for every $v_{i}$ in $v_{1},..,v_{d+1}$ we select a vertex $\prei{v_{i}}$ that has either at least the same $L_{i}$-value as $v_{i}$ if $i\neq d+1$ or an $L_{d+1}$-value that is bigger by at least $1$. That gives the inequality.
        
      \item If $K$ is a result of faults, then it cannot contain a vertex that supports a fault. That is because no arrow connection is made in that case, and hence, if it were to contain a vertex that supports a fault, that vertex would be the only one in $K$, namely $K = H_{K}$. So, we can imagine that all the vertices in $K$ support the identity operator and handle that case as a result of a logical operator.
      \item Where $K$ is the result of a logical operator, we use the observation that a pure logical operator doesn't change the slices (with the exception that the measurement in the phase-basis switches the vertex types, namely $\{X\} \leftrightarrow \{Z\}$, of part of the vertices in the slice, see \Cref{claim:paulipropcon}). So we have that the support of $K$ is contained (up to vertices' types) in the support of $\cup_{j=0}^{k} K_{j}$, and therefore for any $i \in [d+1]$:
        \begin{equation*}
          \begin{split}
            \max \bigl\{ L_{i}(v) : v \in K \bigr\} \le \max \bigl\{ L_{i}(v) : v \in \bigcup_{j=0}^{k}K_{j} \bigr\} 
          \end{split}
        \end{equation*}
    \end{enumerate}
  \end{proof}

  The following claim extends the corollary to slices at the final layer of an \textit{analysis Toom's contour up to step $\tau$}. 
  \begin{claim}
    \label{claim:shrinkan}
For a step $\tau$  let $\mathcal{H}_{\tau}$ be Toom's contour analysis up to step $\tau$. Denote by $t$ the time at which the gate at step $\tau$ is applied. Then, for a slice $K$ and selection for its poles $v_{1}, \ldots, v_{d+1}$ such that $T(K) = t+1$ and $K \neq H_{K}$, the inequality in \Cref{claim:prevslices} becomes:
        \begin{equation*}
          \begin{split}
            \sum_{j=0}^{k}\sum_{i=1}^{d+1} L_{i}(z_{j}^{i}) +  \sum^{k}_{i=1}\Size{F_{i}} \ge \sum_{i=1}^{d+1}L_{i}(v_i) - 2d 
          \end{split}
        \end{equation*}
  \end{claim}
  \begin{proof}
    We allow vertices of $V_{\mathcal{H},X}$ at the final time layer to be connected by an arrow if their counterpart on the skeleton shares a $d^{\prime}-1$ dimensional face. Similarly, two vertices of $V_{\mathcal{H}, Z}$, such that one of them is at the top layer, are connected if they share a $d-d^{\prime}-1$ dimensional face. Therefore, for any $i \in [d]$, the value of $L_{i}$ for $v_{i}$ might be greater by at most 1 than the value of $\prei{v_{i}}$, and the $L_{d+1}$ for $v_{d+1}$ might be greater by at most $\max\{ d^{\prime} - 1, d - d^{\prime} - 1 \}$ than $L_{d+1}(\predd{v_{d+1}})$. Hence, overall, we get that:
    \begin{equation*}
      \begin{split}
        \sum_{i=1}^{d+1}L_{i}\left( \prei{v_{i}} \right) & \ge \sum_{i=1}^{d}\bigl( L_{i}\left( v_{i} \right) - 1 \bigr) + L_{d+1}(v_{d+1}) - \max\{ d^{\prime} - 1, d - d^{\prime}-1\}  \\ 
        & \ge  \Size{K} - 2d
      \end{split}
    \end{equation*}
  \end{proof}

\subsection{Minimal Explanation Argument}
\label{subsection:explantion}

We now bound the number of faults needed to produce a slice in terms of the
slice's geometric size and the number of edges in a subgraph explaining its
history. We construct such a subgraph by joining explanations of predecessor
slices. The key input is that a sweep cycle decreases the relevant syndrome
span by at least one, whereas the other layers do not increase it,
as established by the shrinking argument. This fixed one-unit decrease
is incorporated explicitly below.

We emphasize that the minimal explanation concept was presented in \cite{Berman}, and here we adopt (and extend) the definition of explanation and its existence proof to align with our Toom's contours.
\begin{definition}
  \label{def:explantation}
  Consider a Toom's contour induced by a Toric-encoded circuit. Set
  \begin{equation*}
    \alpha \colonequals 40(d+1)^2, \qquad
    \beta \colonequals 5(d+1), \qquad
    \mathcal{X}_r \colonequals 1-\frac{r(d+1)}{\alpha}
    =1-\frac{r}{40(d+1)} \quad (r=0,1,2,3).
  \end{equation*}
  We interpret the index cyclically, so that
  $\mathcal{X}_{-1}\colonequals\mathcal{X}_3$.

  Fix a slice $K$ of order $r=\operatorname{order}(K)$ and a pole selection
  $v_1,\ldots,v_{d+1}$. Let $W$ be a set of vertices, $A$ a set of arrows,
  and $F$ a set of forks. Write
  \[
    V \colonequals W\cup\{v_1,\ldots,v_{d+1}\}
      \cup\bigcup_{e\in A\cup F}e.
  \]
  We say that $(W,A,F)$ is an \emph{explanation} of $K$ with respect to this
  pole selection if:
  \begin{enumerate}
    \item $V\subseteq\History{K}$ and $(V,A\cup F)$ is connected;
    \item every vertex of $W$ is a leaf of the Toom's contour;
    \item writing $m=|W|$ and
      $s=\sum_{i=1}^{d+1}L_i(v_i)$, we have
      \begin{equation*}
        |A|\leq\alpha(m-\mathcal{X}_r)-\beta s,
        \qquad |F|\leq m-1.
      \end{equation*}
  \end{enumerate}
  The vertices of an explanation may include internal vertices of its
  arrow--fork paths; they need not be either leaves or selected poles.

  For a slice $K$ in the analysis of a Toom's contour up to a step $\tau$,
  we use the same conditions, except that we put $s=\Size{K}$ and replace
  the bound on arrows by
  \begin{equation*}
    |A|\leq\alpha(m+1)-\beta s.
  \end{equation*}
  Whenever useful, a connected explanation may be reduced to a spanning
  tree without increasing either edge count.
\end{definition}

\begin{claim}
  \label{claim:edgesin}
  Let $P_1,\ldots,P_{\mu+1}$ be the Pauli path inducing a Toom's contour.
  Every slice $K$ has an explanation for every selection of its poles.
  In the analysis of the contour up to a step $\tau$, each slice
  has an explanation with poles attaining its size. For a slice in the
  final partial layer, the weaker explanation bound holds for any pole
  selection.
\end{claim}

\begin{proof}
  First consider an ordinary Toom's contour. We induct on
  \[
    h=\Time{K}-\min_{u\in\History{K}}\Time{u}.
  \]
  \begin{enumerate}
    \item \emph{Base case: $h=0$.}
    The history of $K$ contains no arrow. Thus $K=\History{K}=\{u\}$
    for a leaf $u$ created at a fault layer, and
    $r=\operatorname{order}(K)\in\{1,3\}$.
    Take $W=\{u\}$ and $A=F=\varnothing$. Then $m=1$, $s=0$, and
    \[
      0=|A|\leq\alpha(1-\mathcal{X}_r),
      \qquad 0=|F|=m-1.
    \]

    \item \emph{Induction step: $h>0$.}
    Fix poles $v_1,\ldots,v_{d+1}$ of $K$, let
    $r=\operatorname{order}(K)$, and put
    $s=\sum_{i=1}^{d+1}L_i(v_i)$.
    By \Cref{claim:prevslices}, choose predecessor slices
    $K_0,\ldots,K_k$, with respective poles
    $z_j^1,\ldots,z_j^{d+1}$, and connecting forks
    $F_1,\ldots,F_k$. Write
    \[
      s_j\colonequals\sum_{i=1}^{d+1}L_i(z_j^i).
    \]
    The bounds for predecessor slices, together with the fixed one-unit
    improvement in a sweep cycle, imply
    \begin{equation}
      \label{eq:minimal-explanation-predecessor-spans}
      \sum_{j=0}^k s_j+\sum_{j=1}^k\Size{F_j}
      \geq
      \begin{cases}
        s+1,&r=0,\\
        s,&r=1,2,3.
      \end{cases}
    \end{equation}
    By \Cref{claim:maxfork}, each fork has size at most $2d$;
    hence
    \begin{equation}
      \label{eq:minimal-explanation-span-sum}
      \sum_{j=0}^k s_j\geq
      \begin{cases}
        s+1-2dk,&r=0,\\
        s-2dk,&r=1,2,3.
      \end{cases}
    \end{equation}

    Each $K_j$ has smaller induction parameter. Let
    $(W_j,A_j,\mathcal{F}_j)$ be an explanation of $K_j$ with the
    selected poles, and set $m_j=|W_j|$. Its edge counts satisfy
    \[
      |A_j|\leq\alpha(m_j-\mathcal{X}_{r-1})-\beta s_j,
      \qquad |\mathcal{F}_j|\leq m_j-1.
    \]
    Form the parent explanation by taking
    \begin{align*}
      W&\colonequals\bigcup_{j=0}^k W_j,\\
      F&\colonequals\{F_1,\ldots,F_k\}
                         \cup\bigcup_{j=0}^k\mathcal{F}_j,\\
      A&\colonequals\bigl\{\{v_i,\prei{v_i}\}:1\leq i\leq d+1\bigr\}
                         \cup\bigcup_{j=0}^k A_j.
    \end{align*}
    By \Cref{claim:disj}, the predecessor slices have disjoint
    histories, and $W_j\subseteq\History{K_j}$. Thus the sets $W_j$
    are disjoint and
    \[
      m=|W|=\sum_{j=0}^k m_j.
    \]
    The connecting forks and the predecessor arrows join the selected
    poles of $K$ to the connected predecessor explanations, by
    \Cref{claim:prevslices}. All their endpoints lie in $\History{K}$.
    Consequently, the resulting graph is connected and its marked vertices
    $W$ are leaves. Moreover,
    \[
      |F|\leq k+\sum_{j=0}^k(m_j-1)=m-1.
    \]

    For the arrows, \Cref{eq:minimal-explanation-span-sum} yields
    \begin{align*}
      |A|
      &\leq d+1+\sum_{j=0}^k
             \bigl[\alpha(m_j-\mathcal{X}_{r-1})-\beta s_j\bigr]\\
      &\leq \alpha(m-\mathcal{X}_r)-\beta s
          +k(2d\beta-\alpha\mathcal{X}_{r-1})
          +\begin{cases}
              4(d+1)-\beta,&r=0,\\
              0,&r=1,2,3.
            \end{cases}
    \end{align*}
    Here we used
    $\alpha(\mathcal{X}_r-\mathcal{X}_{r-1})=-(d+1)$
    for $r=1,2,3$, and
    $\alpha(\mathcal{X}_0-\mathcal{X}_3)=3(d+1)$.
    The constant term for $r=0$ is nonpositive because
    $\beta=5(d+1)$. Further,
    \[
      \alpha\mathcal{X}_{r-1}
      \geq\alpha\mathcal{X}_3
      =40(d+1)^2-3(d+1)>2d\beta.
    \]
    Both remaining terms are nonpositive, proving
    $|A|\leq\alpha(m-\mathcal{X}_r)-\beta s$.
  \end{enumerate}

  It remains to consider an analysis up to step $\tau$. For slices
  preceding the final layer, apply the ordinary-contour argument with
  poles maximizing each $L_i$; then $s=\Size{K}$, and the ordinary
  bound implies the weaker bound for the analysis. Consider now a slice
  $K$ in the final partial layer.
  If $K=\History{K}$, the single-leaf construction applies. Otherwise,
  repeat the predecessor construction, now using
  \Cref{claim:shrinkan}, which gives
  \[
    \sum_{j=0}^k s_j+\sum_{j=1}^k\Size{F_j}
       \geq s-2d,
    \qquad s=\Size{K}.
  \]
  In particular, $\sum_{j=0}^k s_j\geq s-2d-2dk$.
  The same construction of $W,A,F$ preserves connectedness,
  the disjoint leaf count, and $|F|\leq m-1$. For arrows we obtain
  \begin{align*}
    |A|
    &\leq d+1+\alpha m-\alpha(k+1)\mathcal{X}_{r-1}
                         -\beta(s-2d-2dk)\\
    &=\alpha(m+1)-\beta s
      +\bigl[d+1-\alpha(1+\mathcal{X}_{r-1})+2d\beta\bigr]
      +k(2d\beta-\alpha\mathcal{X}_{r-1})\\
    &\leq\alpha(m+1)-\beta s.
  \end{align*}
  Indeed, $2d\beta-\alpha\mathcal{X}_{r-1}<0$, as already shown,
  and $\alpha\geq d+1+2d\beta$.
\end{proof}

Recall the assumption that each fault cycle is followed by stabilizer measurements (\Cref{def:localstoc}), and thus each fault can be considered a single qubit Pauli. We finish by relating the marked leaves to fault locations.

\begin{corollary}
  \label{coro:edgefaultratio}
  Let $K$ be a slice of size $s$, and let $(W,A,F)$ be an explanation
  (ordinary or at the final layer of a partial analysis). For an ordinary
  slice, choose poles attaining $\Size{K}=s$. Then the
  number of fault locations inducing its marked leaves is at least
  \begin{equation*}
    \frac{1}{2^{\max\{d',d-d'\}}}\;\frac12
    \left(|F|+\alpha^{-1}(|A|+\beta s)\right).
  \end{equation*}
  Moreover, whenever $|A|+|F|>0$, its marked-leaf-to-edge ratio
  satisfies
  \begin{equation*}
    \frac{|W|}{|A|+|F|}\geq\frac{1}{2\alpha}
      =\frac{1}{80(d+1)^2}.
  \end{equation*}
\end{corollary}
\begin{proof}
  Put $m=|W|$. In both kinds of explanation,
  $m\geq |F|+1$. The weaker final-layer arrow bound implies
  \[
    m\geq\alpha^{-1}(|A|+\beta s)-1,
  \]
  and it applies a fortiori to ordinary explanations, since
  $\mathcal{X}_r>0$. Averaging the two inequalities gives
  \[
    m\geq\frac12\bigl(|F|+\alpha^{-1}(|A|+\beta s)\bigr).
  \]
  A single fault can be responsible for at most
  $2^{\max\{d',d-d'\}}$ marked leaves in the relevant syndrome
  sector, yielding the first bound.

  Finally,
  $|A|\leq\alpha(m+1)$ and $|F|\leq m-1$ imply
  \[
    |A|+|F|\leq(\alpha+1)m+\alpha-1\leq 2\alpha m
    \qquad(m\geq1,\ \alpha\geq1),
  \]
  which proves the claimed ratio.
\end{proof}

\section{Proof of the Theorem}

We are ready to prove the main theorem (\Cref{theorem:main}). We start by giving a formal definition of a simulation.

  \begin{definition}[Simulation and Logical Errors]
    \label{def:simu}
    Consider a quantum code $\mathcal{C}$ of length $N$ and dimension $k$. A quantum circuit $D \colon L(\mathcal{H}_{2}^{\otimes N}) \rightarrow L(\mathcal{H}_{2}^{\otimes N})$ will be said to be a referee for $\mathcal{C}$ if, when given a codeword of $\mathcal{C}$, it results in the first $k$ output qubits, the logical state of the given codeword. We say that a Pauli operator $P \in \mathcal{P}^{\otimes N}$ is a \textit{logical error} with respect to $D$ if $DP \neq D$, where the inequality is up to stabilizers of $\mathcal{C}$.

    For quantum circuits $C$ and $\tilde{C}$, a noise channel $\mathcal{N}_{\mathcal{E}}$, and $\delta \in (0,1)$, we say that $\tilde{C}$ \textit{simulates} $C$ with probability at least $1 - \delta$ if
    \begin{enumerate}
      \item
        There is a mapping between the qubits of $C$ and the registers of $\tilde{C}$, along with a referee such that, at the end of running $\tilde{C}$ (on the zero state), decoding the registers back to the logical space results in the same quantum states as those produced by running $C$.
      \item Let $\mathcal{E} \sim \mathcal{N}_{\mathcal{E}}$ be a set of faults. The probability induced by $\mathcal{N}_{\mathcal{E}}$ that at the end of the running of $\tilde{C}[\mathcal{E}]$, one of its quantum registers introduces a logical error (with respect to the referee) is less than $\delta$.
    \end{enumerate}
  \end{definition}

The following lemma is the main probabilistic ingredient; it analyzes the probability of a logical error occurring after the initialization stage, namely when assuming the given input is encoded.

  \begin{lemma}
    \label{lemma:keylemma}
    For any $d, d^{\prime}$, such that $d^{\prime}\ge 2, d-d^{\prime} \ge 2$, there are constants $a, b \in \mathbb{R}^{+}$ and $p_{0} \in (0,1)$ that depend only on $d, d^{\prime}$ and satisfy the following. Consider a quantum circuit $C$ with depth $\mu$ and width $w$, and let $\mathcal{N}_{\mathcal{E}}$ be a local stochastic noise channel with a noise rate $p < p_{0}$. There is a Toric encoded circuit $\tilde{C}$ such that each of its quantum registers is encoded by an $n$-side length, $(d, d^{\prime})$-Toric code, and it simulates $C$ with probability:
\begin{equation*}
  \begin{split}
    q = 1 -   O \left( \operatorname{poly}(w,\mu,n) \right) \left( a p^{b} \right)^{n} 
  \end{split}
\end{equation*}
\end{lemma}

To prove the lemma, we will need the following two claims, which are proved in \cite{Berman, gacs2021newversiontoomsproof}. The first bounds, for a given regular graph, the number of its subgraphs having at most $k$ edges. This claim will be used to infer the number of possible explanation trees for a given slice. The second states that each tree has a subtree with fewer than half of the edges and a leaves-to-edge ratio that is at least half of the ratio of the original tree. This claim will be used to infer that any explanation has a subtree where, under the projection to the finite Toric, its faults are mapped uniquely, meaning no fault is counted twice. 

\begin{claim*}[\Cref{claim:counting_trees}]
  Let $G=(V,E)$ be a graph of maximum degree at most $\Delta+1$, and let $k$ be an integer less than $|E|$. Fix a vertex $r\in V$. There is a constant $\sigma$, depending only on $\Delta$, such that the number of \emph{connected} subgraphs containing $r$ and having fewer than $k$ edges is at most $\sigma^k$.
  \end{claim*}

  \begin{claim*}[\Cref{claim:big_subtree}]
  Let $\gamma,\beta\in\mathbb{R}_{>0}$ and let $n$ be a positive
  integer. For any tree $T$ with $\tau\ge\beta n$ edges and $m$ leaves
  such that $\tau\le\gamma m$, there is a subtree $T'$ of $T$ with a
  number of edges $\tau'<\frac{\beta}{2}n$ and a number of leaves
  $m'\ge\frac{\beta}{4\gamma}n$. The counted leaves of $T'$ can be
  chosen among the original leaves of $T$.
  \end{claim*}
  
The proofs of the claims are given in the appendix. We continue with the proof of the lemma.
  \begin{proof}
    We construct $\tilde{C}$ as follows: for each input wire of $C$, we initialize the logical zero $\ket{\mathbf{0}}$ of the $(d,d^{\prime})$-Toric code. For each of the gates $\mathbf{H}$ and $\mathbf{S}$, we initialize the encoded $\ket{\mathbf{H}}$ and $\ket{\mathbf{S}}$ states, and for each $\mathbf{T}$, we initialize copies of both $\ket{\mathbf{T}}$ and $\ket{\mathbf{S}}$ states. The even layers of $\tilde{C}$ are sweep cycles, while on the odd layers, we set the realization of the logical gates. For $\mathbf{CX}$, we set the transversal realization, and for $\mathbf{H}$, $\mathbf{S}$, and $\mathbf{T}$, we use gate teleportation. Thus the width of $\tilde{C}$ expands to at most $ (w + 2w\mu) n^{2d}$.

    We realize the gate teleportation as described in \Cref{subsection:logical_gates}, so the gate teleportation admits $O(n + \log n)$ and $O(n^{2d})$ depth and width overhead. Thus, the depth of $\tilde{C}$ expands to $\Theta(\mu n)$\footnote{The constant inside the $\Theta$ also accounts for the fact that the realization of a logical gate might contain multiple gate teleportations.}. We will denote the procedure of inferring the logical measurement in the gate teleportation\footnote{This is the codeword checking procedure - $ \mathbf{1}_{x_{S^{\prime}}}$ / $ \mathbf{1}_{x_{S/S^{\prime}}}$ as appears in \Cref{subsection:logical_gates} } by $D$, and we also use $D$ as a realization of measurement in the computational basis and as our referee (\Cref{def:simu}). Denote by $S$ the number of locations, so $S = O(w^{2} \mu^{2} n^{2d+1} )$.
  
    \paragraph{} 
    Consider a set of faults $\mathcal{E}$, and consider the representation of the subjected circuit $\tilde{C}[\mathcal{E}]$ as a distribution over circuits conditioned on measurement outcomes:
    \begin{equation*}
      \begin{split}
        \tilde{C}[\mathcal{E}] = \sum_{\alpha} \prb{\alpha | \mathcal{E} }\tilde{C}[\mathcal{E}, \alpha] 
      \end{split}
    \end{equation*}
    Denote by $\tau_{f}$ the last step of $\tilde{C}[\mathcal{E},\alpha]$, and let $(\mathbf{P},\mathbf{a}) $ be a step-propagation of $\tilde{C}[\mathcal{E},\alpha]$, where $\mathbf{P} \colonequals  P_{1},P_{2},..,P_{\tau_{f}+1}$ and $\mathbf{a} \colonequals \alpha^{1}, \alpha^{2},..,\alpha^{\tau_{f}+1}$ (\Cref{def:propagation}). Thus, if $D \circ \tilde{C}[\mathcal{E}] \neq \tilde{C}$, then there is at least one option of outcomes $\alpha$ for which $D \circ \tilde{C}[\mathcal{E},\alpha] \neq \tilde{C}$, namely: 
    \begin{equation*}
      \begin{split}
        D \circ \tilde{C}[\mathcal{E},\alpha] = D\circ P_{\tau_{f}+1} \tilde{C}[,\alpha^{\tau_{f}+1}]  \neq D \circ \tilde{C}
      \end{split}
    \end{equation*} 
    For this, one of the following two has to occur: either the decoder $D$ fails to decode $P_{\tau_{f}+1}$, or the conditioned circuit $\tilde{C}[,\alpha^{\tau_{f}+1}]$ deviates from the ideal circuit, namely $\tilde{C}[,\alpha^{\tau_{f}+1}] \neq \tilde{C}$. The second occurs only if, along the running, a gate-teleportation fails, meaning the decoding procedure that erases errors from the measured register failed, and therefore the gate applied following the measurement was incorrect. 

    \paragraph{} 

By the barrier argument (\Cref{coro:barrier}), occurrence of both options implies that there is a connected component of either the $X$-syndrome or $Z$-syndrome of at least one of the Paulis in the propagation $P_{1},..,P_{\tau_{f}+1}$ at a size greater than $cn$ (where $c$ is the constant $c^{\prime}$ defined in \Cref{claim:barrierconstant}), and hence, the Pauli-path induced by the propagation has a connected component of the syndrome at a size at least $cn$.

    Since connectivity by faces in the analysis of Toom's contour implies connectivity by arrows, it follows that an occurrence of that event implies the existence of a step $\tau$ such that the analysis of Toom's contour $\mathcal{H}_{\tau}$ of $\tilde{C}[\mathcal{E},\alpha]_{\tau}$ has a slice of size at least $cn$. So it's sufficient to prove that with probability at least $q$, for the sampled faults $\mathcal{E}$, there are no possible outcomes $\alpha$ and a step $\tau$ such that the analysis of Toom's contour $\mathcal{H}_{\tau}$ of $\tilde{C}[\mathcal{E},\alpha]_{\tau}$ has a slice of size at least $cn$.
    \paragraph{} 
    
Denote by $B_{\tau}$ the event where $\tau$ is the first step such that $\mathcal{H}_{\tau}$ has a slice at size greater than $cn$, and by $A_{\tau}$ the event where all explanations of slices (and any of their pole selections) in $\mathcal{H}_{\tau}$ have fewer than $\frac{n}{2d}$ edges. Using the conditional probability formula, we have:
    \begin{equation*}
      \begin{split}
        \prbm{  B_{\tau}    }{ \sim \mathcal{N}_{\mathcal{E}} } & =  \prbm{  B_{\tau}  \cap A_{\tau}  }{ \sim \mathcal{N}_{\mathcal{E}} } + \prbm{  B_{\tau} \cap \bar{A}_{\tau}    }{ \sim \mathcal{N}_{\mathcal{E}} }  \\
  & \le \prbm{  B_{\tau}  \cap A_{\tau}  }{ \sim \mathcal{N}_{\mathcal{E}} } + \prbm{ \bar{A}_{\tau}    }{ \sim \mathcal{N}_{\mathcal{E}} }  \\
      \end{split}
    \end{equation*}
    We begin by bounding the contribution of the right term. According to \Cref{coro:edgefaultratio}, there exists a constant $c_{1}$, depending only on $d$, such that the ratio of leaves to edges in the explanation is at least $c_{1}$. By \Cref{claim:big_subtree}, it follows  that any such explanation with at least $\frac{n}{2d}$ edges contains a subtree with fewer than $\frac{n}{4d}$ edges and at least $\frac{1}{2}c_1 \frac{n}{4d}$ leaves. Using the union bound, we bound the probability that there is a vertex in the base graph that is a root of such a subtree. Since the construction of the $\infty$-lift sets periodic conditions, it's sufficient to ask if there is such a vertex in the restriction to the finite $n$-side length Toric.
  \begin{equation*}
    \begin{split}
      \prbm{ \bar{A}_{\tau} }{ \sim \mathcal{N}_{\mathcal{E}} }  \le  \sum_{ \substack{ v \in V_{o}, \tau^{\prime} \le \tau \\ L_{g}(v) \in [0,n]} } \prbm{ & v \text{ is a root of subtree in } \mathcal{H}_{\tau^{\prime}}  \\ 
      &  \text{ with less than  } \frac{n}{4d} \text{ edges and more than } \frac{1}{2}c_{1} \frac{n}{4d} \text{ leaves}  }{ \sim \mathcal{N}_{\mathcal{E}} } 
    \end{split}
  \end{equation*}
    Now, since any two connected vertices are at a space distance of at most 
    \begin{equation*}
      \begin{split}
        \max\{\Size{\text{arrow}}, \Size{\text{fork}}\}  \le 2d
      \end{split}
    \end{equation*} 
    It follows that all the vertices in the subtree are at a space distance of at most $\frac{1}{2} n$ from each other. Thus, no two of them have the same space coordinate modulo $n$. In particular, projecting the infinite plane back to the finite Toric sends each of the leaves of the subtree to a unique target, namely no fault is counted twice. Therefore the number of faults that induce the explanation is at least:
    \begin{equation*}
      \begin{split}
        \frac{1}{2^{\max\{d^{\prime}, d - d^{\prime}\}}} \frac{1}{2} \frac{1}{2}c_{1} \frac{n}{4d} \ge \frac{1}{2^{d+4}} c_{1} \frac{n}{d} 
      \end{split}
    \end{equation*}
Plugging into the union bounds and combining \Cref{claim:counting_trees} yields:
    \begin{equation*}
      \begin{split}
        \prbm{  \bar{A}_{\tau}  }{ \sim \mathcal{N}_{\mathcal{E}} }  \le O(S) \sigma^{  \frac{n}{4d} } p^{ \frac{c_{1}n}{2^{d+4} d } }\le O(S) \left( \sigma p^{\frac{1}{2^{d + 2}} c_{1}} \right)^{\frac{n}{4d}} 
      \end{split}
    \end{equation*}

    It's left to bound the probability of $B_{\tau} \cap A_{\tau}$. By \Cref{claim:edgesin} any slice, and any pole selection of it, in $\mathcal{H}_{\tau}$, have an explanation. Confining to the event $A_{\tau}$, any explanation has fewer than $\frac{n}{2d}$ edges and therefore, by the same argument as above, the projection of each explanation tree into the finite Toric sends its leaves to unique targets.

Any slice of size at least $cn$ has a pole selection $v_{1}..,v_{d+1}$ such that: 
\begin{equation}
  \begin{split}
    \label{equ:bigspan}
  \sum_{i}^{d+1}L_{i}(v_{i}) \ge cn
  \end{split}
\end{equation}
Consider a slice with size at least $cn$ and a pole selection for it satisfying \Cref{equ:bigspan}. \Cref{claim:edgesin} and \Cref{coro:edgefaultratio} state that there are constants $c_{2},c_{3} > 0$ such that any explanation for the slice and the selection, with $x$ edges, has at least $c_{2} cn + c_{3} x$ leaves. Therefore, we obtain:
    \begin{equation*}
      \begin{split}
        \prbm{  B_{\tau}  \cap A_{\tau}  }{ \sim \mathcal{N}_{\mathcal{E}} } & \le \sum_{ \substack{ v \in V_{o}, x \ge 1 \\ L_{g}(v) \in [0,n]} } \prbm{ v \text{ is a root of explanation with } x \text{ edges} \\ 
      & \ \ \ \ \ \ \ \ \ \ \ \ \ \ \ \  \ \  \ \ \text{ and at least } c_{2}cn + c_{3}x \text{ leaves, } \\
  & \ \ \ \ \ \ \ \ \ \ \ \ \ \ \ \  \ \  \ \ \text{ which are projected to different targets  }  }{ \sim \mathcal{N}_{\mathcal{E}} } 
            \end{split}
          \end{equation*}
          So, for sufficiently small $p$ such that $\sigma p^{c_{3}} < 1$, the probability is bounded by:
          \begin{equation*}
            \begin{split}
              \sum_{x = 1  }^{ \infty } O(S)p^{c_{2}cn} \left( \sigma p^{c_{3} } \right)^{x}  \le O(S) p^{c_{2}c n} 
      \end{split}
    \end{equation*}
    Combining the bounds, we have that there is a constant $c_{4}$ depending only on  $d,d^{\prime}$ such that:
\begin{equation*}
  \begin{split}
    \prbm{  B_{\tau} }{ \sim \mathcal{N}_{\mathcal{E}} } \le  O(S) p^{c_{4} n} 
  \end{split}
\end{equation*}
Now, by the union bound over the time steps, with a probability greater than:
\begin{equation*}
  \begin{split}
  1 -   O \left( S^{2} \right) p^{ c_{4}n } 
  \end{split}
\end{equation*}
No step admits a slice of size at least $cn$ in its corresponding analysis of Toom's contour, so the barrier argument holds; in other words, the propagated errors do not accumulate into a logical error.
\end{proof}
The lemma gives the theorem almost immediately. To complete the proof, we need to plug in the initialization phase, which initializes encoded zeros and resource states subjected to local stochastic noise. We emphasize that the resulting states are subjected to local stochastic noise (in addition to the computation). We neither prove nor describe that phase here; instead, we refer to \cite{grospellier:tel-03364419, gottesman2014faulttolerant, Fawzi_2018}, and \cite{nguyen2025quantumfaulttoleranceconstantspace}.
  \begin{theorem}
    \label{theorem:main}
    There exists a constant $p_{0} \in (0,1)$ such that for any local stochastic noise channel $\mathcal{N}_{\mathcal{E}}$ with rate $p < p_{0}$, and a quantum circuit $C$ with depth $\mu$ and width $w$, there exists a Toric encoded circuit $\tilde{C}$, with depth $\Theta(\mu n + \text{ Polylog}(n) )$ and width $\Theta((w + w\mu) n^{8} )$  that simulates $C$ with probability: 
\begin{equation*}
  \begin{split}
    1 -   O \left( \text{poly}(w,\mu,n) \right)  p^{\Omega(n)} 
  \end{split}
\end{equation*}
and in addition: 
\begin{enumerate}
  \item After a short initialization stage (at depth $\Theta( \text{poly}(\log n))$), each of the qubits is encoded in the $n$-length 4D Toric code.
  \item An error correction cycle takes $O(1)$ depth, and the gates in each correction cycle don't depend on time. In particular, if $C$ is the identity circuit, then after the initialization stage, the layers in $\tilde{C}$ are the same.
\end{enumerate}
  \end{theorem}

\section{The Sweep-rule is Itself a Decoder}
The role of the sweep rule in the fault tolerance construction is to restrain the errors in the system rather than erase them. Here, we show that it can also be used as a decoder. Consequently, if the initial error is distributed according to local stochastic noise, then the execution of noiseless sweep-cycles for $n^{d+1}$ times - the number of vertices times the side's length - erases the errors, with high probability.

The following claim states that an ideal run (not subject to noise) of $n^{d+1}$ sweep-cycles, with the additional assumption that no slice has reached a size greater than $n$, succeeds in erasing the syndrome.
\begin{claim}
  \label{claim:decodestopnoise}
  Let $\ket{\psi}$ be a logical codeword of the $(d,d^{\prime})$-Toric code, with side length $n$, satisfying $d^{\prime},d-d^{\prime}\ge 2$, and let $P \in \mathcal{P}^{(\cdot)}$. Let $C$ be the quantum circuit that takes $P\ket{\psi}$ as an input and performs over it $n^{d+1}$ sweep-cycles. For a possible measurement outcome $\alpha \in \mathbb{F}_{2}^{(\cdot)}$ (of stabilizer measurements), consider the Toom's contour $\mathcal{H}$ induced by $C[,\alpha]$. Assume that no slice of $\mathcal{H}$ has size greater than $n$. Then, at the end of the running of $C[,\alpha]$, the syndrome is erased.
\end{claim}
\begin{proof}
  We will prove that the decoder erases the $\partial^{-}$-boundary corresponding to the $X$-component of $P$, and the proof for the $Z$-component is identical. Let $P^{\infty}$ be the $\infty$-lift of $P$ and let $v_{1}, v_{2}, \ldots$ be the vertices supporting the faces of $\partial^{-}x(P^{\infty})$. During the ideal computation, no new faults are introduced, and therefore $v_{1}, v_{2} \ldots$ are the only leaves in the induced Toom's contour. Without loss of generality, assume that the first $m$ vertices $v_{1},..,v_{m}$ align with the vertices of the finite Toric (namely non-translated vertices), denote them by $V^{\prime}$.

  They set a natural relation over its slices as follows: two slices $K_{1}$ and $K_{2}$, not necessarily with the same time coordinate, will be said to be \textit{history-related}, $K_{1}\sim K_{2}$, if there are leaves $u, w$ such that both $u$ and $w$ are translations of a vertex $v_{i} \in V^{\prime}$ and $u, w$ belong to the histories of $K_{1}, K_{2}$ respectively.
  Thus, for a time $t$ and a leaf $v \in V^{\prime}$, we can define the notation $K_{v,t}$ to be the slice at time $t$ that is reached from leaf $v$ by arrows. We emphasize that $\sim$ is not an equivalence relation, since it doesn't satisfy transitivity. Yet, let us match slices to 'classes' when the map is many-to-many. Here, a class is a grouping of translations of slices which are all history-related. For a vertex $v \in V^{\prime}$, the delegate of the slices reached from vertex $v$ and its translations, namely the translations of slices $\{ K_{v,t} : t \}$, will be denoted by $\mathbf{K}_{v}$. Proving that at the end of time $n^{d+1}$, for any leaf $v$, the slice $K_{v,t}$ is empty, or equivalently is of size zero, and therefore its translations, implies the correctness of the claim.

  For proving this, let us present the following terminology. For leaves $v,u,w$, and a time coordinate $t$, consider the following: 
  \begin{enumerate}
    \item Suppose that $K_{v,t}$ and $K_{u,t+1}$ are slices at times $t$ and $t+1$, such that $K_{v,t}\sim K_{u,t+1}$. If $\Size{K_{u,t+1}} > \Size{K_{v,t}}$, we will say that the class $\mathbf{K}_{v}$ grows at time $t$. Similarly, if $\Size{K_{u,t+1}} < \Size{K_{v,t}}$, then we will say that class $\mathbf{K}_{v}$ shrank at time $t$.
    \item Suppose that $K_{v,t}$ and $K_{u,t}$ are slices at time $t$, and $K_{w,t+1}$ is a slice at time $t+1$, such that $K_{v,t}\not\sim K_{u,t}$, $K_{w,t+1}\sim K_{v,t}$, and $K_{w,t+1}\sim K_{u,t}$. Then we will say that $\textbf{K}_{v}$ and $\textbf{K}_{u}$ merged at time $t$.
    \item The size of a class $\mathbf{K}_{v}$ at time $t$ is $\Size{K_{v,t}}$.
  \end{enumerate}
  Since no new faults are introduced during the computation, we have that in each step, the only option for a class to grow is by merging. Notice that, by definition, classes cannot 'split'. Namely, if $K_{v,t}\in \mathbf{K}_{u}$, then for any time $t^{\prime} \ge t$, it holds that $K_{v,t^{\prime}} \in \textbf{K}_{u}$.
Specifically, if a correction by the sweep rule disconnects a syndrome component into two connected components, they belonged to the same slice (and therefore to the same class) prior to the correction, and thus still belong to the same class after the correction.

Therefore, there can be at most $m-1$ merges. And since the number of classes is bounded by the number of vertices on the finite Toric, we have that $m \le n^{d}$, namely at most $n^{d}-1$ merges can occur. Hence, during the $n^{d+1}$ sweep-cycles, there has to be a continuous sequence of $n$ cycles for which no merge occurs.

During the continuous sequence of those $n$ cycles, each sweep cycle decreases the size of any slice by at least 1. This is followed by \Cref{claim:shrinkslice}, combining that no merging occurred, meaning there is only one predecessor slice. Thus, using the given assumption that no slice in $\mathcal{H}$ reaches a size greater than $n$, namely $\Size{K_{v,t}} \le n$ for any vertex $v$ and time $t$, we have that at the end of that sequence of $n$ cycles, all the slices' sizes shrink to zero; namely, the slices are erased after those cycles.
\end{proof}

\Cref{claim:decodestopnoise} guarantees only that the syndrome is erased; in particular, it says nothing about whether a logical error has or hasn't occurred. To complete the correctness of the decoder, we would like to use the barrier argument. Yet, instead of showing that during the ideal correction no connected component of the syndrome passes the barrier, it will be more convenient to reduce the passing question to the subjected-to-noise setting. We introduce a replacement argument, stating that the probability of an event occurring during a computation where part of its layers are non-noisy is lower than the probability of having the same event in a computation where all its layers are noisy. More precisely, any bound that assumes only that the noise channel is local stochastic captures both cases.
\begin{claim}[Replacement argument]
  \label{claim:replace}
  Let $C$, $\ket{\psi}$, and $P$ be the quantum circuit, the quantum state, and a Pauli as defined in \Cref{claim:decodestopnoise}, with the change that $P$ is distributed according to local stochastic noise $\mathcal{N}_{\mathcal{E}}$ with an error rate $p\in (0,1)$ rather than being fixed.

Denote by $A$ the event in which, at the end of the running of $C$, namely after the $n^{d+1}$ \textbf{ideal} (non-noisy) iterations of the sweep-cycle, the error is not erased, and denote by $B$ the event that, during the application of a subjection of $C$ to $\mathcal{N}_{\mathcal{E}}$, there is a step for which the analysis of Toom's contour up to that step admits a slice of size at least $cn$.

Let $f:(n,d,d^{\prime},p) \rightarrow \mathbb{R}$ be a bound for $\prb{B}$ that holds uniformly over local-stochastic fault distributions of rate $p$. Then, we have:
  \begin{equation}
    \label{equ:replacement}
    \begin{split}
      \prb{A} \le f(n,d,d^{\prime},p)
    \end{split}
  \end{equation}
\end{claim}
\begin{proof} 
A corollary of combining \Cref{claim:decodestopnoise} with \Cref{coro:barrier} (The barrier argument) is that if during the $n^{d+1}$ ideal sweep-cycles, there is no step for which its propagation admits a slice of size at least $cn$ (and therefore no step admits a syndrome's connected component that passes the barrier), then the syndrome is erased and no logical error occurred. So it's sufficient to show that the probability that the condition does not hold is lower than $f(n,d,d^{\prime},p)$. To apply the same bound, construct a space--time fault distribution
$\mathcal N'_{\mathcal E}$ as follows. Sample the initial Pauli error from
$\mathcal N_{\mathcal E}$ and place its support in the initial fault layer;
introduce no faults at later layers. For any prescribed collection $S$ of
space--time locations, the probability that all locations in $S$ are faulty
is zero if $S$ contains a location after initialization. Otherwise, it is at
most $p^{|S|}$ by the local-stochastic assumption on the initial error.
Thus $\mathcal N'_{\mathcal E}$ has local-stochastic rate $p$, and the resulting
noisy-circuit trajectory agrees with the ideal sweep evolution following the
sampled initial error. The uniform contour estimate applies to this specialized
fault distribution, yielding \Cref{equ:replacement}.
\end{proof}
We now conclude that the sweep rule is an ideal decoder under local-stochastic Pauli noise.
  \begin{theorem}
    \label{theorem:idealdec}
    There exists a constant $p_{0} \in (0,1)$ such that for any local stochastic noise channel $\mathcal{N}_{\mathcal{E}}$ with rate $p < p_{0}$, and a quantum state $\rho = \mathcal{N}_{\mathcal{E}}(\ket{\psi}\bra{\psi} )$, where $\ket{\psi}$ is a codeword of the $(d,d^{\prime})$-Toric code satisfying $d^{\prime},d-d^{\prime}\ge 2$, the application of $n^{d+1}$ (ideal) sweep-cycles on $\rho$ results in $\ket{\psi}$ with probability at least:
\begin{equation*}
  \begin{split}
    1 -   O \left( \text{poly}(n) \right)  p^{\Omega(n)} 
  \end{split}
\end{equation*}
  \end{theorem} 
  \begin{proof}
    By the replacement argument (\Cref{claim:replace}), it is sufficient to consider the application of $n^{d+1}$ sweep-cycles subjected to $\mathcal{N}_{\mathcal{E}}$, and bound the probability that during the subjected application there is a step $\tau$ and an analysis of Toom's contour $\mathcal{H}_{\tau}$ such that $\mathcal{H}_{\tau}$ contains a slice of size at least $cn$. Denote by $B_{\tau}$ the event in which step $\tau$ admits a slice of size at least $cn$. The proof of \Cref{lemma:keylemma} bounds the probability of $B_{\tau}$ by the number of locations times $p^{\Omega(n)}$ for $p < p_{0}$. Taking the union bound over the steps gives the desired result.
  \end{proof}

\section{Appendix}

\subsection{Counting Subtrees}

\begin{claim}
  \label{claim:counting_trees}
  Let $G=(V,E)$ be a graph of maximum degree at most $(\Delta+1)$, and let $k$ be an
  integer less than $|E|$. Fix a vertex $r\in V$. There is a constant
  $\alpha$, depending only on $\Delta$, such that the number of
  \emph{connected} subgraphs containing $r$ and having fewer than $k$
  edges is at most $\alpha^k$.
\end{claim}

\begin{proof}
  If $k\le 0$, there are no such subgraphs, so the claim is immediate.
  Set $d=\Delta+1$. For each connected subgraph $H$ containing $r$ and
  having exactly $j$ edges, duplicate every edge of $H$. The resulting
  connected multigraph has even degrees, and hence admits an Eulerian
  circuit starting at $r$. This circuit induces a closed walk of length
  $2j$ in $G$ that traverses precisely the edges of $H$ (each twice).

  Fix an ordering of the edges of $G$ and choose, for each $H$, one such
  walk deterministically, for example the lexicographically first one.
  The correspondence is injective: from the selected walk we recover
  $H$ as the subgraph consisting of all edges traversed by the walk,
  together with $r$ in the case $j=0$.

  Since every vertex has degree at most $d$, there are at most $d^{2j}$ walks
  of length $2j$ starting at $r$. Consequently, the required number
  of subgraphs is at most
  \[
    \sum_{j=0}^{k-1}d^{2j}
    \le \sum_{j=0}^{k-1}(d+1)^{2j}
    \le (d+1)^{2k}.
  \]
  Thus one may take $\alpha=(\Delta+2)^2$, which depends only on
  $\Delta$.
\end{proof}

\subsection{Tree Separator Lemma}
\label{subsection:separator}

Throughout this subsection, a \emph{subtree} means a connected subgraph
of a tree. Whenever we extract a subtree $T'$ from a given tree $T$,
we may count the \emph{original leaves} of $T$ that it contains.
This distinction is useful because a vertex that was not a leaf of $T$
may become a leaf after cutting edges. A single original leaf, viewed
as a one-vertex subtree, counts as one original leaf.

\begin{claim}
  \label{claim:tree_separator}
  Let $T=(V,E)$ be a tree with $m>3$ leaves and $\tau$ edges. Then
  there is a subtree with $m'\in[\frac13 m,\frac23 m]$ leaves and at most
  $\frac{m'}{m}\tau$ edges. Moreover, its leaves can be chosen to be
  leaves of the original tree $T$.
\end{claim}

\begin{proof}
  Assign weight $1$ to every leaf of $T$ and weight $0$ to every other
  vertex. We argue that there is a vertex $v$ such that every connected component of
  $T-v$ contains at most $m/2$ leaves. 

  To see this, consider the following algorithm. Starting at any vertex, move
  towards a component containing more than $m/2$ leaves, if such a
  component exists. A move cannot immediately be reversed, since the
  component on the opposite side then contains fewer than $m/2$ leaves.
  Thus the procedure terminates at the desired vertex.

  Let $a_1,\ldots,a_q$ be the numbers of original leaves in the
  components of $T-v$. They are positive integers, sum to $m$, and
  satisfy $a_i\le m/2$. We can divide these components into two
  nonempty groups, each containing between $m/3$ and $2m/3$ original
  leaves. Indeed, if some $a_i\ge m/3$, take that component as the
  first group. Otherwise, add components to the first group until its
  leaf count first reaches $m/3$; because each summand is smaller than
  $m/3$, the resulting count is smaller than $2m/3$.

  Form two connected subgraphs $T_1,T_2$ by adjoining $v$ to the
  components in the respective groups. Their edge sets partition
  $E$, and their sets of original leaves partition the $m$ leaves of
  $T$. Write $\tau_i=|E(T_i)|$ and let $m_i$ count their original
  leaves. Then
  \[
    \tau_1+\tau_2=\tau,\qquad m_1+m_2=m,
    \qquad \frac{m}{3}\le m_i\le\frac{2m}{3}.
  \]
  At least one of the two subgraphs, say $T_i$, satisfies
  \[
    \frac{\tau_i}{m_i}\le\frac{\tau}{m},
  \]
  because $\tau/m$ is the $m_i/m$-weighted average of the two ratios.

  Finally, replace $T_i$ by the minimal connected subtree spanning
  its $m_i$ original leaves. This can only delete edges and cannot
  delete any of those original leaves. Since $m_i\ge m/3>1$, this
  minimal subtree has exactly those $m_i$ vertices as its leaves.
  It therefore satisfies the asserted leaf and edge bounds.
\end{proof}

The following scale-selection version is sometimes useful.
\begin{corollary}
  Let $T=(V,E)$ be a tree with $m$ leaves and $\tau$ edges. For every
  real number $s$ with $1\le s\le m$, there is a subtree containing
  $m'$ original leaves such that
  \[
    \frac{s}{3}<m'\le s,
    \qquad \tau'\le\frac{m'}{m}\tau,
  \]
  where $\tau'$ denotes its number of edges.
\end{corollary}

\begin{proof}
  If $1\le s<3$, take a one-vertex subtree consisting of any
  original leaf; then $m'=1$ and $\tau'=0$.
  Otherwise, repeatedly apply \Cref{claim:tree_separator}, always
  retaining a subtree with at most the same edge-to-leaf ratio, until
  its number of leaves first becomes at most $s$. The construction
  applies as long as the current number of leaves exceeds $s\ge3$.
  If no step is needed, $m'=m=s$ (since $s\le m$). Otherwise, the
  preceding subtree had more than $s$ leaves and the final one has at
  least one third as many, so $m'>s/3$. In either case $m'\le s$,
  and the edge-to-leaf ratio never increases.
\end{proof}

\begin{claim}
  \label{claim:big_subtree}
  Let $\gamma,\beta\in\mathbb{R}_{>0}$ and let $n$ be a positive
  integer. For any tree $T$ with $\tau\ge\beta n$ edges and $m$ leaves
  such that $\tau\le\gamma m$, there is a subtree $T'$ of $T$ with a
  number of edges $\tau'<\frac{\beta}{2}n$ and a number of leaves
  $m'\ge\frac{\beta}{4\gamma}n$. The counted leaves of $T'$ can be
  chosen among the original leaves of $T$.
\end{claim}

\begin{proof}
  Fix a closed depth-first traversal of $T$, starting at an original
  leaf and crossing every edge exactly twice. It is a cyclic walk of
  length $2\tau$ in which each of the $m$ original leaves occurs
  exactly once among the $2\tau$ cyclic vertex positions.

  Put $b=\beta n$ and $\ell=\lceil b/2\rceil-1$. Since $b>0$, we have
  $\ell\ge0$ and $\ell<b/2$. Consider all $2\tau$ cyclic segments
  consisting of $\ell$ consecutive steps of this traversal. Each
  contains $\ell+1$ consecutive vertex positions. By double counting,
  the average number of original-leaf occurrences in such a segment is
  \[
    \frac{m(\ell+1)}{2\tau}
    =\frac{m\lceil b/2\rceil}{2\tau}
    \ge\frac{mb}{4\tau}
    \ge\frac{b}{4\gamma}.
  \]
  Thus some segment visits at least $b/(4\gamma)$ distinct original
  leaves. The union of the edges traversed by this segment is a
  connected subtree $T'$ containing those leaves and having at most
  $\ell<b/2$ edges, as required. (When $\ell=0$, use the single
  visited vertex, which is an original leaf.)
\end{proof}

\section{Acknowledgment and AI disclosure.}
The author thanks Prof.~Michael Ben-Or for discussions, supervision, and support. ChatGPT assisted with editing, proof checking, and brainstorming. The mathematical ideas and responsibility for the paper remain with the author.

\printbibliography
\end{document}